\documentclass[10pt]{article}

\usepackage{fullpage}
\usepackage{setspace}
\usepackage{parskip}
\usepackage{titlesec}
\usepackage{xcolor}
\usepackage{verbatim}
\usepackage{breakcites}
\usepackage{hyphenat}
\usepackage[version=4]{mhchem}
\usepackage{times}
\usepackage{graphicx}
\usepackage{multirow}
\usepackage{array}
\usepackage{scicite}
\usepackage{color,soul}
\usepackage{listings}

\usepackage{times}

\PassOptionsToPackage{hyphens}{url}
\usepackage[colorlinks=true,urlcolor=blue,
  citecolor=blue,anchorcolor=blue]{hyperref}
\usepackage{etoolbox}

\titlespacing{\section}{0pt}{*3}{*1}
\titlespacing{\subsection}{0pt}{*2}{*0.5}
\titlespacing{\subsubsection}{0pt}{*1.5}{0pt}

\usepackage{authblk}

\usepackage[ngerman,english]{babel}
\usepackage{graphicx}
\usepackage{amsfonts,amsmath,amssymb}
\usepackage{float}
\usepackage{mhchem}
\usepackage{times}

\begin{document}

\title{Classical computational simulation of the FeMo-cofactor model to chemical accuracy and its implications}

\author[1,2]{Huanchen Zhai}
\author[1,3]{Chenghan Li}
\author[1,3]{Xing Zhang}
\author[4]{Zhendong Li}
\author[1,5]{Seunghoon Lee}
\author[1,3]{Garnet Kin-Lic Chan \thanks{gkc1000@gmail.com}}

\affil[1]{Division of Chemistry and Chemical Engineering, California Institute of Technology, Pasadena, California 91125, USA}
\affil[2]{Initiative for Computational Catalysis, Flatiron Institute, 160 Fifth Avenue, New York 10010, New York, USA}
\affil[3]{Marcus Center for Theoretical Chemistry, California Institute of Technology, Pasadena, California 91125, USA}
\affil[4]{Key Laboratory of Theoretical and Computational Photochemistry, Ministry of Education, College of Chemistry, Beijing Normal University, Beijing 100875, China}
\affil[5]{Department of Chemistry, Seoul National University, Seoul 08826, Republic of Korea}

\vspace{-1em}
\date{}

\begingroup
\let\center\flushleft
\let\endcenter\endflushleft
\maketitle
\endgroup

\doublespacing

\newenvironment{sciabstract}{%
\begin{quote} \bf}
{\end{quote}}

\selectlanguage{english}

\begin{sciabstract}
   We use classical computational methods to estimate the ground-state energy to chemical accuracy in a model of the FeMo-cofactor of nitrogenase which is widely studied as a target of quantum computing. Our result relies on the insight that the ground-state problem can be characterized as one of ranking many competing, but largely simple, states. This allows a combination of systematic high-order coupled cluster and density matrix renormalization group calculations together with an extrapolation protocol to obtain an accurate energy. Within the model we identify several spin isomer candidates for the ground-state that are degenerate to chemical accuracy. Beyond this model, we characterize the impact of additional electronic excitations and the cluster and protein geometric fluctuations on the low-lying electronic landscape. We find that many features of the landscape are retained in more detailed representations of nitrogenase, which points to the complexity of spectroscopic interpretations of the electronic structure of the FeMo-cofactor.
\end{sciabstract}

\section*{Introduction}

The main source of reduced nitrogen for living things comes from nitrogenase,
which converts \ce{N2} to \ce{NH3} at the
FeMo-cofactor (FeMo-co)~\cite{beinert1997iron,tanifuji2020metal}.
Because of its role in supporting life, the uncertainty surrounding the catalytic cycle, and its  compositional richness with eight transition metal ions, FeMo-co has fascinated scientists for decades. After much effort~\cite{kirn1992crystallographic,einsle2002nitrogenase,lee2003interstitial}, the complete atomic structure was resolved in Refs.~\cite{spatzal2011evidence,lancaster2011x}.
However, its \emph{electronic} structure, central to reactivity, remains under intense debate.

FeMo-co's complexity, arising from many unpaired electrons, 
has led to suggestions that it lies beyond the reach of classical computational simulation. Consequently, there has been much interest in the potential of quantum algorithms to compute its electronic structure.
Estimating the cost for a quantum computer to obtain the ground-state energy to chemical accuracy ($\sim 1$ kcal/mol) within one or more FeMo-co models~\cite{reiher2017elucidating,li2019electronic-femoco} is a common benchmark of quantum algorithms in quantum chemistry, with numerous resource estimates in the literature~\cite{wan2022randomized,berry2019qubitization,luo2025efficient,low2025fast}.

Here we address how to perform the same task using classical computational methods.
We use a 76 orbital/152 qubit resting state model~\cite{li2019electronic-femoco}, the subject of most quantum resource estimates. Based on insight into the multiple configuration nature of the states, we  devise classical  protocols based on systematic high-order coupled cluster and density matrix renormalization group calculations that yield rigorous or empirical upper bounds to the ground-state energy. Extrapolating these we predict the ground-state energy with an estimated uncertainty on the order of chemical accuracy.
Having performed this long-discussed computational task, we next consider implications beyond the model. We distill a simpler computational procedure which we apply to reveal the electronic landscape in realistic representations of the cofactor. We thus illustrate a path to a precise computational understanding of FeMo-co electronic structure.  

\section*{Results}

\subsection*{FeMo-co models and the structure of the states}

\begin{figure}[!htbp]
\includegraphics[width=0.95\textwidth]
{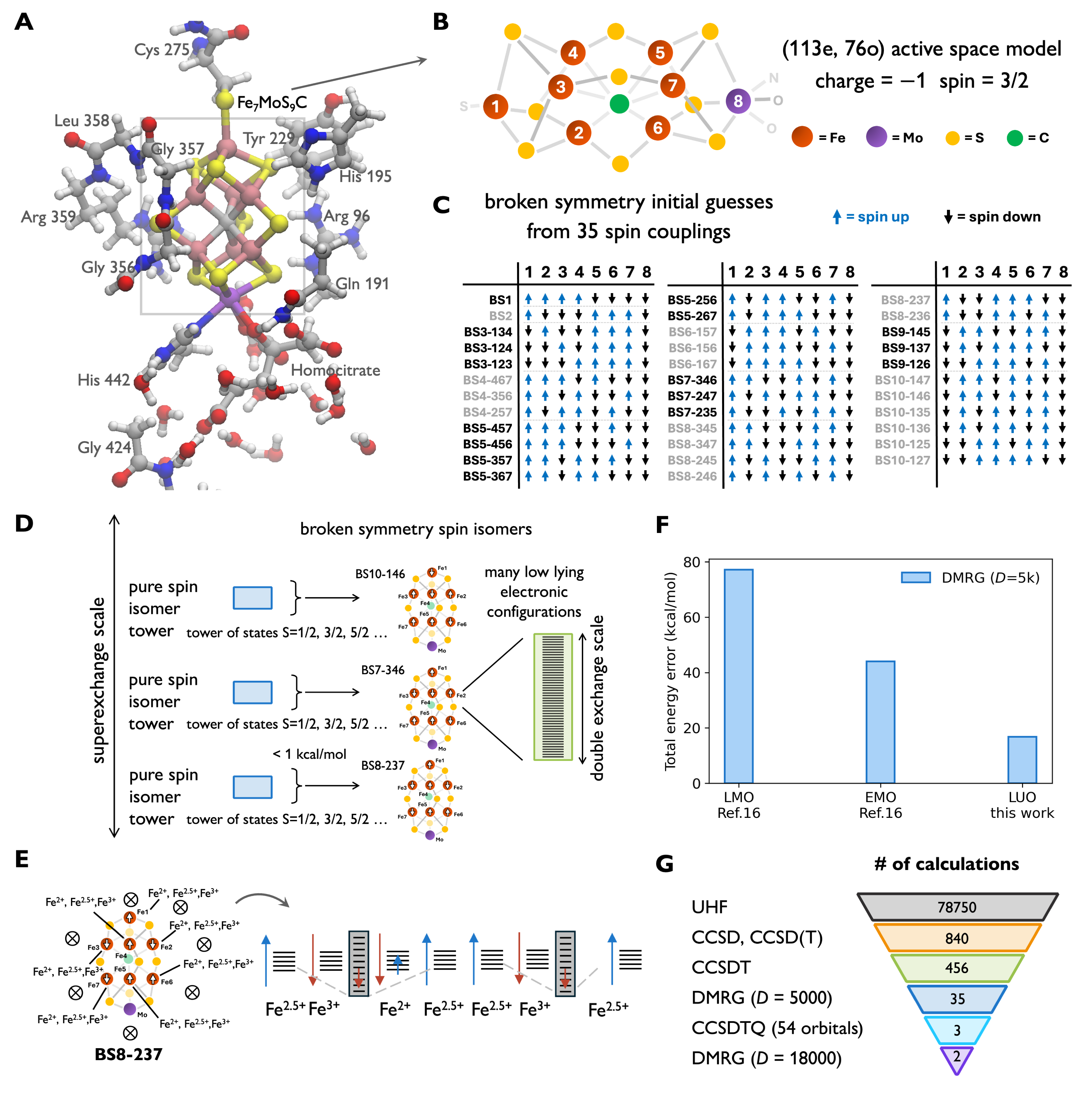}
\centering
\caption{\textbf{Structure, active space model and states of FeMo-co.}
\textbf{A} Geometry of the FeMo-co core and surroundings, built from PDB 3U7Q~\cite{spatzal2011evidence,li2019electronic-femoco}). \textbf{B} Active space model with 113 electrons and 76 spatial orbitals from 22 atoms~\cite{li2019electronic-femoco},
with transition metals labeled by numbers (1-7 correspond to the crystallographic Fe nomenclature). (Note that the model contains more atoms as shown in \textbf{A}, this figure only shows atoms where the active space orbitals are located). \textbf{C} 35 broken symmetry initial guesses representing different spin couplings. 
\textbf{D} Schematic showing the origin of broken symmetry isomers from a tower of pure spin isomers; each broken symmetry isomer further splits due to electron delocalization on the superexchange energy scale. Note that the spin gap in each tower of states is expected to be $< 1$ kcal/mol.
\textbf{E} Schematic showing the interaction between potential electron delocalization and spin coupling in spin isomer BS8-237.
\textbf{F} DMRG calculations with bond dimension 5000 using localized restricted spatial orbitals (LMO), entanglement minimized restricted spatial orbitals (EMO)~\cite{li2025entanglement}, and using localized unrestricted orbitals (LUO, this work). The estimated ground state total energy $-22140.409107$ Hartrees is used as the energy reference.
\textbf{G} Ranking and filtering of low-energy broken-symmetry configurations at different levels of theory.
}\label{fig:1}
\end{figure}

\paragraph{Structure and active space model.} The atomic structure of the FeMo-cofactor is shown in Fig.~\ref{fig:1}A with 
a core structure of $\ce{Fe7MoS9C}$.
We consider only the $E_0$ resting state, where
spectroscopic evidence supports
a cluster total charge of $-1$~\cite{van2020spectroscopy}, and a ground-state with  $S=3/2$~\cite{munck1975nitrogenase}. 

The full structure we use is based on the crystal cutout in Ref.~\cite{bjornsson2017revisiting} and contains the cofactor, several protein residues, and nearby crystal water (Fig.~\ref{fig:1}A, ~\ref{fig:5}A). We study two kinds of electronic models of this structure. 
The first uses
the 113 electron, 76 orbital active space (Fig.~\ref{fig:1}B) of Li, Li, Dattani, Umrigar, and Chan (LLDUC)~\cite{li2019electronic-femoco} 
with a continuum solvation treatment of the remaining protein environment. The active space 
spans the valence orbitals of the atoms in the FeMo-co core, i.e. Fe~3d, Mo~4d, S~3p, and C~2s, 2p orbitals and some ligand 2p orbitals.
We focus on this model because it has commonly served as the Hamiltonian for resource estimates for a quantum computer to compute the ground-state to ``chemical accuracy'' [1.6 milliHartrees (mH) or 1 kcal/mol]~\cite{wan2022randomized,berry2019qubitization,luo2025efficient,low2025fast}. Although this total energy is not physically measurable, it is a simple way to assess algorithmic cost. Our first goal will be to produce the best classical estimate of the total energy of this model.

The LLDUC model is not a  quantitative model of FeMo-co electronic structure: the cofactor involves orbitals beyond the active space, a protein environment, and the atoms are moving. The second type of model we study extends the LLDUC model to account for these effects: to large active spaces (up to {404} orbitals); and to include protein fluctuations, modeled by molecular dynamics.

\paragraph{Classical spin and electronic configurations of FeMo-co.} FeMo-co supports many low-energy states. One origin is the competing spin configurations. In the formal (II) and (III) oxidation states, the local spins of the Fe ions are $S=2$ or $S=5/2$ respectively, and Mo(III) has been suggested to be $S=1/2$ with an unconventional configuration~\cite{bjornsson2014identification,kowalska2019x}. Within a $1/S$ approximate expansion, the large Fe spins can be approximated as classical~\cite{stanley1966high}, 
and 
classical product states of spins (typically collinear) have been used to enumerate low energy broken symmetry states in density functional theory (DFT) calculations~\cite{lovell2002metal,bjornsson2017revisiting,cao2018influence}, and to seed more accurate wavefunction calculations~\cite{li2017spin,li2019electronic,lee2023evaluating}. Assuming collinear spins (non-collinearity is considered in SM~Sec.~\ref{sec:sm-ghf}) and a $S_z=3/2$ ground-state,
there are 35 \textit{spin} isomers (Fig.~\ref{fig:1}C). These enumerate the arrangement of up and down spins, neglecting the distinction between Fe(II) and Fe(III). Further assuming 3-fold symmetry {through the Fe1-Mo} axis, these reduce to 10 isomer families labeled BS$n$ ($n=1,\cdots, 10$); the 35 spin isomers are labeled BS$n$-$ijk$ where $ijk$ labels the iron sites with the minority spin (see SM~Sec.~\ref{sec:sm-bs-name} for other nomenclature)~\cite{bjornsson2017revisiting}. Each broken-symmetry spin isomer can be understood as arising from mixing a tower of pure-spin states for a given isomer (Fig.~\ref{fig:1}D). 

The next source of complexity is the electronic degrees of freedom, due to the additional electron in Fe(II) versus Fe(III). Neglecting mixed valence configurations, each of the 35 spin isomers has 18 assignments of Fe(II) and Fe(III), and for each Fe(II) there are 5 $d$ orbitals for the extra electron. This generates 78750 electronic configurations, 
and more if one then includes mixed valence configurations (Fig.~\ref{fig:1}E). Navigating this complexity is the key challenge in  the electronic structure of this system.

\paragraph{Prior computational and spectroscopic studies of FeMo-co.} 
There have been many prior computational studies of FeMo-co (reviewed in SM~Sec.~\ref{sec:sm-prior}). Most~\cite{benediktsson2017qm,cao2018influence} have used broken-symmetry DFT based on classical spin configurations. While the DFT total energies are not useful for comparison with the models here, the observed relative energy electronic landscape is relevant. Studies e.g. of protonation energies, or the relative stability of spin isomers, find that these are exquisitely sensitive to the choice of density functional~\cite{cao2019extremely,cao2018influence}. Despite this sensitivity, recent BS-DFT studies have focused on the BS7-235 spin isomer as the ground-state~\cite{cao2018influence,benediktsson2017qm}, {because its optimized DFT geometry has been argued to give the best agreement with the X-ray structure}~\cite{benediktsson2017qm}. 

Experimental studies of the FeMo-cofactor (reviewed in SM~Sec.~\ref{sec:sm-prior-expr}) have used electron paramagnetic resonance~\cite{munck1975nitrogenase}, M\"ossbauer spectroscopy~\cite{yoo2000mossbauer}, and X-ray diffraction analysis,~\cite{spatzal2011evidence,lancaster2011x} but there is only very indirect insight into the FeMo-co ground state. In most cases, it is difficult to resolve which part of the signal comes from which ion/unpaired electron, and one requires electronic structure calculations to deconvolve the signal~\cite{bjornsson2017revisiting}. Some recent constraining evidence comes from  site-selective ${}^{57}$Fe labeling of the FeMo-co cluster, which constrains the Fe1 spin to be aligned with the total spin of the ground-state, and further argues, from a mixed valence assignment of Fe1, that a neighboring iron (Fe2, Fe3, or Fe4) should be spin-aligned with Fe1~\cite{badding2023connecting} (although this relies on a picture of resonance delocalization that is not always satisfied~\cite{henthorn2019localized}, see also SM~Sec.~\ref{sec:sm-model-dimer}). 
Another notable study, involving Se substitution coupled to X-ray absorption, suggested that Fe2, Fe6 are more oxidized and antiferromagnetically coupled~\cite{henthorn2019localized}. Finally, spatially resolved anomalous diffraction refinement has been used to infer iron oxidation states, although the connection of the signal to oxidation state is highly complex~\cite{spatzal2016nitrogenase}.

Beyond DFT calculations, FeS cluster electronic structure within active space models has received much attention~\cite{sharma2014low,li2017spin,montgomery2018strong,li2019electronic,li2021expressibility,xiang2024distributed,reiher2017elucidating,brabec2021massively,lee2023evaluating,menczer2024parallel,li2025entanglement}. Such work, often using the density matrix renormalization group (DMRG)~\cite{white1992density}, has highlighted the richness of the electronic landscape in the iron-cubane and P-cluster~\cite{sharma2014low,li2017spin,li2019electronic,li2021expressibility,xiang2024distributed}. However, in the FeMo-co LLDUC model (e.g. Ref.~\cite{lee2023evaluating}) such calculations have been less accurate and fallen far short of the 1 kcal/mol accuracy of the quantum resource estimation task~\cite{reiher2017elucidating,brabec2021massively,lee2023evaluating,menczer2024parallel,li2025entanglement}. For example, Supp. Fig.~8 in Ref.~\cite{lee2023evaluating} estimates a residual error in the variational DMRG energy of $\sim 100$ milliHartrees (63~kcal/mol), far too large to extrapolate the energy to chemical accuracy.

\paragraph{Classical electronic structure methods.} 
The main methods we use here are the DMRG and coupled cluster (CC) theory. We combine them to obtain accurate estimates of the energy and observables. DMRG is a variational method controlled by a parameter (bond dimension) $D$, and the calculation can be made arbitrarily accurate by increasing $D$~\cite{white1992density,white1993density,chan2002highly,mitrushenkov2001quantum,wouters2014density,keller2015efficient,baiardi2020density,brabec2021massively,verstraete2023density}. CC theory provides a non-variational estimate of the ground-state energy using excitations from a starting mean-field state (singles, doubles, and so on, giving CCSD, CCSDT, CCSDTQ, etc.); as the excitation level increases the method becomes exact~\cite{bartlett2007coupled,shavitt2009many}. The resource limiting the accuracy is different in the two methods: for DMRG, it is the entanglement, while for CC it is the closeness to the initial mean-field configuration. Consensus of estimates from the two methods is thus a stringent consistency test.

In this work, we have developed new or customized our existing DMRG and CC implementations. For example, for the CC calculations 
we developed a new memory efficient, parallel, implementation of arbitrary order CC (SM~Sec.~\ref{sec:sm-theory-cc}). However, the main methodological innovation is the special protocol we employ in the calculations, rooted in the nature of the FeMo-co states (details in SM~Sec.~\ref{sec:sm-ref-dep}), which we now discuss.

\paragraph{Qualitative nature of FeMo-co electronic states} 
We perform calculations that
break the $S^2$ spin symmetry of the system. This reduces the amount of entanglement in the modeled states and increases their mean-field character. Symmetry breaking is artifactual in a finite system, but in FeMo-co the many coupled spins generate a small spin gap $\Delta$. The relevant gap is that within the tower of pure-spin states of a \emph{fixed} spin isomer~\cite{tasaki2019long}  (expected to be $<1$~kcal/mol~\cite{li2019electronic}, for additional discussion see SM Sec.~\ref{sec:sm-hierarchy} and~\ref{sec:sm-model-heis}), not the gap between spin isomers, and the pattern of spins in the broken symmetry solution reflects the  pattern of spin-correlations in the tower from which it is derived, see Fig.~\ref{fig:1}D.
If our desired energy precision $> \Delta$, then the symmetry broken solution can represent the ground-state to such precision (see SM~Sec.~\ref{sec:sm-spin} and~\ref{sec:sm-model-heis}), and reflect the true correlations in the state.
The caveat is that (in large systems)
classical calculations initialized as one symmetry broken solution will not (due to ansatz restrictions) ``tunnel'' into another. Thus, we must start from different initial states which then either converge to the (symmetry broken) ground-state or a low-lying excited state.

Beyond spin-excitations, 
we must also carefully consider the orbital/charge degrees of freedom (Fig.~\ref{fig:1}E). The spin and charge are strongly coupled: in a classical unrestricted mean-field, orbitals for spin-up and spin-down  delocalize differently because of the Pauli principle and double exchange~\cite{noodleman1995orbital}.
Consequently, different broken-symmetry states have very different spin-up and spin-down mean-field orbitals.

Prior DMRG calculations on FeS clusters have always used a single (``restricted'') orbital basis, where the spin-charge correlation  must be entirely captured by the DMRG wavefunction.
However, if we allow different orbitals for different spins (see Fig.~\ref{fig:1}F) we find a large improvement in the DMRG energy at a given bond dimension, surpassing all previous choices of orbitals, as much of the spin-charge correlation is captured in the basis itself. Using this unusual ``unrestricted orbital'' DMRG (denoted as UDMRG) formed the basis for progress in this work. Correspondingly, we perform CC with spin-unrestricted HF reference (denoted as UCC).

Even after spin symmetry breaking, the many classical electronic configurations (e.g. 78750 for localized Fe(II) and Fe(III) charges) poses a challenge. For example, if one starts from one local electronic configuration as a reference state for UCC, and the actual ground state is another, the reference is orthogonal to the ground state and UCC will fail (although UDMRG does not suffer from this problem, see SM~Sec.~\ref{sec:sm-ref-dep}). Similarly, the
possibility for delocalized versus localized electrons creates a multiconfigurational problem:
even in a simplified model where delocalized electrons only explore one valence orbital per Fe, the squared overlap of a reference mean-field state with only local Fe charges with the true ground-state decreases by $\sim \frac{1}{2}$ for every delocalized  pair in the exact state. Thus UCC calculations that start from the wrong initial electronic configuration and charge distribution cannot reliably describe the ground-state. However, this multiconfigurational nature is arguably trivial, because it can be removed by properly choosing the initial orbitals. The challenge is one of \emph{multiple competing configurations}, rather than an intrinsic multiconfigurational problem.

The ranking of electronic configurations and the balance between localized and delocalized configurations is intimately tied to the level of electron correlation.
To obtain a reliable UCC description we perform hundreds of UCCSD calculations to rank low energy initial UHF configurations,
then further rank these downselected configurations using higher levels of UCC (up to quadruples, and even beyond, in the LLDUC model) and UDMRG (see SM Sec.~\ref{sec:sm-theory-dmrg} and SM Sec.~\ref{sec:sm-theory-cc}). This filtering funnel is illustrated in Fig.~\ref{fig:1}G. Recognizing that the multiconfigurational problem reduces to a ranking problem was the key to successfully using UCC to describe FeMo-co.

\subsection*{The electronic landscape of the LLDUC model of the FeMo-cofactor}

\begin{figure}[!htbp]
\includegraphics[width=\textwidth]
{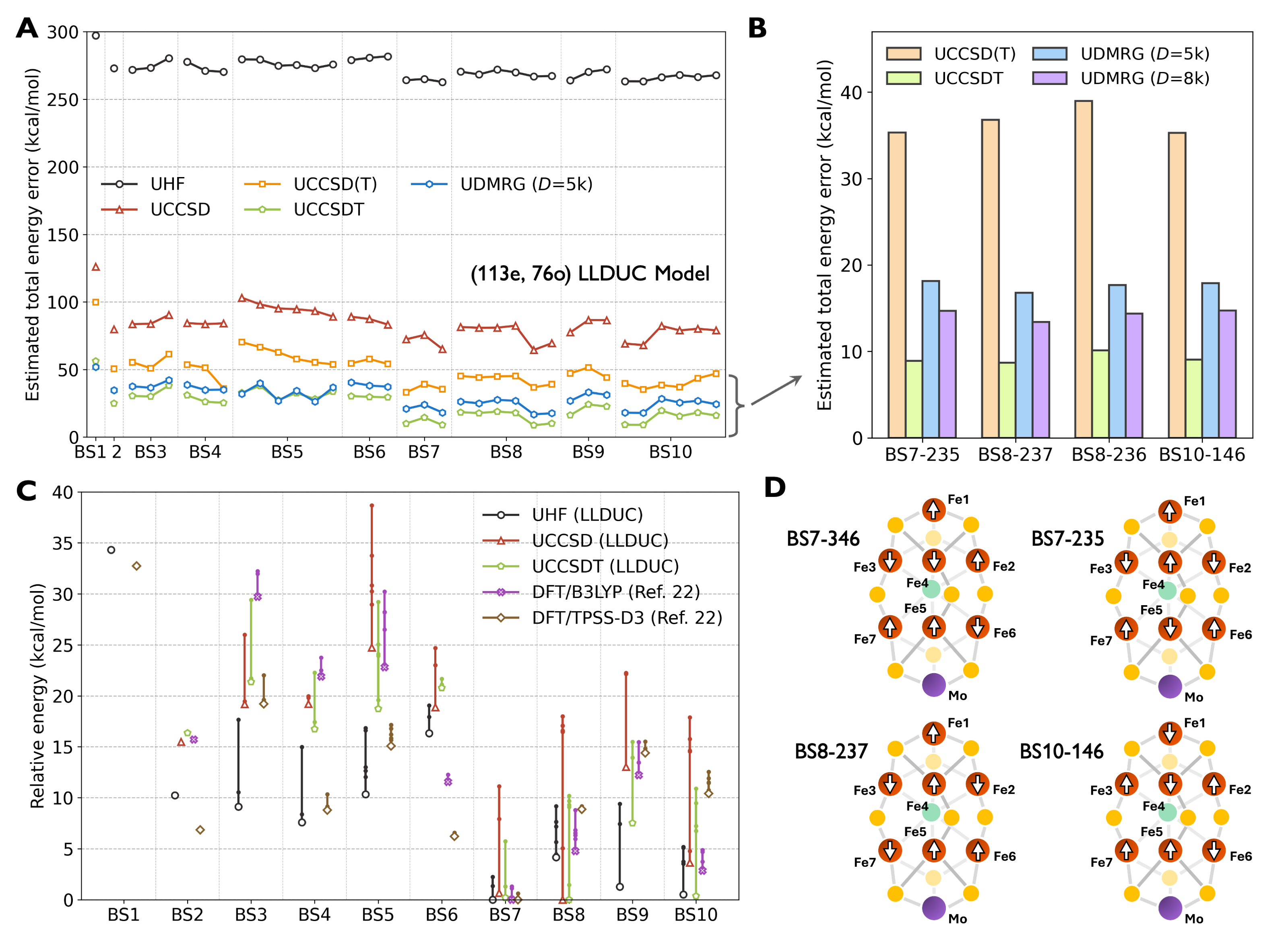}
\centering
\caption{\textbf{Electronic landscape of FeMo-co active space model.}
\textbf{A,B} Energies of different spin isomers computed using UHF, UCCSD, UCCSD(T), UCCSDT, UDMRG with bond dimension $D=5000$ and $8000$, for the (113e, 76o) LLDUC model. The estimated ground state total energy $-22140.409107$ Hartrees is used as the energy reference.
\textbf{C} Relative energies computed using UHF, UCCSD, and UCCSDT for the LLDUC model, compared with DFT results with B3LYP/def2-TZVPD and TPSS-D3/def2-TZVPD from Ref.~\cite{cao2018influence}. The minimum energy across all isomers for each theory is used as the energy reference.
The lowest energy in each BS family is shown as a larger symbol.
\textbf{D} Noodleman's schematic for the BS states BS7-346, BS7-235, BS8-237, and BS10-146 showing the spin coupling~\cite{lovell2001femo,cao2018influence}.
}
\label{fig:2}
\end{figure}

We have carried out unrestricted coupled cluster (UCC) and unrestricted DMRG (UDMRG) calculations using the filtering procedure in SM~Sec.~\ref{sec:sm-theory}.
In Fig.~\ref{fig:2}A we show the energy landscape of the 35 spin isomers in the LLDUC model with unrestricted Hartree-Fock (UHF), UCC up to triples, and UDMRG  with a moderate bond dimension $D=5000$.

We find convergent trends as the correlation level increases. At the highest levels of correlation in the figure, UCCSDT and UDMRG ($D=5000$), the relative orderings between BS families and within them are consistent.
(T) shows a moderately strong reference dependence reflecting the importance of self-consistent UCC amplitudes, see SM~Sec.~\ref{sec:sm-ref-dep}. Additionally, we note that the relative energy of the spin isomers within each broken-symmetry family can be different, mainly due to deviations from $C_3$ symmetry in the geometry, for example, the energy of BS7-247 is much higher than that of the other two spin isomers in the BS7 family.
Nonetheless, all the states correspond to (largely) single reference states (for various diagnostics, see SM~Sec.~\ref{sec:sm-ref-dep} and SM~Sec.~\ref{sec:sm-det-overlap}), confirming our qualitative insight into the states and the effectiveness of the filtering procedure.

It is assumed, and to some extent observed in DFT calculations, that within each BS$n$ class, spin isomers are similar in energy~\cite{bjornsson2017revisiting,cao2018influence}.
We find the variation between spin isomers for the same BS$n$ class to be comparable to the variation between BS$n$ classes. The quantitative variations are sensitive to the level of correlation (see Fig.~\ref{fig:2}C).

In the LLDUC model, while BS7 is the lowest energy class at the mean-field level (similar to as found in prior DFT studies), at the correlated wavefunction level, BS7, BS8 and BS10 are all low in energy, with BS8 becoming almost degenerate with (or lower in energy than)  BS7 as the correlation level increases. BS7 maximizes the number of AFM couplings across Fe-S-Fe bonds, which is sometimes used to justify its low energy in DFT studies~\cite{bjornsson2017revisiting,cao2018influence}, but BS8 and BS10 do not (Fig.~\ref{fig:2}D). The qualitative origin of the relative energies thus requires balancing both exchange and double exchange, where double exchange modifies the charge distribution and delocalization across the cluster. The latter is a large energy scale which we note has strong functional dependence in DFT~\cite{cao2018influence}. 

While there are several spin isomers of BS7, BS8, BS10 that are very close in energy (SM Sec.~\ref{sec:sm-fm-ucc-udmrg-ener}), we next focus on BS7-235 and BS8-237 as candidates for the LLDUC ground-state.

\subsection*{Estimating the ground-state energy of the LLDUC model to chemical accuracy}

\begin{figure}[!htbp]
\includegraphics[width=\textwidth]
{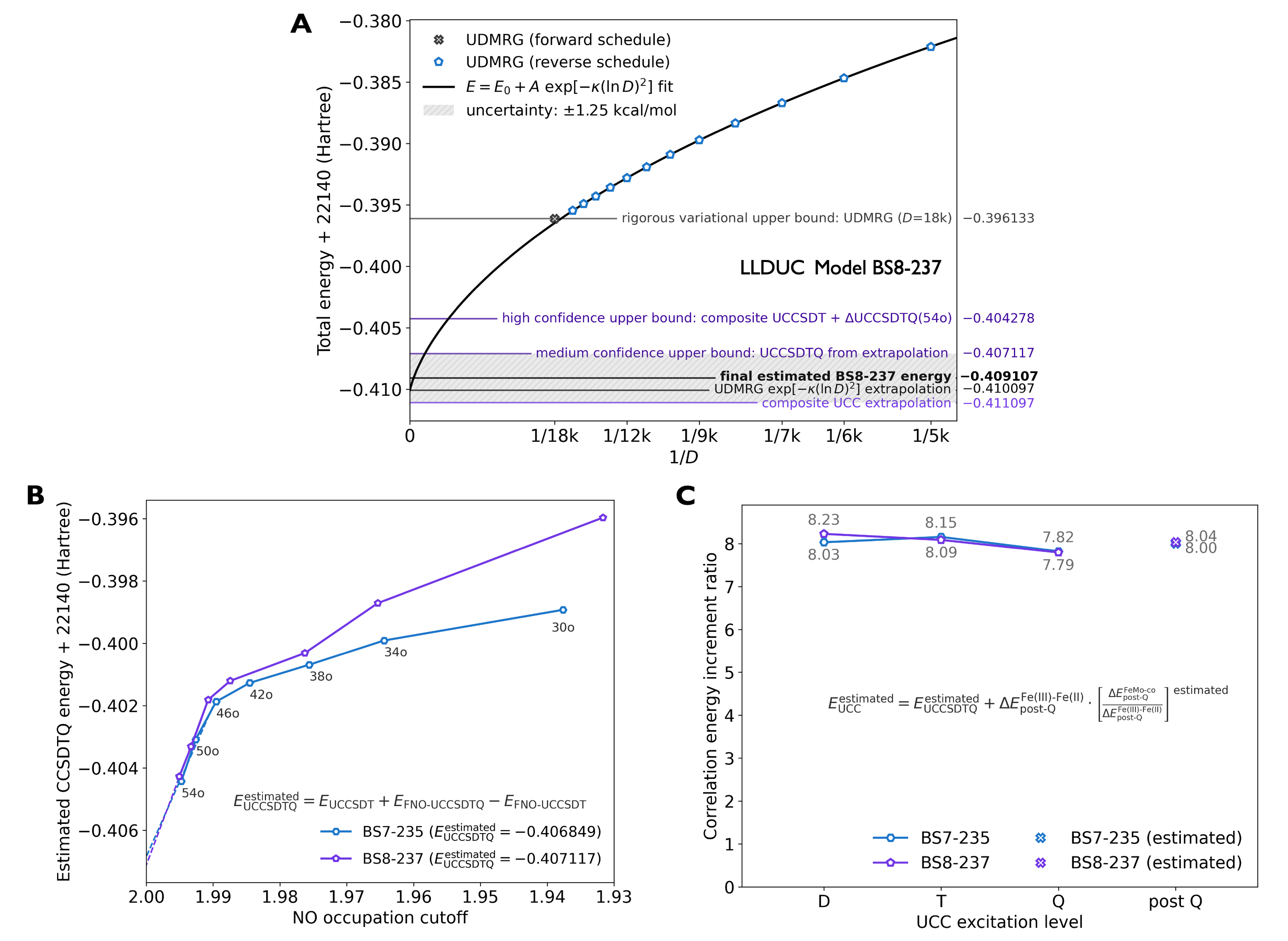}
\centering
\caption{\textbf{Ground-state energy estimation of the FeMo-co active space model.}
\textbf{A} The $\Delta E \sim \exp[-\kappa(\log D)^2]$ fitting of UDMRG energies for the BS8-237 state in the LLDUC model. Note that the $D=18,000$ UDMRG energy from forward schedule is not fully converged, and is not included in the extrapolation.
\textbf{B} Composite UCCSDT and FNO-UCCSDTQ energies and extrapolation (using linear fitting for 46o, 50o, and 54o energies) with respect to UCCSD natural orbital occupation cutoff, for the BS7-235 and BS8-237 states in the LLDUC model.
\textbf{C} Estimation of the exact UCC energy using correlation energy increment ratios between the Fe(III)-Fe(II) dimer
and FeMo-co LLDUC model, for the BS7-235 and BS8-237 states.
}
\label{fig:3}
\end{figure}

To obtain a high accuracy estimate of the ground-state energy of FeMo-co in the LLDUC model, we push the level of the coupled cluster calculations up to quadruples (UCCSDTQ) with a limited estimate of pentuples and beyond, and also extend the UDMRG calculations to a bond dimension of $D=18000$, for BS7-235 and BS8-237. We then separately extrapolate the UDMRG and UCC data to provide multiple estimates of the exact energy as shown in  Fig.~\ref{fig:3}A (total energies are reported here in Hartrees for easier reproducibility). 

Our first prediction of the ground-state energy, based on BS8-237, is $-22140.4101$ Hartrees, from extrapolating the UDMRG energy.
UDMRG energies can typically be extrapolated in one of a few different ways: using the discarded ``weight'' during the variational calculation (which estimates the remaining error in the wavefunction), or using a function of the bond dimension $\Delta E \sim \exp[-\kappa(\log D)^2]$ (see SM~Sec.~\ref{sec:sm-theory-dmrg-extra} and SM~Sec.~\ref{sec:sm-fe4} for an analysis in the iron cubane). Here we use the latter approach, whose fit is shown in Fig.~\ref{fig:3}A.

The UDMRG energy at the largest bond dimension $D=18000$ is a strict variational upper bound, about 8.76 kcal/mol above the estimated ground-state energy. Another upper bound is obtained from UCC calculations.
Using our resources we could not carry out UCCSDTQ calculations in the full LLDUC active space, but instead used a sequence of smaller active subspaces up to 54 orbitals [chosen using a natural orbital cutoff, i.e. frozen natural orbital (FNO)-UCCSDTQ], together with a composite correction from the UCCSDT energy in the full space, to approximate the UCCSDTQ full active space energy
(SM~Sec.~\ref{sec:sm-fno}).
The convergence of this with respect to the cutoff is shown in Fig.~\ref{fig:3}B, where it clearly converges to the full UCCSDTQ from above. Our analysis of post-Q (pentuples  and beyond) contributions from these FNO calculations and from other FeS clusters show it is always either close to zero (i.e. $<$1 kcal/mol) or negative (e.g. SM~Sec.~\ref{sec:sm-fe4}).
Thus the UCCSDTQ composite energy from the largest subspace, namely $-22140.4043$ Hartrees, is (to an uncertainty within chemical accuracy) an upper bound with high confidence. 
We can further extrapolate the FNO-UCCSDTQ results with respect to the cutoff to estimate the full UCCSDTQ energy. Here, 
there is strong curvature (connected to the restoration of spin symmetry, see SM~Sec.~\ref{sec:sm-fno}) that 
complicates the extrapolation. Using the last three points (Fig.~\ref{fig:3}B) for the natural orbital extrapolation yields $-22140.4071$~Hartrees. This is the tightest upper bound we produce, but with medium confidence, due to the (small amount of) extrapolation. 

To predict an exact energy from UCC requires a precise value of the post-Q contributions. We find that in FeMo-co, the D, T, and Q contributions are roughly in a constant ratio $\sim 8$ (Fig.~\ref{fig:3}C) with the corresponding contributions from a mixed-valence Fe-S dimer that serves as a correlation model (SM Sec.~\ref{sec:sm-comp}). Scaling the post-Q correlation of the dimer accordingly yields a FeMo-co post-Q increment of $-0.0040$~Hartrees, for a predicted ground-state energy of $-22140.4111$~Hartrees.

The UCC and UDMRG predicted exact energies are in good agreement, 
and as a consensus value, we can take the average, obtaining 
$E_0 = -22140.4106 \pm 0.0005$ Hartrees (or $\pm 0.31$ kcal/mol). However, as there is no formal theory of these extrapolations, we can also take the more conservative view that the energy lies somewhere between the UCC empirical upper bound and the lowest extrapolated energy, i.e. $[-22140.4071,-22140.4111]$~Hartrees, or 
$E_0=-22140.4091 \pm 0.0020$~Hartrees (or $\pm 1.25$~kcal/mol).
Using a similar procedure (SM~Sec.~\ref{sec:sm-dmrg-extra}) for BS7-C, we obtain
$E_0 = -22140.4089 \pm 0.0021$ Hartrees (or $\pm 1.29$ kcal/mol).
Considering the above estimates and uncertainties, we thus conclude that we predict the ground-state energy with an error $\sim$ chemical accuracy for both spin isomers. Within this accuracy, we find BS7-235 and BS8-237 (and potentially BS8-236, see SM~Sec.~\ref{sec:sm-fno}) to be degenerate.

\subsection*{Cost of classical methods and quantum advantage}

\begin{figure}[!htbp]
\includegraphics[width=\textwidth]
{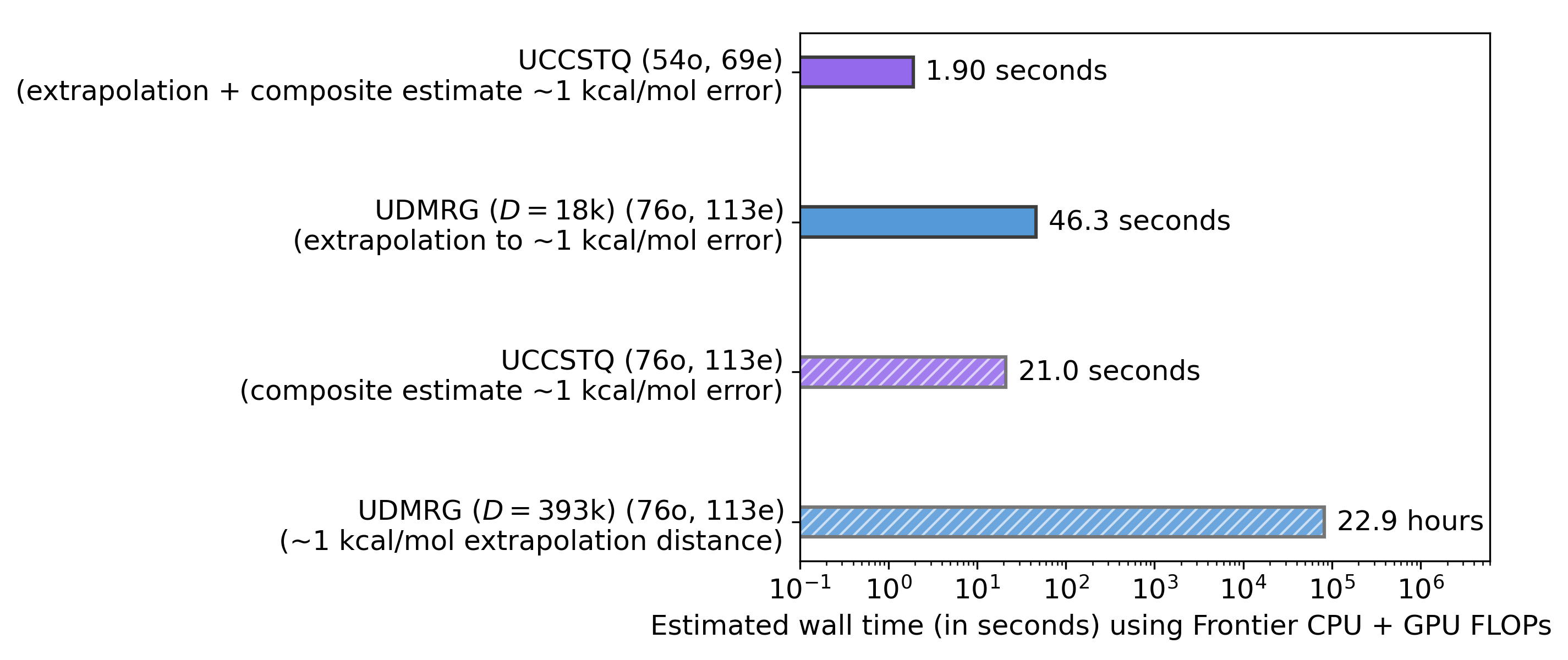}
\centering
\caption{\textbf{Computational cost and scaling of classical methods.}
Hypothetical wall time cost of UCCSDTQ and UDMRG for the LLDUC model assuming efficient use of the total combined CPU and GPU resources on the Frontier supercomputer.
}
\label{fig:4}
\end{figure}

Given this classical estimation of the FeMo-co ground-state energy to chemical accuracy, we now place it in the context of quantum resource estimates. Many discussions of quantum versus classical costs compare the hypothetical runtime of the classical algorithm on a leading supercomputer to the hypothetical quantum execution time. 
While we have not run our calculations using a large supercomputer, we follow the analogous procedure to contextualize our costs. These estimates do not account for important practical factors of implementation, but as these are (usually) ignored in other estimates, we also do not consider them here. 

The most expensive steps in the classical energy estimate arise from generating the data to extrapolate the UCCSDTQ composite upper bound and the UDMRG energies. The largest UCCSDTQ calculation took $\sim 400$ CPU core hours per iteration, and 100 iterations to converge to $10^{-8}$ Hartrees, totaling 40,000 CPU core hours. For the variational UDMRG simulations, a single $D=18,000$ sweep took 317,440 CPU core hours (aggregated over multiple nodes), and we performed two sweeps. Including the cost of extrapolation using the 12 points at $D=16000, 15000, \cdots$, and $5000$, the total UDMRG cost was then 2,768,994 CPU core hours. 

We can also estimate the cost of higher accuracy calculations. A variational UDMRG estimate to an accuracy of 1~kcal/mol {with respect to the extrapolated UDMRG energy requires $D \sim 393,000$. Using the time-scaling $T \sim D^{3.05}$ (see SM~Sec.~\ref{sec:sm-scaling}) this requires $4.08 \times 10^9$ CPU core hours (for two sweeps). 
Similarly, a UCCSDTQ calculation in the full LLDUC space, using the leading scaling (allowing frozen orbitals to be partially occupied, with the observed occupied orbital scaling $T \sim N_{\text{occ}}^{6.58}$, see SM~Sec.~\ref{sec:sm-scaling}), would use $8.94 \times 10^5$ CPU core hours (assuming 100 iterations). 

We now translate the above into the theoretical ideal cost of running the calculations on the Frontier supercomputer, shown in Fig.~\ref{fig:4}. For this we estimated the number of floating point operations (FLOP) associated with the above calculations, and divided them by the number of FLOP achievable per second (from combined CPU + GPU resources) on the Frontier supercomputer (1.102 exaFLOPS). Using this metric, the calculations take from 1.90 seconds for the largest UCCSDTQ calculation in this work, and 46.3 seconds for the UDMRG extrapolations, to 21.0 seconds for the estimated full space UCCSDTQ calculation (assuming no memory limitations and no usage of index-permutation symmetry) and 22.9 hours for two sweeps in a variational UDMRG calculation with $D=393,000$. These can be compared to a recent quantum algorithm resource estimate of 8.6 hours using sum-of-squares spectral amplification and quantum phase estimation to reach 1 kcal/mol accuracy, assuming $O(1)$ overlap with the initial state~\cite{low2025fast}. In SM~Sec.~\ref{sec:sm-det-overlap} we show that the overlap of the filtered broken symmetry determinant with the $D=18,000$ UDMRG state is as large as $0.4468$; this should be interpreted as the overlap with the manifold of eigenstates within 1 kcal/mol of the ground-state.
This comparison is a snapshot in time:  the classical resource cost will certainly decrease in the future, as will the cost of the quantum algorithms. 

\subsection*{Beyond the LLDUC model and comparisons to spectroscopy}

The classical solution of the LLDUC model inspires hope that a precise computational understanding of FeMo-co is possible. 
However, the small scale of the energy landscape raises questions regarding the robustness of our findings when the model limitations are lifted. The two main limitations are (i) lack of electron correlation outside of the 76 orbital active space (dynamical correlation effects), and (ii) geometric fluctuations of the cluster and protein environment. 

\begin{figure}[!htbp]
\includegraphics[width=\textwidth]
{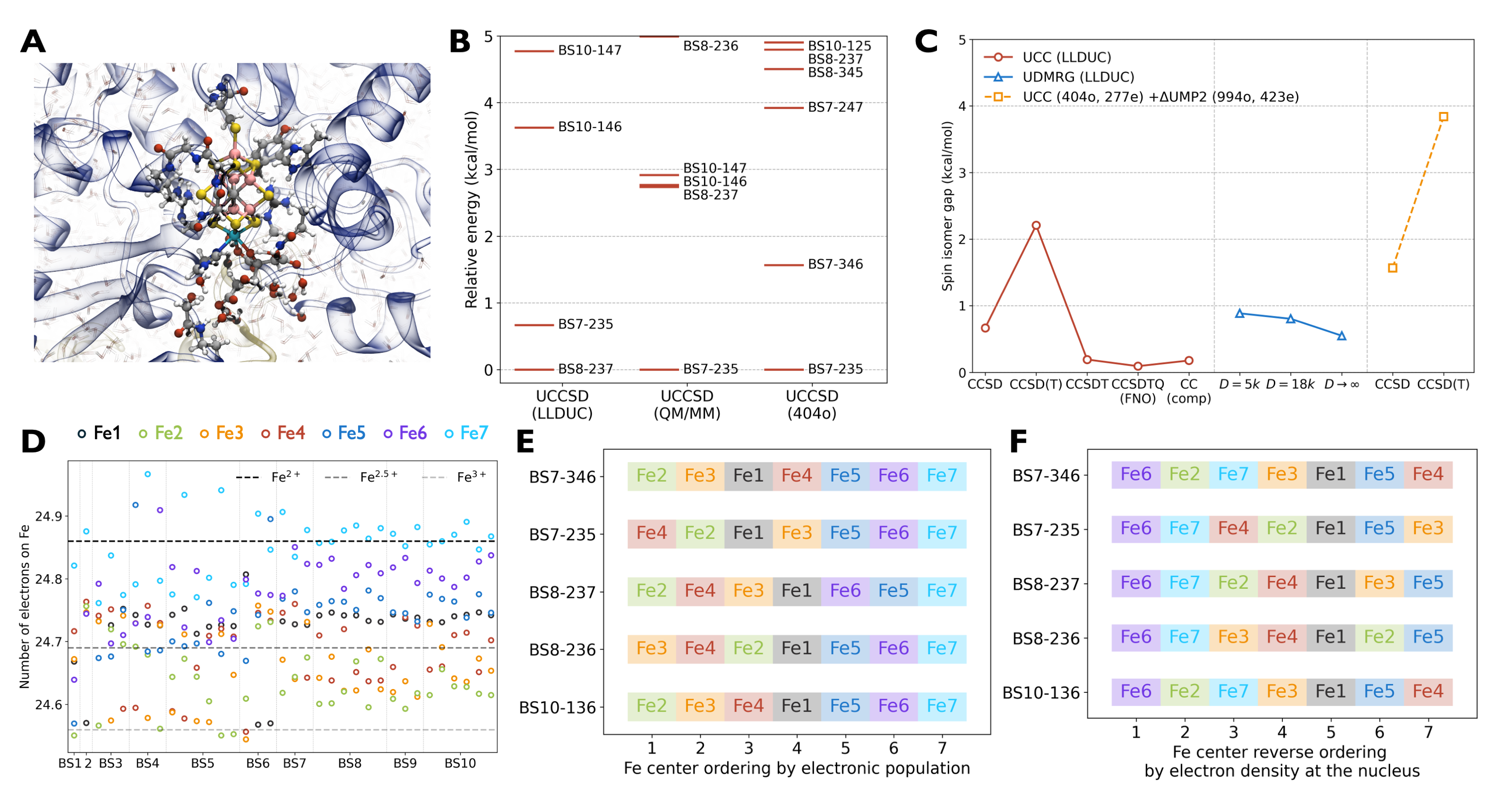}
\centering
\caption{\textbf{Beyond the LLDUC model and Fe oxidation states.}
\textbf{A} Geometry used for the QM/MM simulation (built from PDB 3U7Q~\cite{spatzal2011evidence}). QM and MM atoms are shown as balls and transparent colors, respectively.
\textbf{B} UCCSD relative energies computed using the LLDUC model (with continuum solvation), the 76-orbital active space derived from the QM/MM model with the MD averaged potential, and the larger (404o, 277e) active space with UMP2 composite correction, for low-energy spin isomers.
\textbf{C} Energy gap between two lowest-energy spin isomers computed using different levels of UCC and UDMRG theories, for the LLDUC and the larger (404o, 277e) active space models.
\textbf{D} Number of electrons on Fe from UCCSD in the (285o, 217e) active space using meta-L\"owdin populations without Rydberg contributions. The Fe${}^{3+}$, Fe${}^{2+}$, and Fe${}^{2.5+}$ scales were computed following the procedure in SM~Sec.~\ref{sec:sm-ox-calib}.
\textbf{E} Ordering of Fe centers (oxidized to reduced) in FeMo-co by number of electrons on Fe from UCCSD in the 404-orbital active space, using meta-L\"owdin populations without Rydberg contributions, for low-energy spin isomers (see SM~Sec.~\ref{sec:sm-large-act-ox} for more details). 
\textbf{F} Ordering of Fe centers (oxidized to reduced) in FeMo-co by reverse ordering of the electron density at the Fe nucleus from UCCSD in the 404-orbital active space, for low-energy spin isomers see SM~Sec.~\ref{sec:sm-large-act-ox} for more details).
}
\label{fig:5}
\end{figure}

In the LLDUC model, although reaching a precision of 1~kcal/mol required substantial effort,  qualitative features of the landscape could already be obtained by ranking mean-field solutions based on  relatively affordable broken symmetry coupled cluster methods. For example, BS7 and BS8 correctly appear as the lowest energy spin-isomer classes (see Fig.~\ref{fig:2}C) at the UCCSD level (although the degeneracy of BS7-235, BS8-236, BS8-237 expands to a spread of $\sim 5$~kcal/mol, which may be taken as the ordering uncertainty of the method). 
At the UCCSDT level the ordering of the families is consistent with large bond dimension UDMRG (see SM~Sec.~\ref{sec:sm-fm-ucc-udmrg-ener}).
Consequently, to obtain qualitative insights beyond the LLDUC model, we use a simplified protocol, ranking mean-field solutions at the UCCSD level, with additional energies recomputed at the UCCSD(T) level. For properties beyond the energy, we have also verified this protocol against UCCSDT and UDMRG data (see SM~Sec.~\ref{sec:sm-ref-dep}). 

We first consider the effect of the protein environment and its fluctuations (see Fig.~\ref{fig:5}A). Within a QM/MM setup, we included the heterotetramer of the Mo-Fe protein with water solvation, and using a custom force field, ran molecular dynamics  to equilibrate the MM environment (see SM~Sec.~\ref{sec:sm-qmmm}). From the MD trajectories~\cite{li2025accurate}, we then obtained the average electrostatic environment produced by protein fluctuations.  Fig.~\ref{fig:5}B shows the relative energies of the states in the averaged electrostatic potential for the 76 orbital active space, compared to the LLDUC model with its continuum solvation. The low energy landscape is qualitatively unchanged, and in particular,  BS7, BS8, and BS10 remain the lowest energy spin isomer classes (additional discussion of effects from different cluster geometries is in SM~Sec.~\ref{sec:sm-geom}).

We next consider electron correlation beyond the active space. Starting from the geometry (and continuum solvation treatment) of the LLDUC model, we used a Moller-Plesset second order perturbation theory (MP2)~\cite{pople1976theoretical} natural orbital truncation to construct an active space of size (404o, 277e), localized to have overlap with the FeMo-co core. In the (404o, 277e) active space we carried out a full ranking of the Hartree-Fock references using UCCSD energies (as many as 24 references per spin isomer, for a total of 432 references for BS7 to BS10, see SM~Sec.~\ref{sec:sm-large-act}).
Fig.~\ref{fig:5}B shows the UCCSD relative energies (with composite UMP2 corrections for the natural orbital truncation) of the spin isomers, and again we find the
BS7, BS8, and BS10 to be the lowest in energy. 

As summarized earlier, there is strong experimental evidence from Ref.~\cite{badding2023connecting}
that constrains the Fe1 spin to be aligned with the cluster total spin~\cite{badding2023connecting}. This rules out the BS10 family as the ground-state, but is consistent with BS7 or BS8 as the ground-state. BS8 does not appear as a low-energy spin isomer in most prior density functional calculations (SM~Sec.~\ref{sec:sm-prior-dft}), and hence has not been considered in analysing the experimental data.
Fig.~\ref{fig:5}C shows that, similarly to in the LLDUC model, the gap between the lowest spin isomers is small in the 404 orbital active space: $\sim$1.6~kcal/mol at the UCCSD level.
Thus at room temperature or lower, there is a population of different spin isomers, consistent with the unusual temperature dependence of the M\"ossbauer spectra, which has been posited to arise from a population shift~\cite{badding2023connecting}. 
Although BS8 and BS10 lie somewhat higher in energy in the 404 orbital model, taking into account the $\sim$5 kcal/mol ordering uncertainty of UCCSD, we also cannot rule them out from being thermally accessible after a more accurate correlation treatment.
We do not consider the UCCSD(T) relative spin isomer energies to be more reliable than those of UCCSD due to the (T) reference sensitivity (SM~Sec.~\ref{sec:sm-large-act} lists the UCCSD(T) energies). However, at both the UCCSD and UCCSD(T) level in the 404 orbital active space, we find BS7-235 to be the lowest energy isomer.

Our computations also give insight into the charge structure of the ground-state and current experimental interpretations.
Using a reference compound approach to assign oxidation states from the valence population density (SM~Sec.~\ref{sec:sm-ox-calib}), most of the Fe's in FeMo-co have some mixed valence character (Fig.~\ref{fig:5}D), which may relate to the low-energy scale associated with different spin isomers, as the charge distributions are similar.
BS1-BS6 have a wider spread of oxidation states (and thus more localized charges) than BS7-BS10, which, given the low energies of BS7, BS8, and BS10, suggests that the cluster favours delocalized charges. However, different common measures of assigning oxidation state lead to different conclusions. Figs.~\ref{fig:5}E and \ref{fig:5}F compare the ordering of Fe's (from oxidized to reduced) using the electron density at the nucleus (related to the M\"ossbauer isomer shift) and the Fe valence population. These measures are usually assumed (and often found to be) anticorrelated, but  we do not find a good anticorrelation in the FeMo-co cluster (see  SM~Sec.~\ref{sec:sm-ox-calib} for further discussion). What oxidation state we assign  the Fe centers experimentally and computationally depends on which part of the atomic density on the ion we are probing, and this may resolve potentially conflicting experimental deductions (see SM~Sec.~\ref{sec:sm-prior-expr}).

We focus here on using the valence population to assign oxidation state.
Here, broad features of the ordering are reasonably robust across the spin isomers. For example, across the BS7 and BS8 isomers, Fe2, Fe3, Fe4 are  more oxidized, while Fe5, Fe6, Fe7 are more reduced. The Mo oxidation state is found to be close to Mo(III) (see SM Sec.~\ref{sec:sm-mo-ox-calib}). The fluctuation of the Fe oxidation state across the spin isomers also 
yields an energy scale for shuttling charge across the cofactor. For example, in the low-energy BS7-235 and BS8-237 spin isomers, the charge of Fe2 differs by 0.3. This sensitivity  suggests that the spectroscopic assignment of oxidation state can depend on the conditions under which the cofactor is studied. Finally, the energetic accessibility of multiple redox active sites clearly supports the multi-electron functionality of the cofactor. 

\section*{Discussion}

Our work has addressed several aspects of FeMo cofactor electronic structure. From the viewpoint of the difficulty of ground-state simulation,  our work has produced a classical prediction of the ground-state energy to within an estimated chemical accuracy in a model widely used in quantum resource estimates. This enables quantitative assessment of potential quantum advantage, and establishes the difficulty of the FeMo-co active space problem relative to other computational problems, such as factoring, where classical and quantum costs are known.
Our protocol required extrapolations whose uncertainty can be removed in the future using either improved classical or quantum calculations. The cost estimates we provide for such tasks should be viewed as upper bounds, as we expect further algorithm improvements.
Classical methods succeed here because of insight into the structure of the states. It is expected that future quantum heuristics will use these insights also. 

From the viewpoint of the chemistry and spectroscopy of the FeMo-cofactor, our work highlights the dense electronic manifold of Fe-S clusters first identified in the Fe-cubane, and now confirmed across the different clusters of nitrogenase. The small energy scale 
has a clear impact on interpreting spectroscopic measurements as small changes in different experiments may affect the electronic state being measured. At the same time, our work supports the use of broken symmetry theoretical approaches that capture the physical effect of orbital and spin dependent localization, significantly simplifying the computation of the low-energy manifold. 

For the purposes of understanding the catalytic mechanism of nitrogenase, computing the ground-state energy is only the starting point. 
Importantly, our work indicates that simpler single-reference electronic structure methods, which are practical to use in realistic representations of the cofactor, may suffice. This is because the complexity of the low-energy electronic structure lies not in the description of intrinsically multi-configurational effects, but the correct ranking of many competing, but largely single-configuration, states. This thus provides a feasible path to resolve the ambiguities of current density functional theory studies.

Substantial effort has already been directed towards using lower accuracy electronic structure methods in conjunction with QM/MM to elucidate the nitrogenase reaction~\cite{benediktsson2017qm,cao2018protonation}. Building on our protocols, we now have a viable path forwards where accurate and computationally practical quantum data can be combined with machine learning to improve on such studies~\cite{li2025predictive}. This presents the possibility of future high accuracy computational reaction mechanism modeling, and discovery, in the long-standing nitrogenase problem.

\section*{Methods}

\paragraph{Filtering protocol for the LLDUC model.} To obtain a reliable description of the LLDUC model (especially for the UCC methods), we first performed a thorough sampling of UHF solutions (78750 initial guesses) and selected broken-symmetry states (24 per spin isomer) with low UHF energies. We then computed the UCCSD energies of these 840 UHF solutions and selected 12 to 24 configurations per spin isomer for the UCCSDT calculations, and further higher order calculations were performed on one reference state for BS7-235, BS8-237, and BS8-236. The reference dependence of the UCCSD, UCCSDT, and UDMRG energies and properties are studied in SM Sec.~\ref{sec:sm-ref-dep}.

\paragraph{Broken-symmetry density matrix renormalization group.} We used spin-unrestricted DMRG (UDMRG) with a broken-symmetry orbital basis (split-localized UCCSD natural orbitals computed for the selected reference) to study the low-energy states of the LLDUC model. We determined the orbital ordering by first pairing $\alpha$ and $\beta$ orbitals to maximize overlap within paired orbitals, and used a genetic algorithm to reorder the orbital pairs (SM Sec.~\ref{sec:sm-theory-dmrg}).
We used a sweep based deterministic algorithm to compute the coefficient of the dominant determinant (SM Sec.~\ref{sec:sm-det-overlap}). We implemented UDMRG in the open-source distributed parallelized DMRG code in \textsc{block2} through its Python API~\cite{zhai2021low,zhai2023block2}. We used the $\Delta E \sim \exp[-\kappa(\log D)^2]$ fitting for UDMRG energy extrapolation, based on 12 energies obtained from the reverse schedule with bond dimension 5000 to 16000 (SM Sec.~\ref{sec:sm-dmrg-extra}). We verified the reliability of the $\Delta E \sim \exp[-\kappa(\log D)^2]$ fitting using the smaller 4Fe-4S active space model (SM Sec.~\ref{sec:sm-fe4}).

\paragraph{Broken-symmetry coupled cluster theory.} We computed spin-unrestricted coupled cluster (UCC) solutions based on broken-symmetry references. For the full active space with 76 orbitals, we computed UCCSD, UCCSD(T), and UCCSDT energies. We performed UCCSDTQ in smaller active subspaces (with up to 54 orbitals) for BS7-235, BS8-237, and BS8-236, where the active subspaces were determined by freezing occupied UCCSD natural orbitals (FNO) (SM Sec.~\ref{sec:sm-fno}). For the smallest 30-orbital model, we performed UCCSDTQP to check the accuracy of UCCSDTQ, and estimate pentuples contributions. For subspaces with up to 50 orbitals we performed UDMRG with $D = 8000$ and energy extrapolation to check the accuracy of UCCSDTQ, and we estimated pentuples contributions and beyond using the ratio of correlation energy increments (SM Sec.~\ref{sec:sm-comp}). For the UCC and FNO-UCC calculations we developed a new open-source arbitrary-order CC code \textsc{hast-ucc} based on \textsc{PySCF}~\cite{sun2020recent}, with arbitrary order particle density matrices, where the CC diagrams and intermediates were derived automatically on-the-fly. We implemented a shared-memory parallel tensor contraction library, with special optimizations for the efficient storage and contraction of high-order UCC amplitude tensors (SM Sec.~\ref{sec:sm-theory-cc}).

\paragraph{Recovering the total spin.} We studied the energy gap between the BS families, spin gap, and total spin behavior in UCC and UDMRG in FeMo-co as well as [2Fe-2S] and [4Fe-4S]~\cite{sharma2012spin,li2017spin} (SM Sec.~\ref{sec:sm-spin}). For FeMo-co, we also constructed a toy Heisenberg model using UCCSD(T) data computed for the LLDUC model, assuming $C_3$ geometry and a fixed assignment of oxidation states (SM Sec.~\ref{sec:sm-model-heis}). Through these studies we verified the 
validity of spontaneous symmetry breaking at the chemical accuracy scale and the evolution of broken symmetry spin isomers from the pure spin isomers.

\paragraph{Non-collinear broken-symmetry states.} We investigated the relevance of non-collinear spin symmetry breaking in the LLDUC model at the generalized Hartree Fock (GHF), generalized spin CCSD (GCCSD) and generalized spin CCSD(T) levels of theory. We identified hundreds of unique GHF and GCCSD solutions, and studied the spin configuration similarity between the non-collinear low-energy GHF solutions and representative UHF solutions (SM Sec.~\ref{sec:sm-ghf}).

\paragraph{Active space model based on QM/MM potential.} Using a QM/MM setup~\cite{eastman2023openmm,bannwarth2019gfn2,bjornsson2022ash}, we modeled the heterotetramer of the Mo-Fe protein with water solvation, using the Universal Force Field~\cite{rappe1992uff} for FeMo-co with charges derived from GFN2-xtb~\cite{bannwarth2019gfn2} and the rest of the system described by the Amber14 force field~\cite{maier2015ff14sb} with TIP3P-FB water~\cite{wang2014building}. The electrostatic potential was accumulated from 300~ns MD trajectories (see SM~Sec.~\ref{sec:sm-qmmm}).
We then constructed a 76-orbital active space model with the QM/MM potential, and performed UHF, UCCSD, and UCCSD(T) calculations in the active space. For reference filtering we also projected unique UHF solutions from the LLDUC model as initial guesses (SM Sec.~\ref{sec:sm-qmmm}).

\paragraph{Larger active space models with dynamical correlation.} We first determined a (994o, 423e) active space including all FeMo-co cluster orbitals, based on localized spin-unrestricted natural orbitals computed using the high-spin DFT solution. We then truncated the active space to four smaller subspaces with 117 to 404 orbitals, based on UMP2 natural orbital occupation thresholds. We performed UHF, UCCSD, and UCCSD(T) calculations in the truncated active spaces. For reference filtering, we additionally considered projected unique UHF solutions from other truncated active spaces as the initial guesses. We computed the UMP2 composite correction as the difference between the UMP2 energies in the (994o, 423e) and (404o, 277e) active spaces. To determine the UHF reference for UMP2 in the (994o, 423e) active space we used the projected reference corresponding to the lowest UCCSD energy in the (404o, 277e) active space as the initial guess (SM Sec.~\ref{sec:sm-large-act}).

\paragraph{Oxidation state calibration.} We studied the assignment of Fe and Mo oxidation states based on the electron density at the metal nucleus as well as using meta-L\"owdin population analysis~\cite{sun2014exact}. The latter constructs an orthogonal basis with well defined core, valence, and Rydberg character, and  we considered only the electron population in the core and valence space (namely, without Rydberg contributions). We obtained the electron population scale for different Fe oxidation states using UCCSD performed for different [2Fe-2S] clusters, and checked its transferability to the P-cluster using DFT. For the determination of the electron population scale for Mo(III) we use UCCSD performed on $\ce{[MoFe3S4(SH)6]^{3-}}$~\cite{cook1985electronic,bjornsson2014identification} (SM Sec.~\ref{sec:sm-ox-calib}).

\paragraph{Cluster geometry dependence.} We checked the dependence of the energies, iron electron populations, and nucleus electron densities on the FeMo-co cluster geometry using broken-symmetry DFT (SM Sec.~\ref{sec:sm-geom}).

\section*{Data availability}
Data used in this work are available in the main text and supplementary material. Raw data, input scripts, and reference outputs are available at \url{https://github.com/hczhai/femoco-scripts-2026}.

\section*{Code availability}
Code used in this work can be found in the following open-source repositories. The \textsc{PySCF} code (for mean-field calculations, CCSD, and population analysis) is available at \url{https://github.com/pyscf/pyscf}. 
The \textsc{block2} code (for DMRG) is available at \url{https://github.com/block-hczhai/block2-preview}. The \textsc{hast-ucc} code (for high-order CC) is available at \url{https://github.com/hczhai/hast-ucc}.

\section*{Acknowledgements}
The authors thank Eirik F.~Kjønstad and Sandeep Sharma for helpful discussions and Doug Rees for discussing the complications of interpreting spatially resolved anomalous diffraction. Work by HZ at Caltech was supported by the US National Science Foundation via grant CHE-2102505. Work by CL and XZ was supported by the US Department of Energy, Office of Science, via grant DE-SC0023318. Additional support for software infrastructure development in PySCF was provided by the US National Science Foundation under award no. 2513474. GKC is a Simons Investigator in Physics. The Flatiron Institute is a division of the Simons Foundation. Work by SL was supported by the National Research Foundation of Korea (NRF) grant funded by the Korea government (MSIT) (Grant No. RS-2025-00515475). The computations in this work were run at facilities supported by the Scientific Computing Core at the Flatiron Institute, a division of the Simons Foundation, and at the Resnick High Performance Computing Center, supported by the Resnick Sustainability Institute at Caltech. 

\bibliographystyle{Science}
\bibliography{main}

\clearpage

\part*{Supplementary Materials}

\setcounter{section}{0}
\setcounter{figure}{0}
\setcounter{table}{0}
\setcounter{equation}{0}

\selectlanguage{english}

\renewcommand{\figurename}{Supplementary Figure}
\renewcommand{\tablename}{Supplementary Table}
\renewcommand{\lstlistingname}{Supplementary Listing}

\tableofcontents

\clearpage

\section{Prior work on FeMo-co resting state and nomenclature}
\label{sec:sm-prior}

\subsection{Atomic structure}

In this work, we focus on the resting ($\ce{E_0}$) state of FeMo-cofactor (FeMo-co) models based on the 1.0-\AA\ crystal structure of nitrogenase from \emph{Azotobacter vinelandii} (PDB 3U7Q)~\cite{spatzal2011evidence}. The identification of the FeMo-co atomic structure took almost two decades. In 1992, the first crystal structure was reported, where the central atom was missing and all Fe atoms were three-coordinated~\cite{kirn1992crystallographic}. Ten years later, the central atom was identified as a light atom at 1.16-\AA\ resolution~\cite{einsle2002nitrogenase}, which was finally determined to be a six-coordinated carbon in 2011, by 1.0-\AA\  high-resolution crystallography. Thereafter, the crystal structures for other nitrogenase bacteria have also been reported, while redox potential studies show that there are no significant variations among the FeMo-co proteins expressed by different bacteria~\cite{zanello2019structure}.

\subsection{Spin, charge and oxidation states}
\label{sec:sm-prior-expr}

The total spin of the FeMo-co ground state was determined by electron paramagnetic resonance (EPR) to be $S=3/2$~\cite{munck1975nitrogenase}. The total charge of FeMo-co was experimentally determined to be negative~\cite{huang1993purification}. For the oxidation states of iron and molybdenum, a study in 2000 based on ${}^{57}\ce{Fe}$ M\"ossbauer data supported a $\ce{[Mo^{4+} 3Fe^{3+} 4Fe^{2+}]^{1-}}$ oxidation assignment~\cite{yoo2000mossbauer}. In 2014, Bjornsson et al. proposed the Mo(III) assignment based on high-energy resolution fluorescence detected Mo K-edge X-ray absorption spectroscopy~\cite{bjornsson2014identification}, which has since been supported by density functional calculations and later experiments~\cite{kowalska2019x}. Based on the Mo(III) assignment, total spin, ${}^{57}\ce{Fe}$ M\"ossbauer studies, and further DFT calculations, Bjornsson et al. further proposed an oxidation state assignment of $\ce{[Mo^{3+} 4Fe^{3+} 3Fe^{2+}]^{1-}}$ in 2017~\cite{bjornsson2017revisiting}. We use the $1-$ total charge assignment for models in this work, which has also been used  by the majority of computational studies of the FeMo-co resting state since 2017~\cite{jiang2022quantum,cao2018influence,benediktsson2017qm,li2019electronic-femoco}. Note that compared to the localized formal Fe(II) and Fe(III) charge assignment above, other computations have suggested a picture with more delocalized electrons on the Fe  atoms~\cite{benediktsson2017qm}.

More recent experimental and computational studies have focused on the assignment of oxidation states to individual iron atoms, which is a more challenging problem under active debate. In 2016, spatially resolved anomalous diffraction refinement (SpReAD) analysis of X-ray data 
suggested that Fe1, Fe3, and Fe7 are the most reduced sites, although the interpretation of the signal is generally considered to be complicated~\cite{spatzal2016nitrogenase}. A 2019 study based on selenium K‑edge high resolution X‑ray absorption spectroscopy suggested that Fe2 and Fe6 are more oxidized than Fe3/4/5/7 and form an antiferromagnetic configuration~\cite{henthorn2019localized}. In contrast, a 2020 study based on the classical bond-valence analysis of the crystal structure of FeMo-co suggested Fe1, Fe6 and Fe7 should be considered to be Fe(III), while Fe2/3/4/5 are more mixed-valence~\cite{jin2020bond}. In 2023, analysis of M\"ossbauer data with site-selective ${}^{57}\ce{Fe}$ labeling suggested that Fe1 is aligned with the cluster total spin, has a delocalized $\ce{Fe^{2.5+}}$ oxidation state, and forms a ferromagnetic configuration with at least one of its neighbor irons (Fe2, Fe3, or Fe4)~\cite{badding2023connecting}. Although there is room for overlap in the above assignments, it is also clear that they are not entirely consistent. In the main text, we give some insight into a possible origin of these variations.

\subsection{Density functional studies}
\label{sec:sm-prior-dft}

\subsubsection{Broken-symmetry spin isomers}
\label{sec:sm-bs-name}

Early computational studies of FeMo-co were performed by Noodleman and co-workers~\cite{lovell2001femo,lovell2002metal}, who introduced the BS1-BS10 nomenclature that is widely used in the literature for computational studies of FeMo-co. The notation BS$n$ ($n=1,\cdots,10$) is used to label different spin isomers for FeMo-co low-energy states, assuming the approximate 3-fold symmetry through the Fe1-Mo axis. Each spin isomer corresponds to an assignment of four $\alpha$ (up-oriented) large spins and three $\beta$ (down-oriented) large spins to seven Fe centers. Without the 3-fold symmetry, we have $C^7_3 = 35$ unique spin isomers, where each of them can be labeled by the BS$n$ name and the three integers $ijk$ indicating the iron sites with $\beta$ spin, assuming the total spin projection is aligned with $\alpha$.  Most recent computational studies use the BS$n$-$ijk$ or BS$ijk$ notation.
In this work, we mainly use this convention in the main text, but 
computationally we also use a simple $0,1,\cdots,34$ indexing, and the labeling BS$n$-$X$ ($X=\text{A, B},\cdots,\text{E}$) for brevity in the Supporting Material. 
The correspondence between these notations is summarized in Supplementary Table~\ref{tab:sm-bs-naming}.

We additionally note that the Fe site numbering in the integral and the XYZ coordinate files published with the LLDUC model~\cite{li2019electronic-femoco} is different from the conventional experimental structure numbering of Fe atoms~\cite{spatzal2011evidence,bjornsson2017revisiting}. We used the LLDUC model numbering in the input files of all computations, while for all data presented in the main text and the Supporting Material, we use the experimental structure numbering. The experimental structure labels Fe1, Fe2, Fe3, Fe4, Fe5, Fe6, and Fe7 correspond to Fe1, Fe4, Fe3, Fe2, Fe5, Fe7, and Fe6 in the LLDUC model.

\begin{table}[!htbp]
    \centering
    \caption{Correspondence between the notation for broken-symmetry solutions used in this work and Refs.~\cite{bjornsson2017revisiting,cao2018influence,dance2020computational}. The BS$n$ numbering is the nomenclature introduced in ~\cite{lovell2002metal}. The three numbers $i,j,k$ after BS$n$ are the experimental structure numbering of Fe atoms with $\beta$ spin.}
    \begin{tabular}{
        >{\centering\arraybackslash}p{3cm}
        >{\centering\arraybackslash}p{3cm}
        >{\centering\arraybackslash}p{3cm}
        >{\centering\arraybackslash}p{3cm}
    }
    \hline\hline
    spin isomer index & this work & Ref.~\cite{cao2018influence} & Refs.~\cite{bjornsson2017revisiting,dance2020computational} \\
    \hline
    0 &      BS3-A &      BS3-3 &    BS3-134 \\
    1 &      BS3-B &      BS3-2 &    BS3-124 \\
    2 &      BS9-A &      BS9-3 &    BS9-145 \\
    3 &     BS10-A &     BS10-6 &   BS10-147 \\
    4 &     BS10-B &     BS10-5 &   BS10-146 \\
    5 &      BS3-C &      BS3-1 &    BS3-123 \\
    6 &     BS10-C &     BS10-3 &   BS10-135 \\
    7 &      BS9-B &      BS9-2 &    BS9-137 \\
    8 &     BS10-D &     BS10-4 &   BS10-136 \\
    9 &     BS10-E &     BS10-1 &   BS10-125 \\
   10 &     BS10-F &     BS10-2 &   BS10-127 \\
   11 &      BS9-C &      BS9-1 &    BS9-126 \\
   12 &      BS6-A &      BS6-2 &    BS6-157 \\
   13 &      BS6-B &      BS6-1 &    BS6-156 \\
   14 &      BS6-C &      BS6-3 &    BS6-167 \\
   15 &        BS2 &        BS2 &    BS2-234 \\
   16 &      BS8-A &      BS8-6 &    BS8-345 \\
   17 &      BS8-B &      BS8-4 &    BS8-347 \\
   18 &      BS7-A &      BS7-3 &    BS7-346 \\
   19 &      BS8-C &      BS8-5 &    BS8-245 \\
   20 &      BS7-B &      BS7-2 &    BS7-247 \\
   21 &      BS8-D &      BS8-2 &    BS8-246 \\
   22 &      BS5-A &      BS5-6 &    BS5-457 \\
   23 &      BS5-B &      BS5-5 &    BS5-456 \\
   24 &      BS4-A &      BS4-3 &    BS4-467 \\
   25 &      BS7-C &      BS7-1 &    BS7-235 \\
   26 &      BS8-E &      BS8-3 &    BS8-237 \\
   27 &      BS8-F &      BS8-1 &    BS8-236 \\
   28 &      BS5-C &      BS5-3 &    BS5-357 \\
   29 &      BS4-B &      BS4-2 &    BS4-356 \\
   30 &      BS5-D &      BS5-4 &    BS5-367 \\
   31 &      BS4-C &      BS4-1 &    BS4-257 \\
   32 &      BS5-E &      BS5-1 &    BS5-256 \\
   33 &      BS5-F &      BS5-2 &    BS5-267 \\
   34 &        BS1 &        BS1 &    BS1-567 \\
    \hline\hline
    \end{tabular}
    \label{tab:sm-bs-naming}
\end{table}

\subsubsection{Energy landscape}

In 2017, Benediktsson and Bjornsson performed a QM/MM DFT study for the three BS7 states of the FeMo-co resting state, and found that BS7-A(346) is the most stable spin isomer, with BS7-C(235) and BS7-B(247) being 0.7 and 1.1 kcal/mol higher in energy, respectively~\cite{benediktsson2017qm}. They suggested BS7-C to be the most likely ground state based on a metal–metal distance analysis and comparison with the X-ray structure. Later, Cao and Ryde performed a more comprehensive QM/MM DFT study in 2018 on all 35 broken-symmetry states, for the resting and other intermediates of FeMo-co~\cite{cao2018influence}. For the resting state they found the three BS7 states to be the most stable, and the BS6 states to be the next most stable, being $\sim$ 6.5 kcal/mol higher in energy. The energy variation among the three BS7 states was found to be 0.6 to 1.3 kcal/mol. In their study, BS7-C or BS7-A were found to be the most stable spin isomers when the TPSS-D3 or B3LYP functionals were used, respectively. Another DFT study by Raugei et al. in 2018 (with the COSMO polarizable dielectric continuum treatment for the protein environment not explicitly considered in their model) for the resting state suggested that BS7-A is the most stable, while other spin isomers were 7 to 50 kcal/mol higher in energy~\cite{raugei2018critical}. DFT studies on hydrogenated FeMo-co by Dance in 2019 suggested BS7-A, BS7-B, BS7-C, BS6-A, BS2, BS10-A, and BS10-C to be the low-energy isomers for FeMo-co derivative
structures~\cite{dance2019survey,dance2020computational}.

\subsubsection{DFT functional dependence and protein environment effects}

The 2018 study by Cao and Ryde studied the contribution of different computational factors to the relative energetics. The work suggested that the effects of varying the basis set and the modeling of the protein surroundings contributed up to 2.6 kcal/mol, the treatment of the cluster geometry contributes up to 8.8 kcal/mol, and the effect of the DFT functional (TPSS or B3LYP) is up to 13.9 kcal/mol~\cite{cao2018influence}. Dispersion corrections were found to be important for obtaining accurate energies (up to 13.6 kcal/mol) in a 2021 study by Torbj\"ornsson and Ryde~\cite{torbjornsson2021comparison}. In 2022, Benediktsson and Bjornsson reported a benchmark set of 11 Fe–Fe and Mo–Fe dimeric complexes, and found that the metal–metal distance is very sensitive to the specific choice of DFT functional~\cite{benediktsson2022analysis}. In a 2025 DFT study, Honjo et al. reported that hydrogen bonds and water molecules can significantly contribute to the stability of the LUMO and HOMO, for the resting state of FeMo-co~\cite{honjo2025theoretical}.

In 2023, Van Stappen et al. performed nuclear resonance vibrational spectroscopy experiments for different nitrogenase active sites including FeMo-co, and compared the results with the calculated partial vibrational density of states (PVDOS) using DFT and QM/MM~\cite{van2023structural}. For FeMo-co they considered the three BS7 states. They found that the computed PVDOS is sensitive to the specific choice of the BS state, and a proper simulation may require accounting for the degeneracy of spin isomers. They also reported the importance of the proper modeling of the protein environment for reproducing the experimental vibrational spectrum~\cite{van2023structural}.

\subsection{Beyond density functional theory studies}

Considering the high sensitivity in the functional choice observed in the DFT studies and the small energy difference among BS states, it is desirable to investigate the performance of wavefunction based methods that can go beyond DFT accuracy for nitrogenase systems. Active space models were constructed and studied using density matrix renormalization group (DMRG) for [2Fe-2S] and [4Fe-4S] clusters in 2014~\cite{sharma2014low}, and later for P clusters with different oxidation states in 2019~\cite{li2019electronic}. (54o, 54e) and (67o, 65e) active space models for FeMo-co were proposed by Reiher et. al in 2017 in the context of quantum computing resource estimation~\cite{reiher2017elucidating}. In 2018, Montgomery and Mazziotti studied the correlations in FeMo-co using smaller subspaces of the (54o, 54e) model, using CASSCF and variational 2-RDM methods with up to a (30o, 30e) subspace~\cite{montgomery2018strong}. However, the (54o, 54e) active space model was later shown to be insufficient to capture the open-shell character in the system~\cite{li2019electronic-femoco}. To fix the problem, in 2019 Li et al. proposed an improved (76o, 113e) active space model (the LLDUC model) for FeMo-co and performed preliminary DMRG studies (with bond dimension up to 2000)~\cite{li2019electronic-femoco}. In the 2023 study, Lee et al. revisited the DMRG studies (mainly via the spin-adapted DMRG formalism~\cite{sharma2012spin}) for [2Fe-2S], [4Fe-4S], P cluster (8Fe), and FeMo-co (LLDUC) active space~\cite{li2019electronic-femoco} models with a larger bond dimension (up to 7000 for FeMo-co) and reported the DMRG extrapolation results~\cite{lee2023evaluating}. The results are summarized in Supplementary Figure~\ref{fig:sm-dmrg-extra}. We see that the DMRG extrapolation distance for FeMo-co is significantly larger than that of smaller sized Fe-S clusters and the P cluster. Very recently, in 2025 Li proposed to apply spin-adapted DMRG with entanglement-minimized orbitals and reported the results for FeMo-co (LLDUC model) using a DMRG bond dimension of up to 10,000~\cite{li2025entanglement}.

\begin{figure}[!htbp]
\includegraphics[width=400px]
{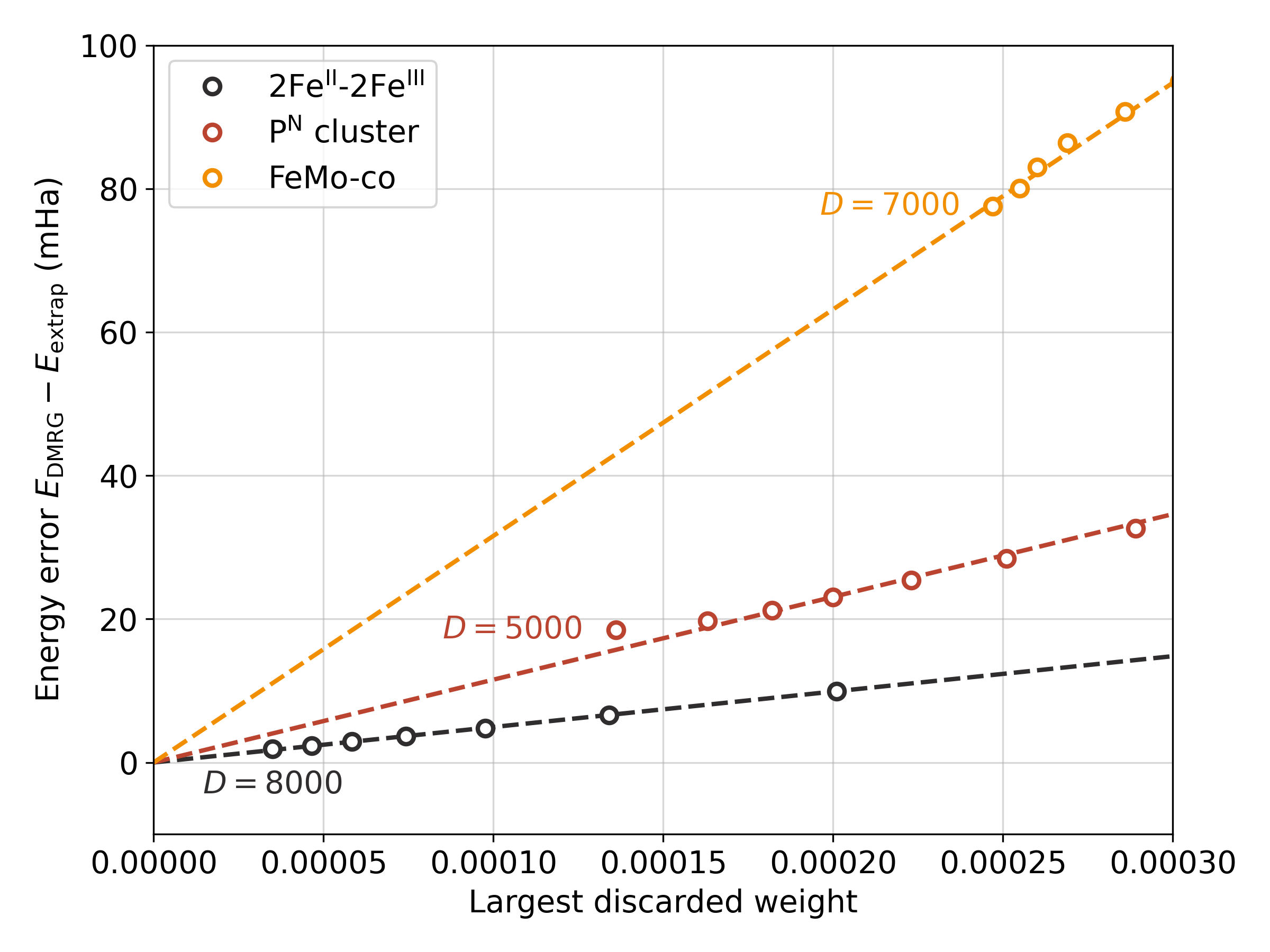}
\centering
\caption{DMRG extrapolation error for $\ce{2Fe^{II}-2Fe^{III}}$ cluster, $\ce{P^N}$ cluster, and FeMo-cofactor (data taken from Ref.~\cite{lee2023evaluating}).}\label{fig:sm-dmrg-extra}
\end{figure}

\section{Theory background}
\label{sec:sm-theory}

\subsection{Density matrix renormalization group}
\label{sec:sm-theory-dmrg}

The density matrix renormalization group (DMRG) approach~\cite{white1992density,white1993density,verstraete2023density} uses the following ansatz for the wavefunction, called a Matrix Product State (MPS)~\cite{chan2016matrix}
\begin{equation}
    |\Psi\rangle = \sum_{\{n\}} \mathbf{A}[1]^{n_1} \mathbf{A}[2]^{n_2}\cdots \mathbf{A}[K]^{n_K} | n_1\ n_2\cdots n_K\rangle
\end{equation}
where $K$ is the number of spatial orbitals, each $n_k$ is the occupation state on each of the orbitals, and each $\mathbf{A}[k]^{n_k} (k=2,\cdots, K-1)$ is a $D\times D$ matrix, and $\mathbf{A}[1]^{n_1}$ and $\mathbf{A}[K]^{n_K}$ are vectors. Given a Hamiltonian, an iterative procedure called a sweep algorithm can be used to optimize the $\mathbf{A}$ matrices and find the approximate ground state. The accuracy of the DMRG energy can be systematically improved by increasing the MPS bond dimension $D$. For quantum chemistry Hamiltonians, one typically uses the spin-adapted formalism of DMRG with spin-restricted (and optionally localized) orbitals, to find the ground state in the correct total spin sector and minimize computational cost~\cite{chan2002highly,sharma2012spin,zhai2021low}.

For the FeMo-co active space model considered in this work, since the mean-field broken symmetry solutions are good approximations to the low-energy spin isomers, we consider a less conventional spin-unrestricted DMRG (UDMRG) procedure based on broken symmetry orbitals. Specifically, for each of the 35 spin isomers for FeMo-co, we start from one of the broken symmetry UHF solutions and perform spin-unrestricted CCSD (UCCSD) to find the spin-unrestricted CCSD natural orbitals (NO). All UHF and UCCSD calculations are done in the same active space as that for UDMRG. We split-localized the UCCSD NO in four parts: $\alpha$ spin occupied, $\alpha$ spin virtual, $\beta$ spin occupied, and $\beta$ spin virtual, using Pipek-Mezey localization~\cite{pipek1989fast}, and used them as the orbitals for UDMRG.

Orbital reordering can have a significant impact on the performance of DMRG~\cite{olivares2015ab}. For the case of UDMRG used in this work, we determine a suitable orbital ordering with the $\alpha\beta\alpha\beta\cdots$ orbital arrangement using a greedy optimization procedure. Specifically, we first find an optimal pairing of $\alpha$ and $\beta$ orbitals by maximizing the sum of the squared overlaps between paired $\alpha$ and $\beta$ orbitals. We then use the standard Genetic Algorithm optimization (GAOPT) procedure~\cite{olivares2015ab} to find a good ordering for the $\alpha\beta$ pairs. To generate a good initial guess for UDMRG, we use the fractional occupation number of the UCCSD NO as a hint for determining the initial quantum number distribution in the MPS~\cite{zhai2023block2}. The UDMRG procedure is implemented using the Python API of \textsc{block2}~\cite{zhai2021low,zhai2023block2} interfaced with \textsc{PySCF}~\cite{sun2018pyscf,sun2020recent}, with support for hybrid distributed and shared-memory parallelization.

\subsubsection{Theory of DMRG extrapolation}
\label{sec:sm-theory-dmrg-extra}

As the quality of DMRG calculations can be improved systematically by the bond dimension $D$, it is common to perform an extrapolation of the energy (or other observables) over a series of DMRG calculations with different bond dimensions. There are a variety of extrapolation techniques in the literature. The most common is to extrapolate with respect to the discarded weight $w(D)$, the compression error of the DMRG matrix product state in the variational two-site algorithm. In the limit of large $D$, one can justify the behaviour $\Delta E(D) \propto w(D)$ where $\Delta E$ is the residual energy error~\cite{legeza1996accuracy}. However, as 
seen in Sec.~\ref{sec:sm-dmrg-extra}, we are not yet in this asymptotic regime in our calculations on the FeMo-co LLDUC model.

An alternative is to directly extrapolate $\Delta E$ as a function of the bond dimension $D$. One form that has been used is
\begin{equation}
    \Delta E \sim e^{-\mathrm{const} (\log D)^\alpha} \label{eq:extrapweights}
\end{equation}
where $\alpha=2$. This can be derived from the asymptotic behaviour of $w(D)$ for large $D$ and using $\Delta E(D) \propto w(D)$. Under the condition that the entanglement Hamiltonian (logarithm of the system density matrix in DMRG) corresponds to a set of weakly interacting quasiparticles, one obtains a discarded weight of the form Eq.~\ref{eq:extrapweights}, where $\alpha > 1$ is a system specific parameter~\cite{okunishi1999universal,chan2002distribution,chan2002highly}. This form has previously been used for energy extrapolation in spin-adapted DMRG calculations on Fe dimers~\cite{lee2023evaluating} and to motivate an extrapolation of the overlap between a small bond dimension MPS and the exact state in Ref.~\cite{berry2025rapid} also in Fe-S clusters. 

Note that while the theoretical derivation of the bond dimension extrapolation also requires $\Delta E \propto w(D)$, as an empirical extrapolation relation, the bond dimension and weight extrapolations may in practice have different regimes of applicability. We use Eq.~\ref{eq:extrapweights} to carry out bond dimension extrapolation in this work.

\subsection{Coupled cluster theory}
\label{sec:sm-theory-cc}

In the (single reference) coupled cluster (CC) theory we consider the following exponential ansatz~\cite{bartlett2007coupled,shavitt2009many}
\begin{equation}
    |\Psi\rangle = \mathrm{e}^{\hat{T}} |\Phi_0\rangle
\end{equation}
where $|\Phi_0\rangle$ is a spin-restricted or spin-unrestricted mean-field state, called the reference state in CC theory, and
\begin{equation}
    \hat{T} = \hat{T}_1 + \hat{T}_2 + \cdots =\sum_{ia,\sigma} T_{ia,\sigma}\ c^\dagger_{a\sigma} c_{i\sigma}
    + \frac{1}{(2!)^2}\sum_{ijab,\sigma\sigma'}T_{ijab,\sigma\sigma'}\ c^\dagger_{a\sigma} 
    c^\dagger_{b\sigma'} c_{j\sigma'} c_{i\sigma}+\cdots
\end{equation}
is known as the cluster operator, which can be expanded in contributions from single ($\hat{T}_1$), double ($\hat{T}_2$), triple ($\hat{T}_3$), quadruple ($\hat{T}_4$), etc. excitations. Indices $i,j, \cdots$ and $a,b, \cdots$ are used to label occupied and unoccupied orbitals in the reference state $|\Phi_0\rangle$, respectively. For efficient classical simulations, $\hat{T}$ is truncated to include only low-order excitations. For example, in the so-called Coupled Cluster Singles and Doubles (CCSD), we have $\hat{T} = \hat{T}_1 + \hat{T}_2$. We can additionally include an approximate correction for the triples, which gives the  commonly used CCSD(T) method.

It has been shown that CC with a spin-unrestricted broken-symmetry reference can provide accurate exchange couplings in bridged transition metal dimers~\cite{schurkus2020theoretical}, while CC with conventional spin-restricted references may potentially have convergence issues for these cases. In our previous work, we have also shown that broken-symmetry CCSD(T) can provide reliable relative energies for the protonation of dimeric models of nitrogenase iron--sulfur clusters~\cite{zhai2023multireference}.

For iron--sulfur models studied in this work, one of the difficulties is the large number of broken-symmetry mean-field solutions (typically with different UHF energies). We consider 24 low-energy UHF references (for each spin isomer in FeMo-co) and perform CCSD and CCSD(T) for each of these references. For more expensive theories including CCSDT and DMRG we consider one or a few references with the lowest CCSD energies. The reference dependence of the CC energy and density matrices will be discussed in Supplementary Sec.~\ref{sec:sm-ref-dep}. For some unfavorable UHF references, broken-symmetry CC can also have convergence difficulties, especially with full triples. In this work we use a large DIIS space (12 to 36) together with normal CC self-consistent iterations (as implemented in \textsc{PySCF}) to alleviate the convergence problem.

For spin-unrestricted CCSDT and CCSDTQ, the large tensor size of the CC amplitude tensor can create a memory bottleneck for systems considered in this work. We developed a memory-efficient arbitrary-order spin-unrestricted CC implementation with support for the contraction of CC tensors in the packed storage format, where only index-symmetry unique elements are kept in each anti-symmetric tensor. The code is parallelized using multi-threading and interfaced with \textsc{PySCF}. To further save the memory cost of CCSDTQ in some large FNO spaces, we also used fully or partially out of core storage for DIIS vectors.

\section{Ground state estimation for FeMo-co active space models}

\subsection{Minimal active space (76o, 113e) LLDUC model}

\label{sec:sm-llduc}

The LLDUC model was defined in Ref.~\cite{li2019electronic-femoco}. The integral (FCIDUMP) file and geometry can be found in \url{https://github.com/zhendongli2008/Active-space-model-for-FeMoco}. The XYZ coordinates (in Angstroms) of the geometry are listed in Supplementary Listing~\ref{list:xyz}.

\begin{lstlisting}[caption={The XYZ coordinates (in Angstroms) of the geometry used in the LLDUC model.}, label=list:xyz]
225

Fe 10.018000 -7.568000 56.109001
Fe 12.528000 -8.326000 55.692001
Fe 11.277000 -6.902000 53.853001
Fe 11.990000 -5.771000 56.166000
Fe 14.820000 -7.601000 54.671001
Fe 13.550000 -6.198000 52.848999
Fe 14.226000 -5.072000 55.094002
Mo 16.132000 -5.638000 53.306000
C 13.044000 -6.665000 54.723999
S 11.719000 -7.459000 57.646000
S 16.169001 -6.043000 55.618000
S 10.001000 -5.522000 55.134998
S 13.162000 -3.960000 56.627998
S 14.407000 -9.544000 55.750999
S 14.337000 -4.109000 53.098000
S 10.718000 -8.998000 54.466999
S 15.203000 -7.626000 52.453999
S 11.674000 -6.465000 51.712002
S 8.197000 -8.169000 57.320999
C 13.298000 -2.024000 50.055000
N 12.366000 -2.944000 50.620998
C 11.087000 -2.665000 50.879002
N 10.465000 -1.648000 50.319000
N 10.437000 -3.395000 51.761002
C 15.000000 -1.023000 59.625000
C 16.169001 -0.765000 58.679001
C 17.408001 -1.557000 59.002998
O 17.870001 -1.613000 60.158001
N 18.021999 -2.150000 57.976002
C 8.265000 -1.347000 60.679001
C 9.399000 -1.996000 59.963001
N 9.933000 -3.152000 60.439999
C 10.065000 -1.670000 58.792999
C 10.903000 -3.530000 59.608002
N 10.987000 -2.647000 58.599998
C 8.211000 -4.616000 49.907001
C 7.638000 -4.371000 51.291000
C 6.881000 -3.235000 51.567001
C 7.762000 -5.325000 52.312000
C 6.315000 -3.036000 52.811001
C 7.180000 -5.134000 53.570000
C 6.480000 -3.971000 53.827999
O 5.954000 -3.723000 55.074001
N 5.234000 -10.420000 55.743999
C 6.175000 -9.825000 56.689999
C 5.492000 -9.417000 57.998001
O 4.993000 -8.311000 58.167999
C 6.936000 -8.649000 56.117001
N 5.537000 -10.342000 58.952000
C 18.375999 -9.779000 57.895000
O 18.938999 -9.535000 58.952999
N 17.193001 -9.276000 57.611000
C 16.604000 -8.174000 58.390999
C 15.700000 -8.561000 59.548000
O 16.066999 -8.490000 60.719002
N 14.482000 -8.960000 59.175999
C 13.413000 -9.170000 60.129002
C 12.448000 -10.321000 59.819000
O 11.585000 -10.619000 60.638000
N 12.608000 -10.986000 58.669998
C 11.669000 -12.033000 58.242001
C 12.397000 -13.165000 57.548000
O 12.311000 -14.342000 57.938000
N 13.074000 -12.829000 56.459999
C 13.702000 -13.811000 55.596001
C 14.324000 -13.116000 54.386002
C 13.275000 -12.471000 53.491001
C 13.911000 -11.621000 52.422001
N 12.878000 -11.123000 51.516998
C 13.118000 -10.372000 50.460999
N 14.374000 -10.095000 50.109001
N 12.094000 -9.945000 49.705002
O 22.684000 -7.070000 47.084999
C 18.180000 -7.369000 50.759998
C 18.711000 -7.463000 52.162998
N 18.159000 -6.788000 53.228001
C 19.879999 -8.011000 52.627998
C 18.976999 -6.904000 54.275002
N 20.003000 -7.668000 53.919998
C 17.480000 -1.176000 54.216000
C 17.136999 -1.687000 52.838001
C 17.757999 -3.064000 52.515999
C 19.298000 -2.979000 52.311001
C 19.787001 -4.275000 51.637001
C 21.243001 -4.658999 51.755000
C 17.111000 -3.651000 51.271999
O 17.582001 -2.027000 55.126999
O 17.659000 0.041000 54.360001
O 21.753999 -5.267001 50.771999
O 21.844000 -4.408001 52.838001
O 16.628000 -4.810000 51.337002
O 17.166000 -2.989000 50.195999
O 17.490000 -3.979000 53.599998
O 15.888000 -4.907000 46.811001
O 19.811001 -2.775000 44.562000
O 21.295000 -1.675000 56.474998
O 14.797000 -5.542000 49.339001
O 20.999001 -7.558000 49.209999
O 21.018000 -4.234000 48.202000
O 18.273001 -3.799000 47.834000
O 24.264000 -5.111000 53.854000
O 18.688999 -0.721000 49.472000
O 20.888000 -4.190000 55.507000
O 19.867001 0.773000 55.806000
O 19.983999 2.139000 49.183998
O 19.601999 0.653000 51.790001
H 14.253909 -2.551439 49.902245
H 13.485333 -1.148637 50.707842
H 12.976245 -1.666277 49.058362
H 12.756517 -3.749057 51.133080
H 10.948382 -1.006864 49.696509
H 9.418611 -3.350350 51.840380
H 10.851817 -4.301103 52.064489
H 14.176828 -0.309351 59.447134
H 15.333467 -0.925316 60.673517
H 14.594556 -2.042045 59.487892
H 15.865440 -0.939269 57.629728
H 16.465566 0.302217 58.741955
H 17.747578 -2.051739 56.983900
H 18.886142 -2.654352 58.166876
H 7.943033 -0.415330 60.180924
H 8.542154 -1.100693 61.721569
H 7.393391 -2.026762 60.730390
H 11.631869 -2.750219 57.800192
H 9.950464 -0.842408 58.089483
H 11.523221 -4.427460 59.671513
H 8.474673 -3.670967 49.398976
H 7.479231 -5.140569 49.262567
H 9.119194 -5.242955 49.957698
H 6.731721 -2.478799 50.782532
H 8.369443 -6.224645 52.146371
H 5.741910 -2.119104 53.011911
H 7.324840 -5.870775 54.367803
H 5.546271 -2.840425 55.063551
H 5.750165 -10.676231 54.895864
H 4.564723 -9.698382 55.448024
H 6.922962 -10.606213 56.936088
H 6.237039 -7.808570 55.941648
H 7.401622 -8.936745 55.157491
H 6.037121 -11.218482 58.832310
H 5.164447 -10.123761 59.874931
H 18.781655 -10.466135 57.105626
H 16.704789 -9.500843 56.734897
H 16.043744 -7.544498 57.677036
H 17.429698 -7.586986 58.824307
H 14.219156 -8.808225 58.189655
H 12.783401 -8.263772 60.234252
H 13.872680 -9.367031 61.112602
H 13.256381 -10.608473 57.970429
H 11.174293 -12.420930 59.144023
H 10.895271 -11.603903 57.570696
H 13.174849 -11.839973 56.192474
H 14.473224 -14.370958 56.160710
H 12.945500 -14.559086 55.282907
H 15.032972 -12.341634 54.741837
H 14.913117 -13.854269 53.810118
H 12.641772 -13.254547 53.028674
H 12.605520 -11.821374 54.090680
H 14.436696 -10.768534 52.906517
H 14.648390 -12.199351 51.832233
H 11.938028 -11.063509 51.914966
H 14.531875 -9.489172 49.305241
H 15.069805 -9.986501 50.852478
H 12.200422 -9.064588 49.201927
H 23.601459 -6.857982 47.330021
H 22.249074 -6.194096 47.039440
H 18.955188 -7.698977 50.048433
H 17.905390 -6.332504 50.518737
H 17.264032 -7.973340 50.626379
H 20.628332 -8.602807 52.103837
H 18.853442 -6.425326 55.245736
H 16.033435 -1.787606 52.792646
H 17.426546 -0.972948 52.052632
H 19.528063 -2.102232 51.681236
H 19.759202 -2.868484 53.309073
H 19.247123 -5.115528 52.106350
H 19.502098 -4.271284 50.571493
H 16.780810 -4.624223 47.121885
H 15.433213 -5.151276 47.644086
H 19.980766 -3.723341 44.435067
H 19.321145 -2.766364 45.406260
H 21.109337 -1.750623 57.425786
H 20.759935 -0.882264 56.201074
H 15.501236 -5.269415 49.977894
H 14.144043 -5.981686 49.917945
H 21.703906 -7.489930 48.526249
H 21.149790 -6.782293 49.795777
H 21.232006 -4.616723 49.099190
H 21.241122 -3.291069 48.291377
H 19.195384 -4.089918 48.024181
H 17.876980 -3.527801 48.695535
H 24.697333 -5.483950 53.066151
H 23.382051 -4.837434 53.479899
H 18.953931 -0.321612 50.327834
H 18.115647 -1.481067 49.723219
H 20.992458 -3.241281 55.753084
H 21.160164 -4.239411 54.564410
H 20.330760 1.130055 55.030296
H 18.979135 0.500481 55.434638
H 19.940635 1.936802 50.139503
H 19.518351 1.363575 48.815374
H 20.354371 0.130718 52.119534
H 18.942524 0.622904 52.521728
H 20.831763 -7.852887 54.530690
H  9.545331 -1.367341 50.646139
H 11.144726 -10.137694 50.018567
C 23.292999 -8.105000 51.974998
C 24.523001 -7.559000 51.272999
O 25.587000 -8.169000 51.252998
N 24.365000 -6.351000 50.728001
C 25.365000 -5.756000 49.866001
C 24.761000 -4.521000 49.146000
O 22.260000 -7.893000 55.609001
H 17.491148 -3.347688 54.445403
H 22.481544 -8.230248 51.233996
H 22.943402 -7.372156 52.723626
H 23.534473 -9.061675 52.460204
H 23.421240 -5.901816 50.778903
H 26.261472 -5.456441 50.446366
H 25.720615 -6.508632 49.132117
H 24.470248 -3.748936 49.879990
H 23.848896 -4.798400 48.587383
H 25.488867 -4.081178 48.443117
H 21.937137 -7.439342 56.409331
H 22.855842 -7.224751 55.210211
\end{lstlisting}

\subsubsection{Non-collinear spin solutions}
\label{sec:sm-ghf}

Since the broken-symmetry UHF solutions have been shown to be useful in describing the low-energy landscape of FeMo-co, one may ask whether further spin symmetry breaking can help. For this purpose we computed the general Hartree-Fock (GHF) solutions and the CCSD and CCSD(T) solutions with the GHF solution as references (termed GCCSD and GCCSD(T), respectively), for the LLDUC model of FeMo-co.
We note that the current implementation of GHF in \textsc{PySCF} can only reliably explore GHF solutions in the real domain. To sample the \emph{real} GHF space, for each of the 35 UHF spin isomers, we considered 2250 possible UHF initial density matrices, but with a small amount of noise to connect the $\alpha$ and $\beta$ spin manifolds. We solve for the GHF solution using the Newton solver implemented in \textsc{PySCF}, and identified 286 unique GHF solutions for the LLDUC model. The number of unique GHF solutions is significantly smaller than the number of unique UHF solutions, mainly because the 35 UHF spin isomer are no longer ``disconnected'' in the optimization manifold, and many of them optimize to the same broken-symmetry solution in the GHF space.

The best GHF energy we obtained is $-22139.997279$ Hartrees, which is 4.3 kcal/mol lower than the best UHF energy ($-22139.990392$ Hartrees). We computed GCCSD and GCCSD(T) energies using the 286 unique GHF solutions as references, and 221 CCSD calculations converged within 400 iterations. In Supplementary Table ~\ref{tab:sm-ghf-ener} we list the GHF, GCCSD, and GCCSD(T) energies for 16 references with the lowest GCCSD energies. The comparison between the UCCSD and GCCSD energies for the low-energy references is shown in Supplementary Figure~\ref{fig:sm-ghf-ener}. The best GCCSD energy is 3.9 kcal/mol lower than the best UCCSD energy. In Supplementary Figure~\ref{fig:sm-ghf-spin} we plot the spin orientation in the Fe and Mo atoms (with the spin orientation of Fe1 fixed on the left). We found that half of these low-energy GHF states are small perturbations or copies of the UHF BS states BS7-C, BS8-E and BS8-F. Other low-energy GHF states show clear non-collinear spin configurations, particularly in the right cubane (Fe5, Fe6, and Fe7). We additionally note that the UHF BS states BS7-C, BS8-E and BS8-F have exactly the same spin configuration for Fe1, Fe2, Fe3, and Fe4. As a result, the non-collinear spin arrangement found in the low-energy GHF solutions (such as GHF-64) may be related to the very small energy gap among UHF BS states BS7-C and BS8-E.

Overall, we have not exhaustively studied the GHF solution space to the extent we have characterized the UHF solution space in this work, and further work on GHF solutions is desirable in the future.

\begin{table}[!htbp]
    \centering
    \caption{Broken-symmetry GHF, GCCSD and GCCSD(T) energies (in Hartrees) computed using different references for the LLDUC model. The GHF references are indexed by their ranking in the GHF energies, counting from zero. The low-energy UHF BS states and their UHF, UCCSD, and UCCSD(T) energies are also listed for comparison.}
    \begin{tabular}{
        >{\centering\arraybackslash}p{3.2cm}|
        >{\centering\arraybackslash}p{2.7cm}
        >{\centering\arraybackslash}p{2.7cm}
        >{\centering\arraybackslash}p{2.7cm}
        >{\centering\arraybackslash}p{2.5cm}
    }
    \hline\hline
    reference & $E_{\mathrm{HF}}$& $E_{\mathrm{CCSD}}$ & $E_{\mathrm{CCSD(T)}}$ & similar to \\
    \hline
    GHF-61 &   -22139.984067 &   -22140.312627 &   -22140.355756 &  BS7-C (235) \\
    GHF-64 &   -22139.983413 &   -22140.308508 &   -22140.353316 &              \\
    GHF-21 &   -22139.990392 &   -22140.305389 &   -22140.352783 &  BS7-C (235) \\
    GHF-63 &   -22139.983618 &   -22140.304696 &   -22140.352173 &              \\
    GHF-44 &   -22139.986013 &   -22140.304318 &   -22140.352781 &  BS7-C (235) \\
    GHF-77 &   -22139.981522 &   -22140.301346 &   -22140.348974 &              \\
    GHF-62 &   -22139.983774 &   -22140.301272 &   -22140.349397 &  BS8-E (237) \\
    GHF-47 &   -22139.985676 &   -22140.300503 &   -22140.350474 &              \\
    GHF-33 &   -22139.987906 &   -22140.300021 &   -22140.349813 &              \\
    GHF-66 &   -22139.983280 &   -22140.298387 &   -22140.346968 &  BS8-F (236) \\
   GHF-153 &   -22139.971545 &   -22140.296702 &   -22140.355347 &  BS7-C (235) \\
    GHF-88 &   -22139.979837 &   -22140.296366 &   -22140.345343 &              \\
   GHF-129 &   -22139.973956 &   -22140.296346 &   -22140.359345 &              \\
   GHF-187 &   -22139.967960 &   -22140.296047 &   -22140.350789 &              \\
   GHF-159 &   -22139.971136 &   -22140.295958 &   -22140.349573 &              \\
   GHF-135 &   -22139.973441 &   -22140.295794 &   -22140.350849 &  BS7-C (235) \\
\hline
     BS7-C &   -22139.990392 &   -22140.305389 &   -22140.352783 & \\
     BS8-E &   -22139.981193 &   -22140.306453 &   -22140.350447 & \\
     BS8-F &   -22139.983280 &   -22140.298387 &   -22140.346968 & \\
    \hline\hline
    \end{tabular}
    \label{tab:sm-ghf-ener}
\end{table}

\begin{figure}[!htbp]
  \centering
  \includegraphics[width=0.85\linewidth]{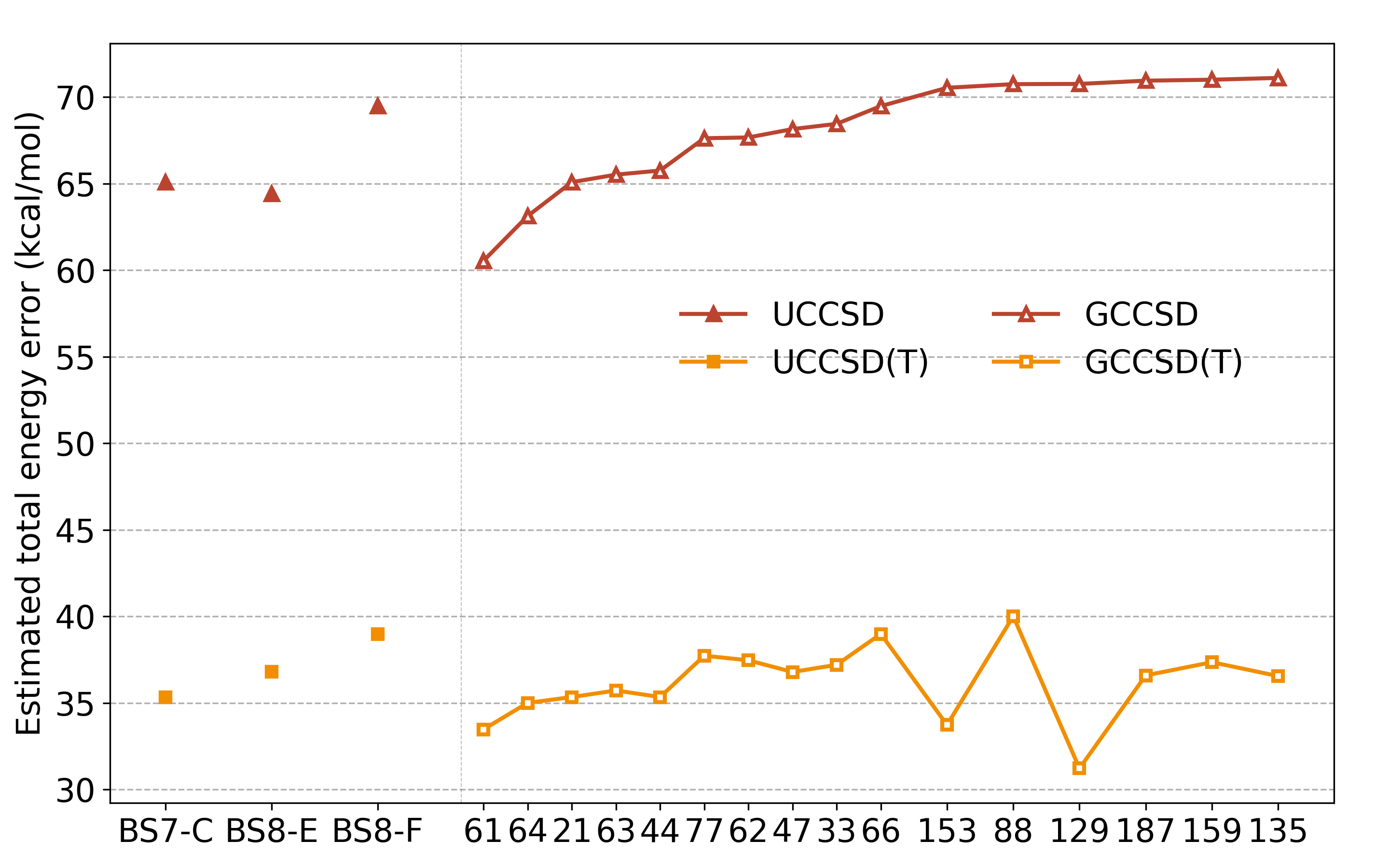}
  \caption{Energy landscape of FeMo-co LLDUC model computed using GCCSD and GCCSD(T), compared with UHF counterparts. Integers are GHF reference indices based on ranking of GHF energies.}
  \label{fig:sm-ghf-ener}
\end{figure}

\begin{figure}[!htbp]
  \centering
  \includegraphics[width=\linewidth]{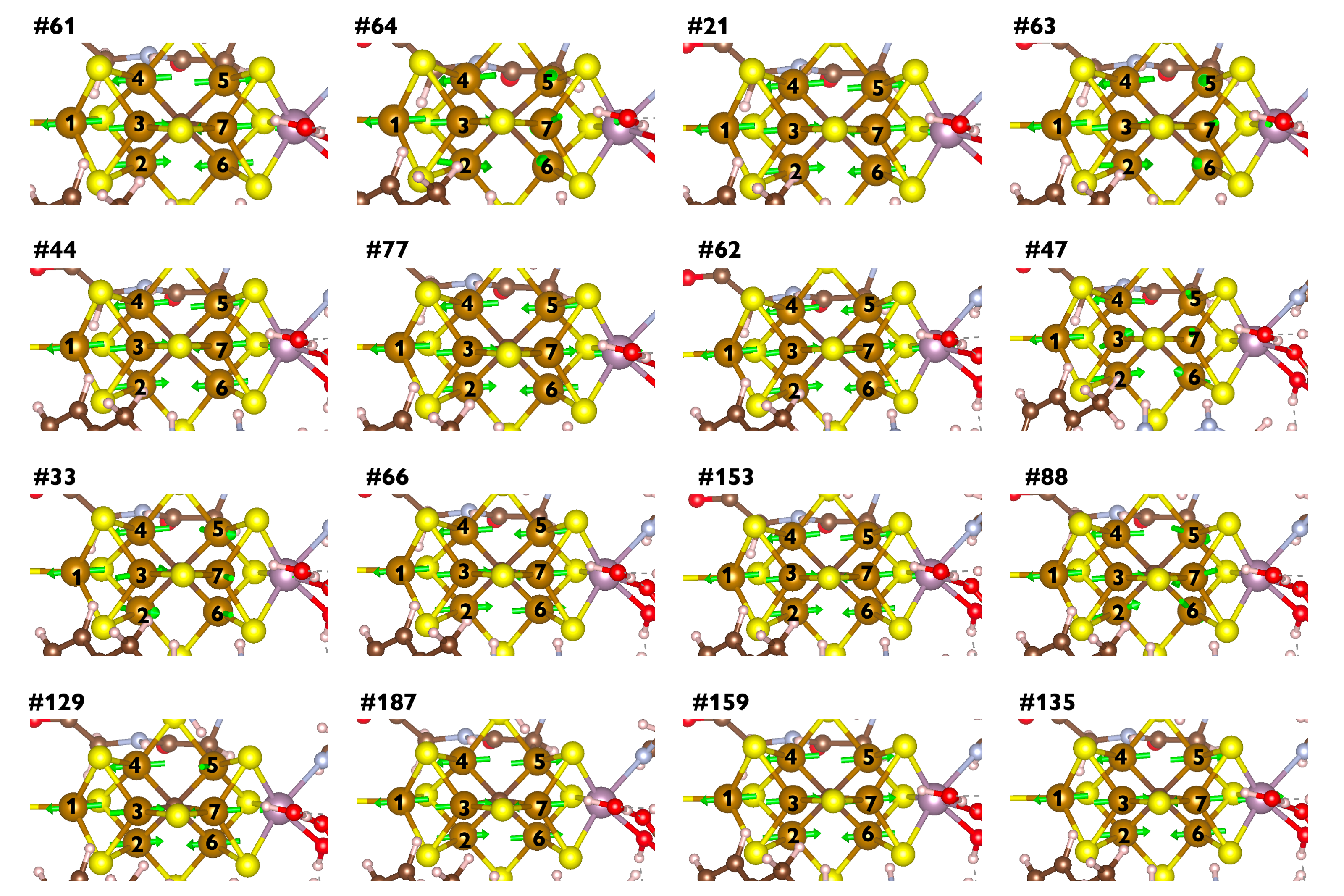}
  \caption{The spin orientation computed using the GHF density matrix for 16 GHF references with the lowest GCCSD energies. The spin orientation of Fe1 is fixed on the left.}
  \label{fig:sm-ghf-spin}
\end{figure}

\subsubsection{Reference dependence}
\label{sec:sm-ref-dep}

As discussed in Supplementary Sec.~\ref{sec:sm-theory}, we use broken-symmetry mean-field solutions as references for CC and for constructing split-localized broken-symmetry orbitals for DMRG. For each of the 35 spin isomers, there are different ways of assigning the Fe(II) and Fe(III) oxidation states and the doubly occupied orbital in Fe(II). As a result, one would expect that there are many ways to initialize the UHF self-consistent field procedure, and we may end up with different UHF solutions (local minima) for each spin isomer. In this section, we discuss the manifold of these UHF solutions and the consequences of using them as references in broken-symmetry CC and DMRG calculations, respectively.

Note that for each given UHF broken-symmetry reference, it is also possible to have multiple CC or DMRG solutions, since the CC and DMRG wavefunctions are obtained from some type of self-consistent iterative algorithm. However, in practice, we do not find this to be a problem for the current study. As long as the algorithm converges, we always end up with a well-defined CC or DMRG solution for each considered UHF reference. A detailed investigation of local minima and convergence behavior in the broken-symmetry CC/DMRG solution manifold is beyond the scope of this work.

\textbf{Enumeration of UHF solutions.} Although the Fe atoms in a converged UHF solution are typically mixed-valence, for the purpose of sampling the low-energy UHF solutions for the LLDUC model, we use  density matrices with collinear spins and classical integer Fe(II) and Fe(III) charges~\cite{li2019electronic} as the initial guesses for UHF, based on simplicity and ease of reproducibility considerations. As the LLDUC model space is relatively small (compared to the larger active space models used in later sections), and the number of such initial classical density matrices is large ($5^3\times 18=2250$ for each spin isomer), we find that this procedure works very well for identifying the UHF solution with the lowest energy for each spin isomer, as indicated by the relatively flat UHF curve shown in Fig. 2A. For each spin isomer, we found that many of the 2250 initial guesses optimized to the same UHF solution, but the number of unique UHF solutions is still significant, as listed in Supplementary Table~\ref{tab:sm-uhf-cnt}. The total number of unique UHF solutions we found for this model is 12300, more than the square of the number of unique UHF solutions we found for mixed valence [4Fe-4S], where we found 26 unique UHF solutions (see Supplementary Sec.~\ref{sec:sm-fe4}). We label the different UHF solutions within each spin isomer as BS$n$-$X$-$m$ where $m = 0,1,2,\cdots$ indicates the ordering by UHF energies. Note that the number of unique UHF solutions can also heavily depend on the Newton optimization solver and the counting procedure.

\begin{table}[!htbp]
    \centering
    \caption{Number of unique UHF solutions found for each spin isomer in the LLDUC model.}
    \begin{tabular}{
        >{\centering\arraybackslash}p{1.95cm}
        >{\centering\arraybackslash}p{1.17cm}|
        >{\centering\arraybackslash}p{1.95cm}
        >{\centering\arraybackslash}p{1.17cm}|
        >{\centering\arraybackslash}p{1.95cm}
        >{\centering\arraybackslash}p{1.17cm}|
        >{\centering\arraybackslash}p{1.95cm}
        >{\centering\arraybackslash}p{1.17cm}
    }
    \hline\hline
    spin isomer & \# UHF & spin isomer & \# UHF &
    spin isomer & \# UHF & spin isomer & \# UHF \\
    \hline
       BS1 &   258 &      BS5-B &   313 &      BS7-B &   390 &      BS9-B &   423 \\
       BS2 &   446 &      BS5-C &   280 &      BS7-C &   312 &      BS9-C &   448 \\
     BS3-A &   286 &      BS5-D &   243 &      BS8-A &   391 &     BS10-A &   285 \\
     BS3-B &   218 &      BS5-E &   360 &      BS8-B &   417 &     BS10-B &   310 \\
     BS3-C &   308 &      BS5-F &   294 &      BS8-C &   392 &     BS10-C &   389 \\
     BS4-A &   364 &      BS6-A &   484 &      BS8-D &   490 &     BS10-D &   378 \\
     BS4-B &   228 &      BS6-B &   401 &      BS8-E &   290 &     BS10-E &   370 \\
     BS4-C &   359 &      BS6-C &   342 &      BS8-F &   408 &     BS10-F &   331 \\
     BS5-A &   379 &      BS7-A &   376 &      BS9-A &   337 & \textbf{total} & \textbf{12300} \\
    \hline\hline
    \end{tabular}
    \label{tab:sm-uhf-cnt}
\end{table}

For active space models larger than the LLDUC model, the sampling based on only the integer assignment of charges in the electronic configurations
may not be sufficient. We additionally consider the projection of unique UHF solutions between different active space models. Details are discussed in Supplementary Sec.~\ref{sec:sm-large-act}.

\textbf{Reference dependence: CC energies.} In Supplementary Figure~\ref{fig:sm-ref-dep-cc} we show the dependence of CCSD, CCSD(T), and CCSDT energies on the specific choice of low-energy UHF references, for some selected spin isomers. As broken-symmetry CC calculations are expensive, to reduce the computational cost we performed CCSD and CCSD(T) for 24 references with the lowest UHF energies. For CCSDT we considered 12-24 references with the lowest UHF energies. We can see that for BS10-F, BS6-A, and BS8-E, the reference with the lowest UHF energy (labeled as BS10-F-0, BS6-A-0, and BS8-E-0, respectively) is not the same as the reference with the lowest UCCSD or UCCSDT energy, which indicates that there is a non-trivial reference dependence for UCCSD solutions. The references with the lowest CC energy are BS10-F-20, BS6-A-13, and BS8-E-4 for UCCSD, and BS10-F-12, BS6-A-4, and BS8-E-23 for UCCSDT, respectively. In particular, in some cases the best UCCSD/UCCSDT solution is found with a relatively high energy UHF reference. Nevertheless, it seems to be sufficient to sample a finite number of low-energy UHF references to find CC solutions that are \emph{close enough} to the best CC solution. It can also be seen from Supplementary Figure~\ref{fig:sm-ref-dep-cc} that the UCCSDT energies mostly track the relative UCCSD energies of different references within each given spin isomer, but with a smaller spread of energies, indicating a weaker reference dependence at higher excitation levels of CC. At a very high accuracy level (mHa), the CCSD and CCSDT energy ordering can be slightly different. But this reordering is actually within the energy uncertainty of the CCSDT approximation itself. 

The CCSD(T) energies also  track the CCSD energies as the reference is varied for a given spin isomer. The T1 diagnostic is on average large ($\sim 0.1$) indicating that (T) may not be entirely reliable.
An ideal method should yield the same energy regardless of reference, for a given spin isomer.
Comparing the range of energies across different references (for each given spin isomer), we see that it increases going from CCSD to CCSD(T), highlighting the stronger reference dependence of the (T) correction. For this reason, although the (T) correction does not seem pathological in the LLDUC model, we do not consider it to be more reliable than CCSD (and indeed it may be more unreliable) for giving the ordering and energy differences of spin isomers.
In the case of BS8-E-15, the effect of the large (T) correction is partially removed in UCCSDT, again suggesting the importance of the inclusion of full triples for achieving  quantitative accuracy for such systems.

Based on the above observation of the reference dependence, in this work, we performed UCCSD and UCCSD(T) for 24 references with the lowest UHF energy for each spin isomer, and report the minimum energy obtained at each level of theory. For UCCSDT calculations we consider 12 references with the lowest UHF energies (including the reference with the lowest UCCSD energy), for each spin isomer (except for BS7-C, BS8-E, and BS8-F, for which 24 references are considered).

\begin{figure}[!htbp]
  \includegraphics[width=\linewidth]{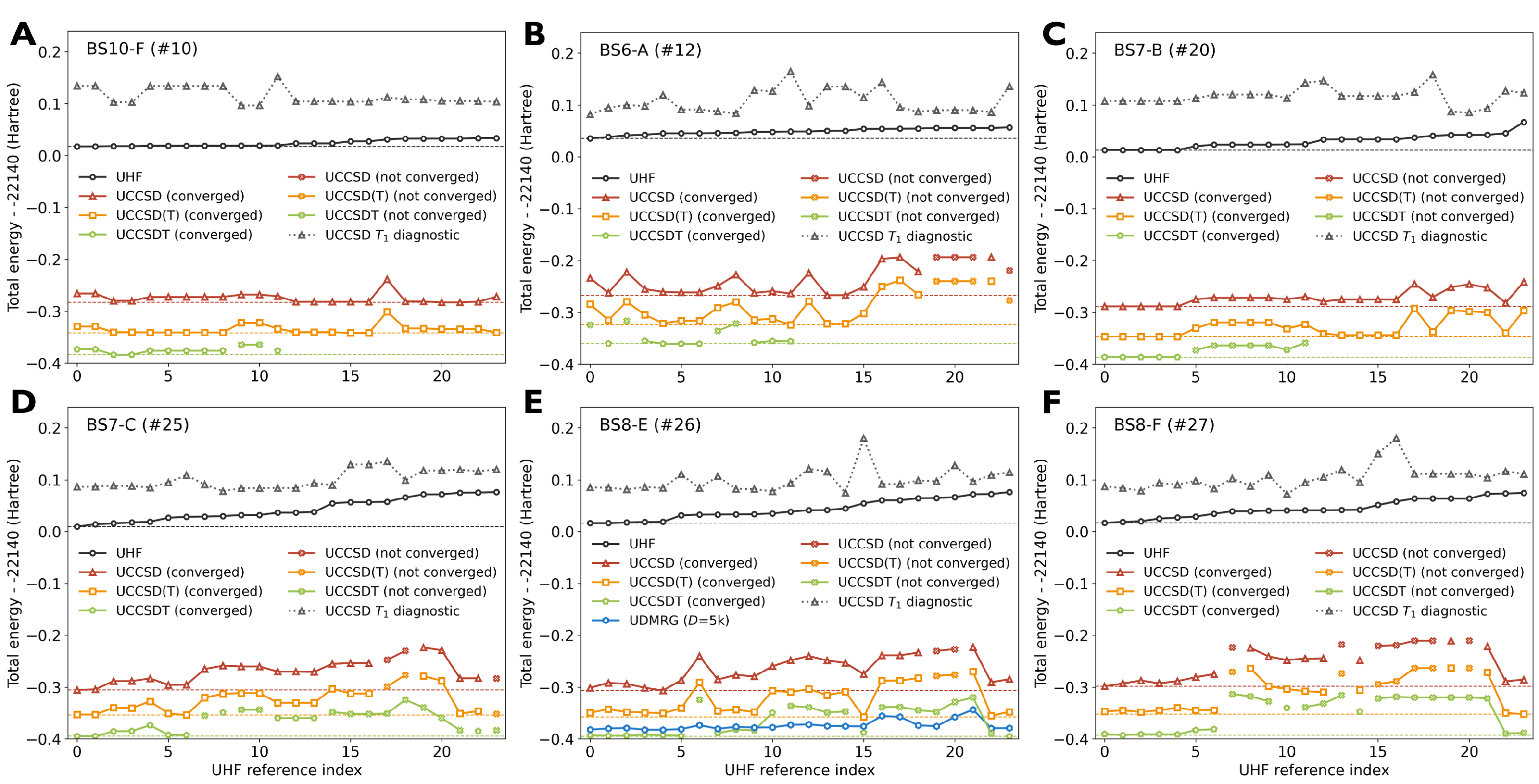}
  \caption{UHF, UCCSD, UCCSD(T), and UCCSDT energies and UCCSD $T_1$ diagnostic for different references in selected spin isomers (A) BS10-F, (B) BS6-A, (C) BS7-B, (D) BS7-C, (E) BS8-E, and (F) BS8-F, in the LLDUC model. For BS8-E, UDMRG energies (with bond dimension 5000) for the references with low CCSD energies are also shown. Dashed lines indicate the minimal energy obtained for the considered references at each level of theory.}
  \label{fig:sm-ref-dep-cc}
\end{figure}

\textbf{Reference dependence: DMRG energies.} The dependence of DMRG energies on the UHF references is computed for BS8-E (for 24 references with the lowest UHF energies and bond dimension 5000) and shown in Supplementary Figure~\ref{fig:sm-ref-dep-cc}E. Compared to UCCSD, we see that UDMRG is less dependent on the particular choice of reference. This is expected because the DMRG ansatz is more flexible for capturing multireference effects in these systems; for example, it can freely reoccupy different orbitals. The UHF reference affects the DMRG calculations only indirectly via the dependence of the UCCSD natural orbitals on the reference, and the initial quantum number distribution which is determined from the UCCSD natural occupations. Because of the reduced dependence on the UHF reference, for the large bond dimension DMRG computations performed in this work, we only considered one reference with the best UHF energy for BS7-C and BS8-E, respectively.

\textbf{Reference dependence: CC and DMRG properties.} We additionally study the reference dependence of the state properties (electron density at Fe nucleus and electron population on Fe) using the one-particle density matrices computed using UCCSD and UDMRG. The results for BS8-E are shown in Supplementary Figure~\ref{fig:sm-ref-dep-prop}. We see that UDMRG nuclear densities are visibly insensitive with respect to the change of reference, and for the electron population on Fe, its reference dependence is also much smaller than that of UCCSD. For both UCCSD and UDMRG, the ordering of the valence density is mostly unchanged when different references are used, while the ordering of the electron density at Fe nucleus (related to the M\"ossbauer isomer shift) is more robust with respect to the change of reference.

\begin{figure}[!htbp]
  \includegraphics[width=0.90\linewidth]{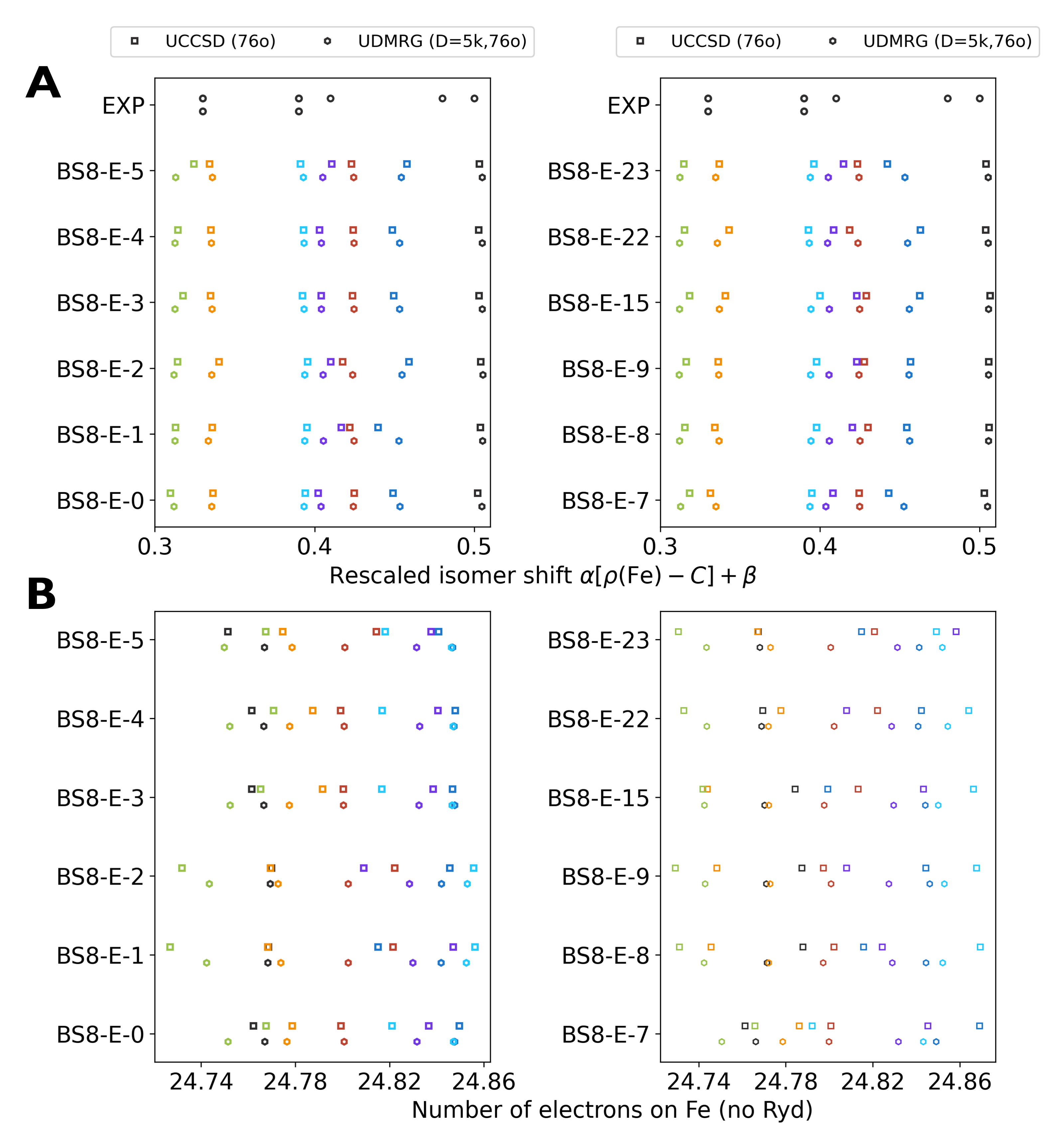}
  \centering
  \caption{UCCSD and UDMRG (A) rescaled isomer shift (computed using the electron density at the Fe nucleus) and (B) electron population on Fe (computed using meta-L\"owdin populations without Rydberg contributions) for different references in spin isomer BS8-E in the LLDUC model.}
  \label{fig:sm-ref-dep-prop}
\end{figure}

\subsubsection{UCC and UDMRG energies}
\label{sec:sm-fm-ucc-udmrg-ener}

In Supplementary Table~\ref{tab:sm-fm-ucc-udmrg-ener} we list the UHF, UCC, and UDMRG energies with the best reference for each spin isomer of the FeMo-co LLDUC model. The UHF energy is the minimum among all sampled references, the UCCSD energy is the minimum among 24 references with the lowest UHF energies, the UCCSD(T) energies are obtained using the same reference as the one for UCCSD, and the UCCSDT energy is the minimum among 12 references with the lowest UHF energies and the reference with the lowest UCCSD energy. Additionally, for BS7-C, BS8-E, and BS8-F we computed UCCSDT energies for all 24 low-UHF-energy references and took the minimum. In Supplementary Figure~\ref{fig:sm-llduc-bs-levels} we show the spin isomer relative energies computed using different theories. We see the gap between low-energy spin isomers (especially between BS7-C and BS8-E) decreases when the correlation level increases, showing the importance of a high accuracy treatment of electron correlation in the LLDUC model.

\begin{table}[!htbp]
    \scriptsize
    \centering
    \caption{UHF, UCC, and UDMRG energies with the best reference for each spin isomer of the FeMo-co LLDUC model. Energies are in Hartrees, shifted by -22140.0 Hartrees.}
    \begin{tabular}{
        >{\centering\arraybackslash}p{1.6cm}|
        >{\centering\arraybackslash}p{1.3cm}
        >{\centering\arraybackslash}p{1.3cm}
        >{\centering\arraybackslash}p{1.3cm}
        >{\centering\arraybackslash}p{1.3cm}
        >{\centering\arraybackslash}p{1.3cm}
        >{\centering\arraybackslash}p{1.3cm}
        >{\centering\arraybackslash}p{2.7cm}
    }
    \hline\hline
    spin isomer & $E_{\mathrm{UHF}}$ & $E_{\mathrm{UCCSD}}$ & $E_{\mathrm{UCCSD(T)}}$ & $E_{\mathrm{UCCSDT}}$ & $E_{\mathrm{UDMRG}}^{D=5000}$ & $E_{\mathrm{UDMRG}}^{D=8000}$ & CCSD $T_1$ diagnostic \\
    \hline
     BS3-A &   0.024129 &  -0.275898 &  -0.320929 &  -0.360651 &  -0.349340 &            &  0.08332 \\
     BS3-B &   0.026386 &  -0.275446 &  -0.327962 &  -0.361219 &  -0.350934 &            &  0.07482 \\
     BS9-A &   0.011646 &  -0.285676 &  -0.333989 &  -0.383278 &  -0.366400 &            &  0.09446 \\
    BS10-A &   0.010505 &  -0.298846 &  -0.346108 &  -0.394618 &  -0.380355 &            &  0.09817 \\
    BS10-B &   0.010424 &  -0.300672 &  -0.352890 &  -0.394707 &  -0.380576 &  -0.385609 &  0.09116 \\
     BS3-C &   0.037725 &  -0.265054 &  -0.311283 &  -0.348382 &  -0.341913 &            &  0.08548 \\
    BS10-C &   0.015230 &  -0.277971 &  -0.347828 &  -0.377909 &  -0.363839 &            &  0.11822 \\
     BS9-B &   0.021474 &  -0.271003 &  -0.327126 &  -0.370578 &  -0.356186 &            &  0.11668 \\
    BS10-D &   0.017884 &  -0.283249 &  -0.350147 &  -0.384475 &  -0.368422 &            &  0.10524 \\
    BS10-E &   0.015563 &  -0.281324 &  -0.339846 &  -0.380163 &  -0.366238 &            &  0.08382 \\
    BS10-F &   0.017734 &  -0.283147 &  -0.334302 &  -0.383722 &  -0.370353 &            &  0.10529 \\
     BS9-C &   0.024564 &  -0.271147 &  -0.338945 &  -0.373028 &  -0.359321 &            &  0.09221 \\
     BS6-A &   0.035623 &  -0.267123 &  -0.322240 &  -0.360736 &  -0.344611 &            &  0.13550 \\
     BS6-B &   0.038193 &  -0.269801 &  -0.317138 &  -0.361635 &  -0.348310 &            &  0.09217 \\
     BS6-C &   0.039963 &  -0.276379 &  -0.322975 &  -0.362158 &  -0.349788 &            &  0.08872 \\
       BS2 &   0.025921 &  -0.281779 &  -0.328731 &  -0.369197 &  -0.353921 &            &  0.11196 \\
     BS8-A &   0.021796 &  -0.279251 &  -0.337185 &  -0.379831 &  -0.367156 &            &  0.09591 \\
     BS8-B &   0.018615 &  -0.280213 &  -0.338863 &  -0.380751 &  -0.369281 &            &  0.09218 \\
     BS7-A &   0.011770 &  -0.293826 &  -0.356408 &  -0.393178 &  -0.375853 &            &  0.08827 \\
     BS8-C &   0.024204 &  -0.280004 &  -0.337721 &  -0.379041 &  -0.365101 &            &  0.10271 \\
     BS7-B &   0.013159 &  -0.288743 &  -0.346790 &  -0.386094 &  -0.370890 &            &  0.10770 \\
     BS8-D &   0.021050 &  -0.277791 &  -0.337159 &  -0.380429 &  -0.366410 &            &  0.10414 \\
     BS5-A &   0.036424 &  -0.244800 &  -0.297193 &  -0.356859 &  -0.358046 &            &  0.10624 \\
     BS5-B &   0.036072 &  -0.252662 &  -0.303041 &  -0.348706 &  -0.345641 &            &  0.10365 \\
     BS4-A &   0.033503 &  -0.274622 &  -0.323695 &  -0.359774 &  -0.347444 &            &  0.08711 \\
     BS7-C &   0.009608 &  -0.305389 &  -0.352783 &  -0.394946 &  -0.380177 &  -0.385656 &  0.08666 \\
     BS8-E &   0.016226 &  -0.306453 &  -0.350447 &  -0.395255 &  -0.382348 &  -0.387713 &  0.08429 \\
     BS8-F &   0.016720 &  -0.298387 &  -0.346968 &  -0.392966 &  -0.380937 &  -0.386193 &  0.08716 \\
     BS5-C &   0.028775 &  -0.257313 &  -0.309178 &  -0.365435 &  -0.366255 &            &  0.10850 \\
     BS4-B &   0.022972 &  -0.275849 &  -0.327340 &  -0.367484 &  -0.353577 &            &  0.10940 \\
     BS5-D &   0.029747 &  -0.258257 &  -0.317180 &  -0.357113 &  -0.354539 &            &  0.08308 \\
     BS4-C &   0.021727 &  -0.274881 &  -0.351995 &  -0.368593 &  -0.353227 &            &  0.11247 \\
     BS5-E &   0.026097 &  -0.260294 &  -0.321079 &  -0.364067 &  -0.367336 &            &  0.08462 \\
     BS5-F &   0.030319 &  -0.267021 &  -0.323392 &  -0.355363 &  -0.350530 &            &  0.08016 \\
       BS1 &   0.064277 &  -0.208181 &  -0.250076 &  -0.319433 &  -0.326568 &            &  0.09848 \\
    \hline\hline
    \end{tabular}
    \label{tab:sm-fm-ucc-udmrg-ener}
\end{table}

\begin{figure}[!htbp]
  \includegraphics[width=\linewidth]{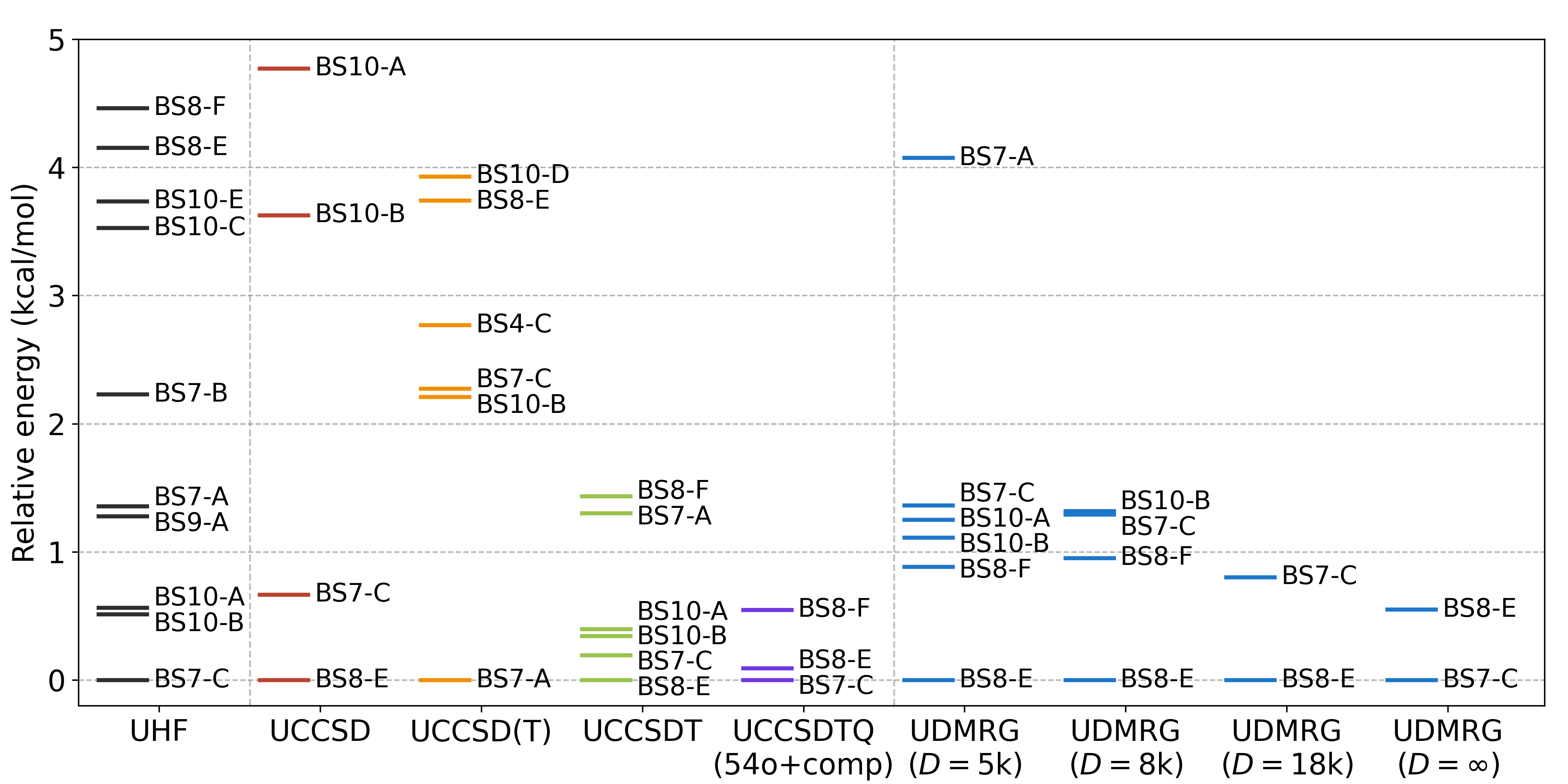}
  \centering
  \caption{UHF, UCC, and UDMRG relative energies computed for the LLDUC model for low-energy spin isomers. Note that only 4, 3, and 2 low-energy spin isomers were computed at UDMRG($D=8000$), UCCSDTQ(54o), and UDMRG($D=18000$) levels of theory, respectively.}
  \label{fig:sm-llduc-bs-levels}
\end{figure}

\subsubsection{Recovering the total spin}
\label{sec:sm-spin}

While we have shown that broken-symmetry CC and DMRG theories are effective at finding low-energy states for FeMo-co, the states do not have pure spin. (As discussed in Sec.~\ref{sec:sm-hierarchy}, these broken-symmetry states live in the manifold of a ``tower of states'' for each spin isomer). In this section, we perform a detailed analysis on the the effect of the broken-symmetry approximation, and how the pure total spin is recovered at large excitation order (for CC) and for large MPS bond dimension (for DMRG).

\textbf{The [2Fe-2S] 20-orbital active space model.} We first study the correlation between $\langle S^2\rangle$ and the energy error for UCC and UDMRG solutions, using the minimal active space model with (20o, 30e) and (20o, 31e) for Fe(III)--Fe(III) and Fe(III)--Fe(II) dimers, respectively. The active space model was defined in Ref.~\cite{li2017spin} and the integral file can be obtained from \url{https://github.com/zhendongli2008/Active-space-model-for-Iron-Sulfur-Clusters/blob/main/Fe2S2_and_Fe4S4/Fe2S2/fe2s2}. Note that the core energy -4976.26532397 Hartrees for this active space model is not included in the reported total energy. For this model we can obtain the exact energy for the pure spin state using spin-adapted DMRG (SA-DMRG), based on which the energy error for UCC (up to UCCSDTQP) and UDMRG (with different MPS bond dimensions) can be computed. The results are listed in Supplementary Table~\ref{tab:sm-fe2-std-cc}-\ref{tab:sm-fe2-mix-dmrg} and are plotted in Supplementary Figure~\ref{fig:sm-fe2-ssq}. We can see that when we increase the UHF/CC excitation order, $\langle S \rangle$ approaches the pure spin $S$ value monotonically. For UDMRG, when the MPS bond dimension $D$ is sufficiently large, $\langle S \rangle$ converges to the pure spin value. From Supplementary Figure~\ref{fig:sm-fe2-ssq}C and Supplementary Figure~\ref{fig:sm-fe2-ssq}F we see that for UDMRG the spin contamination quickly decreases once the energy error is below a threshold ($10^{-3}$ Hartrees for the [2Fe-2S] case). Therefore, the energy scale $10^{-3}$, which is related to the spin gap of the problem, can be viewed as the error caused by the broken-symmetry scheme for this system, and is removed once the accuracy of the theory itself is sufficient to distinguish different states at this scale, or in the energy extrapolation procedure (to infinite MPS bond dimension or infinite CC excitation order).

\begin{table}[!htbp]
    \centering
    \caption{UHF, UCC, and the exact (estimated using large bond dimension spin-adapted DMRG) energies, $\langle S^2\rangle, 2\langle S\rangle$, and $2S_z$ for a (20o, 30e) active space model of the Fe(III)--Fe(III) dimer.}
    \begin{tabular}{
        >{\centering\arraybackslash}p{3.5cm}|
        >{\centering\arraybackslash}p{2.8cm}
        >{\centering\arraybackslash}p{1.5cm}
        >{\centering\arraybackslash}p{1.5cm}
        >{\centering\arraybackslash}p{1.9cm}
        >{\centering\arraybackslash}p{1.9cm}
    }
    \hline\hline
    theory & energy (Ha) & $\langle S^2\rangle$ &
    $2\langle S\rangle$ & $2S_z$ (Fe1)  & $2S_z$ (Fe2) \\
    \hline
             UHF &  -116.512692 &   4.893150 &   3.535703 &  -4.197729 &   4.200378 \\
           UCCSD &  -116.580470 &   3.601338 &   2.924965 &  -3.916799 &   3.919647 \\
        UCCSD(T) &  -116.596038 &   3.104838 &   2.663243 &  -3.818287 &   3.821016 \\
          UCCSDT &  -116.600883 &   2.719797 &   2.446620 &  -3.763482 &   3.766248 \\
         UCCSDTQ &  -116.604181 &   1.600034 &   1.720319 &  -3.418594 &   3.421149 \\
        UCCSDTQP &  -116.604756 &   1.077378 &   1.304238 &  -2.931590 &   2.933808 \\
SA-DMRG[$D$=12k] &  -116.605609 &   0.000000 &   0.000000 &            &            \\
    \hline\hline
    \end{tabular}
    \label{tab:sm-fe2-std-cc}
\end{table}

\begin{table}[!htbp]
    \centering
    \caption{UHF UCC, and the exact (estimated using large bond dimension spin-adapted DMRG) energies, $\langle S^2\rangle, 2\langle S\rangle$, and $2S_z$ for a (20o, 31e) active space model of the Fe(III)--Fe(II) dimer.}
    \begin{tabular}{
        >{\centering\arraybackslash}p{3.5cm}|
        >{\centering\arraybackslash}p{2.8cm}
        >{\centering\arraybackslash}p{1.5cm}
        >{\centering\arraybackslash}p{1.5cm}
        >{\centering\arraybackslash}p{1.9cm}
        >{\centering\arraybackslash}p{1.9cm}
    }
    \hline\hline
    theory & energy (Ha) & $\langle S^2\rangle$ &
    $2\langle S\rangle$ & $2S_z$ (Fe1)  & $2S_z$ (Fe2) \\
    \hline
             UHF &  -116.322099 &   4.660419 &   3.431893 &  -3.773833 &   4.125092 \\
           UCCSD &  -116.361322 &   3.294955 &   2.765610 &  -3.603649 &   3.918227 \\
        UCCSD(T) &  -116.367095 &   2.942437 &   2.573479 &  -3.537574 &   3.866800 \\
          UCCSDT &  -116.372305 &   2.276939 &   2.179270 &  -3.332917 &   3.733398 \\
         UCCSDTQ &  -116.373827 &   1.489036 &   1.637450 &  -2.739205 &   3.252087 \\
        UCCSDTQP &  -116.373996 &   1.197270 &   1.406051 &  -2.212663 &   2.777618 \\
SA-DMRG[$D$=12k] &  -116.374322 &   0.750000 &   1.000000 &            &            \\
    \hline\hline
    \end{tabular}
    \label{tab:sm-fe2-mix-cc}
\end{table}

\begin{table}[!htbp]
    \centering
    \caption{UDMRG energies, $\langle S^2\rangle, 2\langle S\rangle$, and $2S_z$ at different bond dimensions for a (20o, 30e) active space model of the Fe(III)--Fe(III) dimer.}
    \begin{tabular}{
        >{\centering\arraybackslash}p{3.0cm}|
        >{\centering\arraybackslash}p{3cm}
        >{\centering\arraybackslash}p{1.5cm}
        >{\centering\arraybackslash}p{1.5cm}
        >{\centering\arraybackslash}p{1.9cm}
        >{\centering\arraybackslash}p{1.9cm}
    }
    \hline\hline
    MPS $D$ & energy (Ha) & $\langle S^2\rangle$ &
    $2\langle S\rangle$ & $2S_z$ (Fe1)  & $2S_z$ (Fe2) \\
    \hline
   100 &  -116.598906 &   3.595044 &   2.921757 &  -3.869613 &   3.870381 \\
   200 &  -116.601798 &   2.561228 &   2.353343 &  -3.734341 &   3.734903 \\
   400 &  -116.603444 &   1.551455 &   1.684366 &  -3.409451 &   3.411074 \\
   600 &  -116.604077 &   1.075703 &   1.302783 &  -3.082867 &   3.084524 \\
   800 &  -116.604421 &   0.789053 &   1.038679 &  -2.762896 &   2.764400 \\
  1000 &  -116.604571 &   0.492156 &   0.722970 &  -2.321323 &   2.322487 \\
  1200 &  -116.604751 &   0.135284 &   0.241425 &  -1.330488 &   1.331083 \\
  1600 &  -116.605170 &   0.014958 &   0.029482 &  -0.439387 &   0.439701 \\
  2000 &  -116.605377 &   0.010609 &   0.020997 &  -0.373072 &   0.373298 \\
  2400 &  -116.605476 &   0.011585 &   0.022907 &  -0.391477 &   0.391762 \\
    \hline\hline
    \end{tabular}
    \label{tab:sm-fe2-std-dmrg}
\end{table}

\begin{table}[!htbp]
    \centering
    \caption{UDMRG energies, $\langle S^2\rangle, 2\langle S\rangle$, and $2S_z$ at different bond dimensions for a (20o, 31e) active space model of the Fe(III)--Fe(II) dimer.}
    \begin{tabular}{
        >{\centering\arraybackslash}p{3.0cm}|
        >{\centering\arraybackslash}p{3cm}
        >{\centering\arraybackslash}p{1.5cm}
        >{\centering\arraybackslash}p{1.5cm}
        >{\centering\arraybackslash}p{1.9cm}
        >{\centering\arraybackslash}p{1.9cm}
    }
    \hline\hline
    MPS $D$ & energy (Ha) & $\langle S^2\rangle$ &
    $2\langle S\rangle$ & $2S_z$ (Fe1)  & $2S_z$ (Fe2) \\
    \hline
   100 &  -116.371550 &   2.284151 &   2.183803 &  -3.332811 &   3.738070 \\
   200 &  -116.372867 &   1.557705 &   1.689018 &  -2.942814 &   3.408071 \\
   400 &  -116.373636 &   1.017338 &   1.251522 &  -2.306630 &   2.858374 \\
   600 &  -116.373982 &   0.804693 &   1.053965 &  -1.736648 &   2.363313 \\
   800 &  -116.374146 &   0.758738 &   1.008719 &  -1.401038 &   2.072533 \\
  1000 &  -116.374230 &   0.752147 &   1.002146 &  -1.278340 &   1.966404 \\
  1200 &  -116.374273 &   0.751040 &   1.001040 &  -1.238854 &   1.932182 \\
  1600 &  -116.374306 &   0.750770 &   1.000770 &  -1.225353 &   1.920452 \\
  2000 &  -116.374317 &   0.750908 &   1.000908 &  -1.231956 &   1.926105 \\
  2400 &  -116.374321 &   0.751185 &   1.001184 &  -1.243982 &   1.936448 \\
    \hline\hline
    \end{tabular}
    \label{tab:sm-fe2-mix-dmrg}
\end{table}

\begin{figure}[!htbp]
  \includegraphics[width=\linewidth]{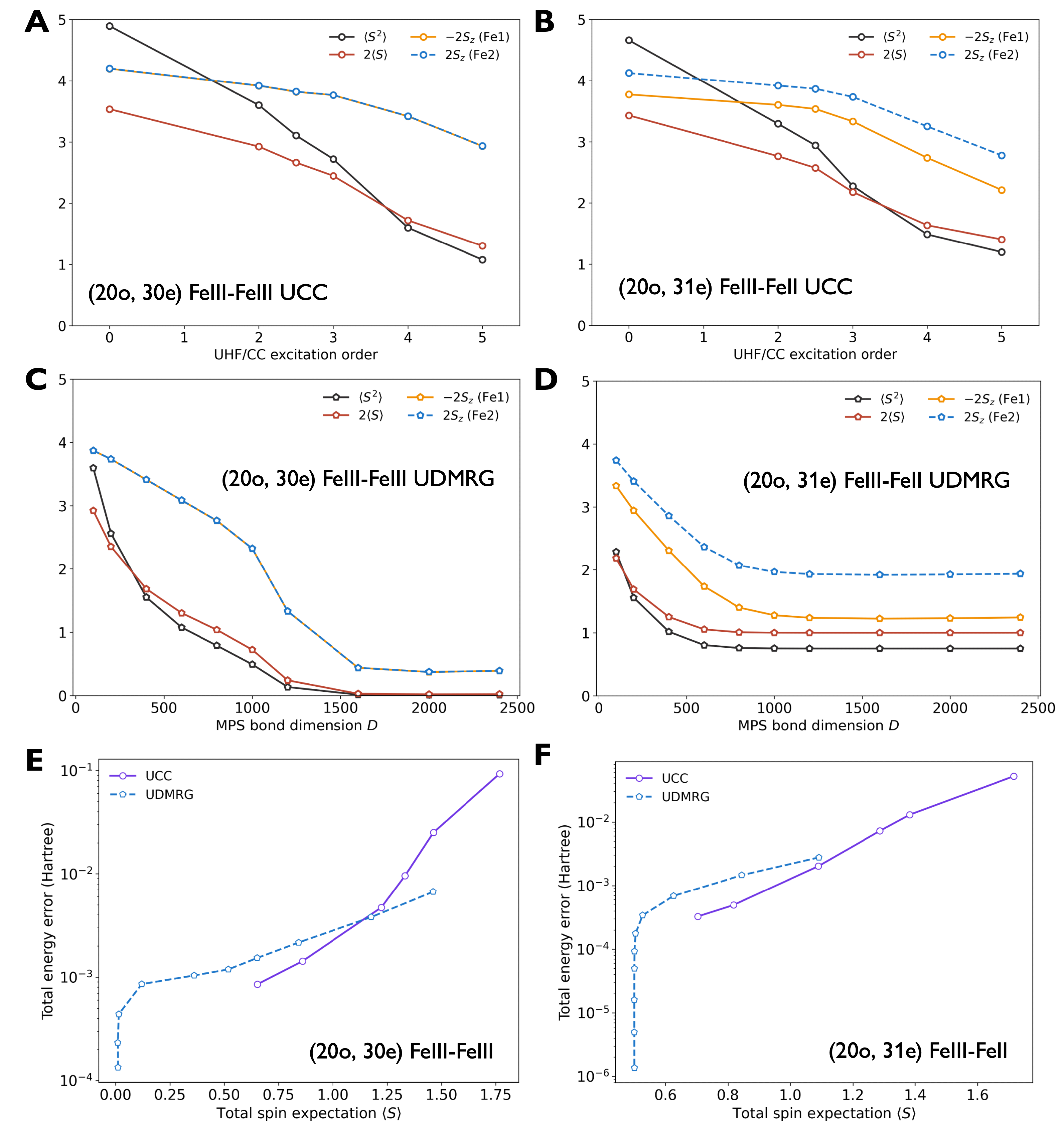}
  \caption{
  \textbf{A,B} $\langle S^2\rangle, 2\langle S\rangle$, and $2S_z$ for the (20o, 30e) and (20o, 31e) active space models of the [2Fe2S] dimer as a function of CC excitation order (with UHF = 0, UCCSD(T) = 2.5).
  \textbf{C,D} $\langle S^2\rangle, 2\langle S\rangle$, and $2S_z$ for the (20o, 30e) and (20o, 31e) active space models of the [2Fe2S] dimer as a function of MPS bond dimension $D$.
  \textbf{E} Correlation between energy error and $\langle S\rangle$ for the Fe(III)--Fe(III) model.
  \textbf{F} Correlation between energy error and $\langle S\rangle$ for the Fe(III)--Fe(II) model.
  }
  \label{fig:sm-fe2-ssq}
\end{figure}

\textbf{The FeMo-co LLDUC model.} We further study $\langle S^2 \rangle$ for the FeMo-co LLDUC model. We list the results in Supplementary Tables~\ref{tab:sm-fm-cc-ssq} and~\ref{tab:sm-fm-dmrg-ssq}. The correlation between $\langle S \rangle$ and the FeMo-co LLDUC model total energy (computed using UHF, UCCSD, UCCSDT, and UDMRG) is shown in Supplementary Figure~\ref{fig:sm-fm-ssq}. We see that the $\langle S \rangle$ computed at UCCSDT and UDMRG ($D = 11000)$ is between 5/2 and 3 [these do not correspond to the highest levels of theory used for the energy (FNO-UCCSDTQ and UDMRG ($D=18000$)]. As $\langle S\rangle$ is converging from above, we conclude the ground state $S \leq 5/2$, which is consistent with the experimental $S = 3/2$ for this system. However, we do not have a guarantee that the LLDUC model shares the same ground-state $S$ as the true system.

\begin{table}[!htbp]
    \centering
    \caption{UHF, UCCSD, and UCCSDT energies (in Hartrees, shifted by -22140.0 Hartrees) and $\langle S^2 \rangle$ of BS7-C, BS8-E, and BS8-F states computed for the LLDUC model.}
    \begin{tabular}{
        >{\centering\arraybackslash}p{2.2cm}|
        >{\centering\arraybackslash}p{1.8cm}
        >{\centering\arraybackslash}p{1.8cm}
        >{\centering\arraybackslash}p{1.8cm}
        >{\centering\arraybackslash}p{1.8cm}
        >{\centering\arraybackslash}p{1.8cm}
        >{\centering\arraybackslash}p{1.8cm}
    }
    \hline\hline
    spin isomer & $E_\mathrm{UHF}$ & $\langle S^2\rangle_\mathrm{UHF}$ & $E_\mathrm{UCCSD}$ & $\langle S^2\rangle_\mathrm{UCCSD}$ & $E_\mathrm{UCCSDT}$ & $\langle S^2\rangle_\mathrm{UCCSDT}$ \\
    \hline
     BS7-C &   0.009608 &    17.88 &  -0.305389 &    11.92 &  -0.394946 &     9.31 \\
     BS8-E &   0.016226 &    17.90 &  -0.306453 &    12.62 &  -0.395255 &     8.22 \\
     BS8-F &   0.016720 &    17.96 &  -0.298387 &    12.56 &  -0.392966 &    10.43 \\
    \hline\hline
    \end{tabular}
    \label{tab:sm-fm-cc-ssq}
\end{table}

\begin{table}[!htbp]
    \centering
    \caption{UDMRG energies (in Hartrees, shifted by -22140.0 Hartrees) and $\langle S^2 \rangle$ of BS8-E computed for the LLDUC model.}
    \begin{tabular}{
        >{\centering\arraybackslash}p{2.5cm}
        >{\centering\arraybackslash}p{2cm}
        >{\centering\arraybackslash}p{2cm}|
        >{\centering\arraybackslash}p{2.5cm}
        >{\centering\arraybackslash}p{2cm}
        >{\centering\arraybackslash}p{2cm}
    }
    \hline\hline
    MPS $D$ & $E_\mathrm{UDMRG}$ & $\langle S^2\rangle_\mathrm{UDMRG}$ & MPS $D$ & $E_\mathrm{UDMRG}$ & $\langle S^2\rangle_\mathrm{UDMRG}$ \\
    \hline
   5000 &    -0.382131 &    12.27 &    8000 &    -0.388369 &    11.60 \\
   6000 &    -0.384697 &    12.02 &    9000 &    -0.389741 &    11.42 \\
   7000 &    -0.386720 &    11.80 &   11000 &    -0.391926 &    11.20 \\
    \hline\hline
    \end{tabular}
    \label{tab:sm-fm-dmrg-ssq}
\end{table}

\begin{figure}[!htbp]
  \includegraphics[width=0.75\linewidth]{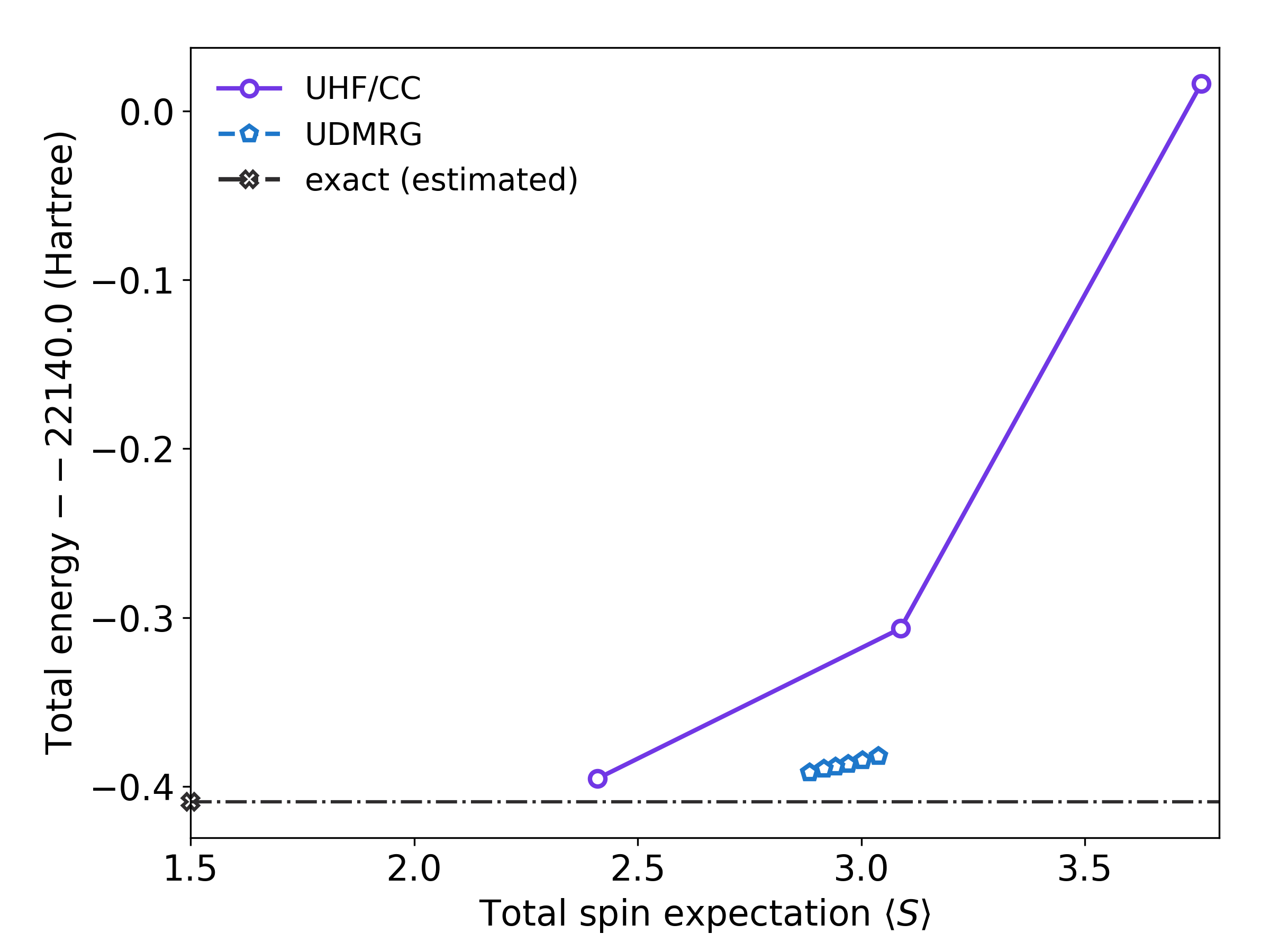}
  \centering
  \caption{
  Correlation between shifted total energy and $\langle S\rangle$ of BS8-E computed for the LLDUC model.
  }
  \label{fig:sm-fm-ssq}
\end{figure}

\subsubsection{Frozen natural orbital UCC and UDMRG}
\label{sec:sm-fno}

The large difference between the UCCSD(T) and UCCSDT energies indicates that the full quadruple (Q) contribution can be significant for the LLDUC model. Since UCCSDTQ is too expensive to compute (with our resources) for the complete 76-orbital LLDUC model, we perform UCCSDTQ in a series of smaller active subspaces defined using frozen CCSD natural orbitals (FNO, with spin symmetry breaking). In the FNO model, we keep all 18 and 21 (for $\alpha$ and $\beta$ spin, respectively) virtual orbitals in all FNO subspaces. For the occupied orbitals, we first compute the broken-symmetry UCCSD natural orbitals (for a given spin isomer) by separately diagonalizing the occupied-occupied block of the UCCSD one-particle density matrix for $\alpha$ and $\beta$ spins, and then freeze $22, 26, 30, \cdots, 46$ $\alpha$ and $\beta$ occupied natural orbitals with the highest natural orbital occupation. The ``FNO occupation cutoff'' is then defined as the sum of the $\alpha$ and $\beta$ spin natural orbital cutoffs (see Supplementary Table~\ref{tab:sm-fno-occ}).

We also perform UDMRG (up to MPS bond dimension 8000) within each of the FNO subspaces, and use the extrapolated UDMRG energies to estimate the error in UCCSDTQ. The results are listed in Supplementary Table~\ref{tab:sm-fno-error}. In the smaller subspaces (e.g. $\leq 38$o for BS7-C, BS8-F, and $\leq 46$o for BS8-E), the extrapolated UDMRG number lies below the UCCSDTQ number by at most 0.99~mHa. In the larger subspaces, the UDMRG number lies above the UCCSDTQ number by at most 0.95~mHa. 
For the smallest FNO space with (30o, 21e), we also computed UCCSDTQP energies and the results are listed in Supplementary Table~\ref{tab:sm-fno-ccsdtqp}. The largest energy difference between the extrapolated UDMRG and UCCSDTQP energies in the (30o, 21e) is 0.38 mHa (with the UDMRG number lying below the UCCSDTQP energy). 

There are two interpretations of the data in the FNO space. The first is that the UCCSDTQ energies become non-variational in the larger active spaces. Supporting this is the fact that the UCCSDTQP energy for (30o, 21e) actually lies slightly (0.009~mHa) above the UCCSDTQ energy for BS8-E, suggesting non-uniform convergence with excitation level, although the UDMRG estimate for the exact number still supports the variationality of the UCCSDTQ energy.
The second is that the extrapolated DMRG energies lie above the true energy in the larger active spaces. Since the extrapolation distance for DMRG increases in the larger active spaces, the accuracy of the extrapolation is clearly lower. Without further evidence, it is not possible to pick between these scenarios with high confidence, other than to say that in all cases the data supports the UCCSDTQ energy being either variational, or very close (within 1 kcal/mol) of the `exact' estimate. In the iron cubane (SM Sec.~\ref{sec:sm-fe4}), where very accurate variational DMRG numbers can be produced, we show that the UCCSDTQ energy is indeed variational and that the extrapolated UDMRG number can sometimes be above this variational estimate.

As shown in Fig. 3B, we see a large curvature in the FNO-UCCSDTQ composite corrected UCCSDT energy versus FNO occupation cutoff curve, which prevents the composite energy from being simply extrapolated using linear regression. This may be partially explained by the non-linear change in $\langle S^2\rangle$ when considering different FNO subspaces (see Supplementary Table~\ref{tab:sm-fno-ssq}). In fact, from Supplementary Figure~\ref{fig:sm-fm-fno-ssq} we see a similar non-linear trend for the $\langle S\rangle$ computed for UCCSD and UCCSDT in different FNO subspaces. As each FNO active space series is defined for a specific spin isomer (to compute the UCCSD natural orbitals), the energy of states with other spin couplings will be significantly increased in the small FNO spaces, and the near-degeneracy among the spin isomers will be recovered only when the FNO space is very close to the full 76-orbital space.

\begin{table}[!htbp]
    \small
    \centering
    \caption{Natural orbital cutoff in FNO subspaces for BS7-C, BS8-E, and BS8-F states for the LLDUC model.}
    \begin{tabular}{
        >{\centering\arraybackslash}p{1.8cm}|
        >{\centering\arraybackslash}p{1.59cm}
        >{\centering\arraybackslash}p{1.59cm}
        >{\centering\arraybackslash}p{1.59cm}
        >{\centering\arraybackslash}p{1.59cm}
        >{\centering\arraybackslash}p{1.59cm}
        >{\centering\arraybackslash}p{1.59cm}
        >{\centering\arraybackslash}p{1.59cm}
    }
    \hline\hline
    spin isomer & (30o, 21e) & (34o, 29e) & (38o, 37e) & (42o, 45e) &
    (46o, 53e) & (50o, 61e) & (54o, 69e) \\
    \hline
     BS7-C &   1.937625 &   1.964417 &   1.975624 &   1.984529 &   1.989548 &   1.992606 &   1.994763 \\
     BS8-E &   1.931639 &   1.965347 &   1.976265 &   1.987435 &   1.990742 &   1.993289 &   1.995066 \\
     BS8-F &   1.934736 &   1.968874 &   1.977565 &   1.985572 &   1.991067 &   1.993575 &   1.995543 \\
    \hline\hline
    \end{tabular}
    \label{tab:sm-fno-occ}
\end{table}

\begin{table}[!htbp]
    \scriptsize
    \centering
    \caption{UHF, UCC and UDMRG energies in the FNO subspaces for BS7-C, BS8-E, and BS8-F states for the LLDUC model. Energies are in Hartrees shifted by $-22140.0$ Hartrees. Extrapolated UDMRG energies are computed using the $\exp[-\kappa(\log D)^2]$ fitting based on data with MPS bond dimensions $D = 7500, 7000, \cdots, 5000$.}
    \begin{tabular}{
        >{\centering\arraybackslash}p{1.3cm}|
        >{\centering\arraybackslash}p{1.8cm}
        >{\centering\arraybackslash}p{1.1cm}
        >{\centering\arraybackslash}p{1.15cm}
        >{\centering\arraybackslash}p{1.15cm}
        >{\centering\arraybackslash}p{1.15cm}
        >{\centering\arraybackslash}p{1.15cm}
        >{\centering\arraybackslash}p{1.15cm}
        >{\centering\arraybackslash}p{3.0cm}
    }
    \hline\hline
    spin isomer & FNO subspace & $E_{\mathrm{UHF}}$ & $E_{\mathrm{UCCSD}}$ & $E_{\mathrm{UCCSDT}}$ & $E_{\mathrm{UCCSDTQ}}$ & $E_{\mathrm{UDMRG}}^{D=8000}$ & $E_{\mathrm{UDMRG}}^{\mathrm{extrap}}$ & $E_{\mathrm{UCCSDTQ}} - E_{\mathrm{UDMRG}}^{\mathrm{extrap}}$ \\
    \hline
    BS7-C & (30o, 21e) &   0.013987 &  -0.102385 &  -0.123010 &  -0.126991 &  -0.127483 &  -0.127539 &  +0.000548 \\
           & (34o, 29e) &   0.013987 &  -0.154096 &  -0.188616 &  -0.193582 &  -0.193110 &  -0.193973 &  +0.000391 \\
           & (38o, 37e) &   0.013987 &  -0.190847 &  -0.235962 &  -0.241705 &  -0.240057 &  -0.241741 &  +0.000036 \\
           & (42o, 45e) &   0.013987 &  -0.215455 &  -0.268514 &  -0.274840 &  -0.270758 &  -0.274277 &  -0.000563 \\
           & (46o, 53e) &   0.013987 &  -0.237776 &  -0.298481 &  -0.305408 &  -0.300172 &  -0.304977 &  -0.000431 \\
           & (50o, 61e) &   0.013987 &  -0.255100 &  -0.321765 &  -0.329905 &  -0.319909 &  -0.328953 &  -0.000952 \\
           & (54o, 69e) &   0.013987 &  -0.269585 &  -0.342146 &  -0.351627 &            &            &            \\
     \hline
     BS8-E & (30o, 21e) &   0.076451 &  -0.098580 &  -0.136016 &  -0.136728 &  -0.136941 &  -0.137102 &  +0.000374 \\
           & (34o, 29e) &   0.076451 &  -0.145703 &  -0.196855 &  -0.200313 &  -0.200097 &  -0.201260 &  +0.000947 \\
           & (38o, 37e) &   0.076451 &  -0.181494 &  -0.243404 &  -0.248464 &  -0.246454 &  -0.249325 &  +0.000861 \\
           & (42o, 45e) &   0.076451 &  -0.206176 &  -0.276008 &  -0.281959 &  -0.278102 &  -0.282951 &  +0.000992 \\
           & (46o, 53e) &   0.076451 &  -0.226300 &  -0.303998 &  -0.310556 &  -0.302324 &  -0.311081 &  +0.000524 \\
           & (50o, 61e) &   0.076451 &  -0.242257 &  -0.327246 &  -0.335307 &  -0.323274 &  -0.335004 &  -0.000303 \\
           & (54o, 69e) &   0.076451 &  -0.255650 &  -0.347246 &  -0.356269 &            &            &            \\
     \hline
     BS8-F & (30o, 21e) &   0.018614 &  -0.112553 &  -0.141566 &  -0.146057 &  -0.146808 &  -0.146984 &  +0.000927 \\
           & (34o, 29e) &   0.018614 &  -0.158841 &  -0.203066 &  -0.209615 &  -0.209593 &  -0.210390 &  +0.000775 \\
           & (38o, 37e) &   0.018614 &  -0.190547 &  -0.245325 &  -0.252449 &  -0.251102 &  -0.252766 &  +0.000317 \\
           & (42o, 45e) &   0.018614 &  -0.215290 &  -0.279408 &  -0.287242 &  -0.282822 &  -0.287116 &  -0.000125 \\
           & (46o, 53e) &   0.018614 &  -0.235811 &  -0.306788 &  -0.315083 &  -0.306934 &  -0.315018 &  -0.000064 \\
           & (50o, 61e) &   0.018614 &  -0.252355 &  -0.329961 &  -0.339335 &  -0.328617 &  -0.338639 &  -0.000696 \\
           & (54o, 69e) &   0.018614 &  -0.265288 &  -0.348697 &  -0.359283 &            &            &            \\
    \hline\hline
    \end{tabular}
    \label{tab:sm-fno-error}
\end{table}

\begin{table}[!htbp]
    \small
    \centering
    \caption{UCCSDTQP and UDMRG energies in the FNO subspace (30o, 21e) for BS7-C, BS8-E, and BS8-F states for the LLDUC model. Energies are in Hartrees shifted by $-22140.0$ Hartrees. Extrapolated UDMRG energies are computed using the $\exp[-\kappa(\log D)^2]$ fitting based on data with MPS bond dimensions $D = 7500, 7000, \cdots, 5000$.}
    \begin{tabular}{
        >{\centering\arraybackslash}p{1.8cm}|
        >{\centering\arraybackslash}p{2.3cm}
        >{\centering\arraybackslash}p{2.0cm}
        >{\centering\arraybackslash}p{1.7cm}
        >{\centering\arraybackslash}p{1.7cm}
        >{\centering\arraybackslash}p{3.8cm}
    }
    \hline\hline
    spin isomer & FNO subspace & $E_{\mathrm{UCCSDTQP}}$ & $E_{\mathrm{UDMRG}}^{D=8000}$ & $E_{\mathrm{UDMRG}}^{\mathrm{extrap}}$ & $E_{\mathrm{UCCSDTQP}} - E_{\mathrm{UDMRG}}^{\mathrm{extrap}}$ \\
    \hline
     BS7-C & (30o, 21e) &  -0.127496 &  -0.127483 &  -0.127539 &  +0.000043 \\
     BS8-E & (30o, 21e) &  -0.136719 &  -0.136941 &  -0.137102 &  +0.000383 \\
     BS8-F & (30o, 21e) &  -0.146846 &  -0.146808 &  -0.146984 &  +0.000138 \\
    \hline\hline
    \end{tabular}
    \label{tab:sm-fno-ccsdtqp}
\end{table}

\begin{table}[!htbp]
    \small
    \centering
    \caption{UCC and UDMRG $\langle S^2\rangle$ in the FNO subspaces for BS7-C, BS8-E, and BS8-F states for the LLDUC model.}
    \begin{tabular}{
        >{\centering\arraybackslash}p{1.8cm}|
        >{\centering\arraybackslash}p{2.3cm}
        >{\centering\arraybackslash}p{1.6cm}
        >{\centering\arraybackslash}p{1.7cm}
        >{\centering\arraybackslash}p{2.0cm}
        >{\centering\arraybackslash}p{1.9cm}
        >{\centering\arraybackslash}p{1.9cm}
    }
    \hline\hline
    spin isomer & FNO subspace & $\langle S^2 \rangle_{\mathrm{UCCSD}}$ & $\langle S^2 \rangle_{\mathrm{UCCSDT}}$ & $\langle S^2 \rangle_{\mathrm{UCCSDTQ}}$ & $\langle S^2 \rangle_{\mathrm{UDMRG}}^{D=5000}$ & $\langle S^2 \rangle_{\mathrm{UDMRG}}^{D=8000}$ \\
    \hline
     BS7-C & (30o, 21e) &    16.70 &    16.66 &    16.66 &    16.66 &    16.66 \\
           & (34o, 29e) &    16.52 &    16.46 &          &    16.49 &    16.48 \\
           & (38o, 37e) &    16.15 &    16.01 &          &    16.07 &    16.05 \\
           & (42o, 45e) &    15.79 &    15.59 &          &    15.71 &    15.67 \\
           & (46o, 53e) &    15.30 &    14.95 &          &    15.15 &    15.08 \\
           & (50o, 61e) &    14.80 &    14.27 &          &    14.74 &    14.62 \\
           & (54o, 69e) &    14.14 &    13.31 &          &          &          \\
     \hline
     BS8-E & (30o, 21e) &    16.77 &    16.68 &    16.52 &    16.52 &    16.52 \\
           & (34o, 29e) &    16.53 &    16.41 &          &    16.34 &    16.33 \\
           & (38o, 37e) &    16.16 &    16.02 &          &    16.12 &    16.09 \\
           & (42o, 45e) &    15.87 &    15.65 &          &    15.86 &    15.82 \\
           & (46o, 53e) &    15.46 &    15.13 &          &    15.61 &    15.53 \\
           & (50o, 61e) &    14.89 &    14.30 &          &    15.20 &    15.08 \\
           & (54o, 69e) &    14.13 &    12.94 &          &          &          \\
     \hline
     BS8-F & (30o, 21e) &    16.74 &    16.63 &    16.66 &    16.67 &    16.67 \\
           & (34o, 29e) &    16.48 &    16.35 &          &    16.44 &    16.43 \\
           & (38o, 37e) &    16.11 &    15.96 &          &    16.09 &    16.08 \\
           & (42o, 45e) &    15.76 &    15.53 &          &    15.77 &    15.72 \\
           & (46o, 53e) &    15.42 &    15.10 &          &    15.49 &    15.40 \\
           & (50o, 61e) &    15.08 &    14.57 &          &    15.11 &    14.98 \\
           & (54o, 69e) &    14.73 &    14.03 &          &          &          \\
    \hline\hline
    \end{tabular}
    \label{tab:sm-fno-ssq}
\end{table}

\begin{figure}[!htbp]
  \includegraphics[width=0.7\linewidth]{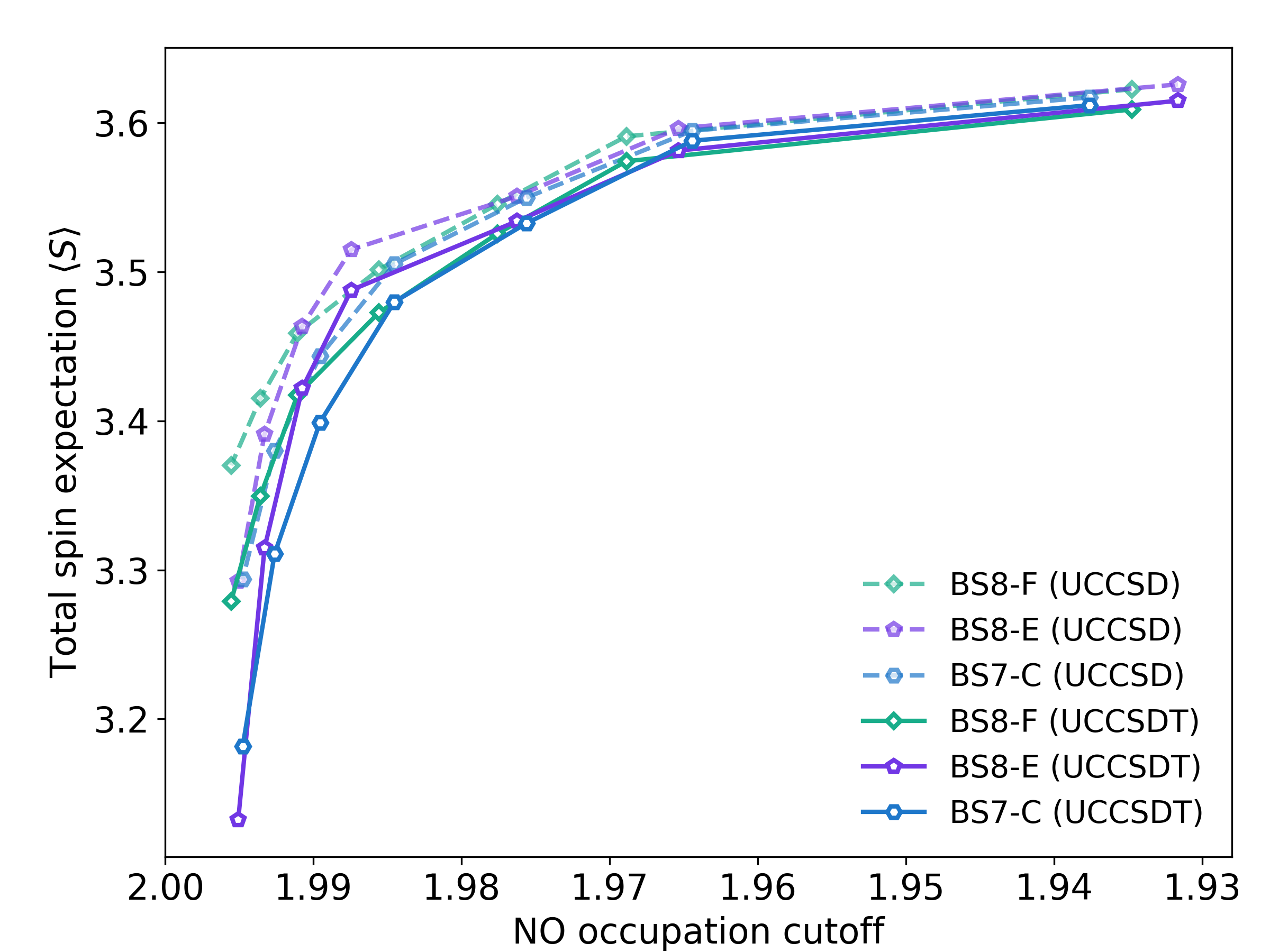}
  \centering
  \caption{UHF/CCSD and UHF/CCSDT $\langle S\rangle$ in the FNO subspaces for BS7-C, BS8-E, and BS8-F states for the LLDUC model.}
  \label{fig:sm-fm-fno-ssq}
\end{figure}

\textbf{FNO-UCCSDTQ composite correction.} The FNO-UCCSDTQ composite energy is computed as
\begin{equation}
    E_{\mathrm{compsosite}} = E_{\mathrm{UCCSDT}} + E_{\mathrm{FNO-UCCSDTQ}}
        - E_{\mathrm{FNO-UCCSDT}}
\end{equation}

We list the FNO-UCCSD, FNO-UCCSDT, FNO-UCCSDTQ, full space UCCSD, UCCSDT, and the composite energies for BS7-C, BS8-E, and BS8-F in Supplementary Table~\ref{tab:sm-fno-composite}. Extrapolation was done using linear fitting for the 46o, 50o, and 54o energies with respect to UCCSD natural orbital occupation cutoff.
Note that the UCCSD energies listed in Supplementary Table~\ref{tab:sm-fno-composite} were computed using the same reference as that of UCCSDT.

\begin{table}[!htbp]
    \centering
    \caption{UCCSD, UCCSDT and UCCSDTQ energies in the FNO subspaces, and full space UCCSD, UCCSDT and composite energies for BS7-C, BS8-E, and BS8-F states for the LLDUC model. Energies are in Hartrees shifted by $-22140.0$ Hartrees. Extrapolation was done using linear fitting for 46o, 50o, and 54o energies with respect to UCCSD natural orbital occupation cutoff. The listed UCCSD energies use the same reference as that of UCCSDT, namely the reference for the best full space UCCSDT energy.}
    \begin{tabular}{
        >{\centering\arraybackslash}p{2.0cm}|
        >{\centering\arraybackslash}p{2.6cm}
        >{\centering\arraybackslash}p{2.1cm}
        >{\centering\arraybackslash}p{2.1cm}
        >{\centering\arraybackslash}p{2.5cm}
        >{\centering\arraybackslash}p{2.1cm}
    }
    \hline\hline
    spin isomer & FNO subspace & $E_{\mathrm{UCCSD}}$ & $E_{\mathrm{UCCSDT}}$ & $E_{\mathrm{UCCSDTQ}}$ & $E_{\mathrm{composite}}$ \\
    \hline
     BS7-C & (30o, 21e) &  -0.102385 &  -0.123010 &  -0.126991 &  -0.398928 \\
           & (34o, 29e) &  -0.154096 &  -0.188616 &  -0.193582 &  -0.399912 \\
           & (38o, 37e) &  -0.190847 &  -0.235962 &  -0.241705 &  -0.400689 \\
           & (42o, 45e) &  -0.215455 &  -0.268514 &  -0.274840 &  -0.401272 \\
           & (46o, 53e) &  -0.237776 &  -0.298481 &  -0.305408 &  -0.401873 \\
           & (50o, 61e) &  -0.255100 &  -0.321765 &  -0.329905 &  -0.403087 \\
           & (54o, 69e) &  -0.269585 &  -0.342146 &  -0.351627 &  -0.404427 \\
           & (76o, 113e) &  -0.304318 &  -0.394946 & \textbf{extrapolated} & \textbf{-0.406849} \\
     \hline
     BS8-E & (30o, 21e) &  -0.098580 &  -0.136016 &  -0.136728 &  -0.395967 \\
           & (34o, 29e) &  -0.145703 &  -0.196855 &  -0.200313 &  -0.398713 \\
           & (38o, 37e) &  -0.181494 &  -0.243404 &  -0.248464 &  -0.400315 \\
           & (42o, 45e) &  -0.206176 &  -0.276008 &  -0.281959 &  -0.401206 \\
           & (46o, 53e) &  -0.226300 &  -0.303998 &  -0.310556 &  -0.401813 \\
           & (50o, 61e) &  -0.242257 &  -0.327246 &  -0.335307 &  -0.403316 \\
           & (54o, 69e) &  -0.255650 &  -0.347246 &  -0.356269 &  -0.404278 \\
           & (76o, 113e) &  -0.284755 &  -0.395255 & \textbf{extrapolated} & \textbf{-0.407117} \\
     \hline
     BS8-F & (30o, 21e) &  -0.112553 &  -0.141566 &  -0.146057 &  -0.397456 \\
           & (34o, 29e) &  -0.158841 &  -0.203066 &  -0.209615 &  -0.399515 \\
           & (38o, 37e) &  -0.190547 &  -0.245325 &  -0.252449 &  -0.400091 \\
           & (42o, 45e) &  -0.215290 &  -0.279408 &  -0.287242 &  -0.400800 \\
           & (46o, 53e) &  -0.235811 &  -0.306788 &  -0.315083 &  -0.401261 \\
           & (50o, 61e) &  -0.252355 &  -0.329961 &  -0.339335 &  -0.402340 \\
           & (54o, 69e) &  -0.265288 &  -0.348697 &  -0.359283 &  -0.403552 \\
           & (76o, 113e) &  -0.292974 &  -0.392966 & \textbf{extrapolated} & \textbf{-0.405741} \\
    \hline\hline
    \end{tabular}
    \label{tab:sm-fno-composite}
\end{table}

\subsubsection{Overlap with dominant determinants}
\label{sec:sm-det-overlap}

Given that the bond dimension $D=18000$ MPS obtained from UDMRG is a good approximate ground state of the LLDUC model, we can compute an estimate of the overlap of the dominant broken symmetry determinant with this  state. Using an efficient sweep-like algorithm~\cite{zhai2023block2,lee2021externally}, we can extract all determinants (defined in the split-localized broken-symmetry CCSD natural orbital basis) with the absolute value of the overlap larger than a given value. The overlap with dominant determinants (with absolute value larger than 0.05) are listed in Supplementary Table~\ref{tab:sm-mps-overlap}. We find that the overlaps with the dominant determinant are 0.4738 and 0.4468 for BS7-C and BS8-E, respectively. The next largest overlap is below 0.1 in both states.

\begin{table}[!htbp]
    \centering
    \caption{Overlap between the bond dimension $D=18000$ MPS obtained from UDMRG and the dominant determinants (with absolute value larger than 0.05).}
    \begin{tabular}{
        >{\centering\arraybackslash}p{2.0cm}|
        >{\centering\arraybackslash}p{2.0cm}
        >{\centering\arraybackslash}p{2.0cm}
        >{\centering\arraybackslash}p{2.0cm}
        >{\centering\arraybackslash}p{2.0cm}
        >{\centering\arraybackslash}p{2.0cm}
    }
    \hline\hline
    spin isomer & \multicolumn{5}{c}{determinant overlap} \\
    \hline
     BS7-C &  -0.473757 &  -0.081447 &  -0.078313 &  -0.058582 &   0.053784 \\
     \hline
     BS8-E &  -0.446836 &   0.099686 &  -0.075533 &   0.074390 &  -0.063381 \\
           &  -0.061440 &   0.056716 &  -0.055592 &   0.054155 &  -0.052423 \\
    \hline\hline
    \end{tabular}
    \label{tab:sm-mps-overlap}
\end{table}

\subsubsection{Computational cost and scaling}
\label{sec:sm-scaling}

In this section, we summarize the actual scaling and computational cost of the UCC and UDMRG methods used in this work. Performance analysis can help us establish an estimate of the classical simulation time to reach chemical accuracy for the LLDUC model. Timings are measured on AMD EPYC 9474F CPUs with 96 CPU cores and 1.5TB memory per node, supported by the Scientific Computing Core at the Flatiron Institute.

\textbf{UDMRG.} We perform UDMRG for the BS7-C and BS8-E states using a normal forward DMRG schedule up to a large MPS bond dimension ($D = 18000$). After that, to collect data necessary for extrapolation, we additionally perform the reverse schedule with MPS bond dimensions $D = 16000, 15000, \cdots$, and $5000$. To make sure that the energy at each of the bond dimensions is fully converged, we do four sweeps for each of the reverse schedule bond dimensions. The computational cost and timings are listed in Supplementary Table~\ref{tab:sm-dmrg-cost}. Note that forward schedule sweeps are in general more expensive than the reverse schedule sweeps, because in the forward schedule we need to apply perturbative noise to reduce the chance of getting stuck in local minima when increasing the bond dimension. The scaling of the UDMRG computational cost for BS8-E is shown in Supplementary Figure~\ref{fig:sm-dmrg-cc-cost}A.

\begin{table}[!htbp]
    \small
    \centering
    \caption{Computational cost of UDMRG for the LLDUC model.}
    \begin{tabular}{
        >{\centering\arraybackslash}p{1.8cm}|
        >{\centering\arraybackslash}p{2.2cm}
        >{\centering\arraybackslash}p{1.0cm}
        >{\centering\arraybackslash}p{0.9cm}
        >{\centering\arraybackslash}p{0.9cm}
        >{\centering\arraybackslash}p{2.0cm}
        >{\centering\arraybackslash}p{2.0cm}
        >{\centering\arraybackslash}p{2.0cm}
    }
    \hline\hline
    spin isomer & schedule type & $D$ & $N_{\mathrm{nodes}}$ & $N_{\mathrm{sweeps}}$ & FLOPs/sweep & wall time/sweep & CPU hours/sweep \\
    \hline
    BS7-C &    forward &   18000 &   16 &    2 & $4.01\times 10^{18}$ & 160.2 hours &    246091 \\
           &    reverse &   16000 &   16 &    4 & $2.78\times 10^{18}$ &  72.7 hours &    111639 \\
           &    reverse &   15000 &   16 &    4 & $2.33\times 10^{18}$ &  62.1 hours &     95411 \\
           &    reverse &   14000 &   16 &    4 & $1.91\times 10^{18}$ &  57.4 hours &     88202 \\
           &    reverse &   13000 &   16 &    4 & $1.54\times 10^{18}$ &  39.9 hours &     61298 \\
           &    reverse &   12000 &   16 &    4 & $1.22\times 10^{18}$ &  31.5 hours &     48397 \\
           &    reverse &   11000 &   16 &    4 & $9.49\times 10^{17}$ &  30.4 hours &     46642 \\
           &    reverse &   10000 &   16 &    4 & $7.22\times 10^{17}$ &  18.7 hours &     28732 \\
           &    reverse &    9000 &   16 &    4 & $5.34\times 10^{17}$ &  13.7 hours &     21052 \\
           &    reverse &    8000 &   16 &    4 & $3.78\times 10^{17}$ &  9.72 hours &     14925 \\
           &    reverse &    7000 &   16 &    4 & $2.58\times 10^{17}$ &  7.76 hours &     11917 \\
           &    reverse &    6000 &   16 &    4 & $1.64\times 10^{17}$ &  5.07 hours &      7787 \\
           &    reverse &    5000 &   16 &    4 & $9.86\times 10^{16}$ &  3.14 hours &      4825 \\
           & & & & & \multicolumn{2}{r}{\textbf{total CPU hours:}} & \textbf{2655489} \\
     \hline
     BS8-E &    forward &   18000 &   24 &    2 & $5.00\times 10^{18}$ & 137.8 hours &    317440 \\
           &    reverse &   16000 &   16 &    4 & $2.85\times 10^{18}$ &  73.6 hours &    113066 \\
           &    reverse &   15000 &   16 &    4 & $2.38\times 10^{18}$ &  74.1 hours &    113812 \\
           &    reverse &   14000 &   16 &    4 & $1.95\times 10^{18}$ &  51.4 hours &     79026 \\
           &    reverse &   13000 &   16 &    4 & $1.57\times 10^{18}$ &  41.0 hours &     62985 \\
           &    reverse &   12000 &   16 &    4 & $1.25\times 10^{18}$ &  34.1 hours &     52374 \\
           &    reverse &   11000 &   12 &    4 & $8.37\times 10^{17}$ &  29.5 hours &     33928 \\
           &    reverse &   10000 &   12 &    4 & $6.40\times 10^{17}$ &  22.7 hours &     26117 \\
           &    reverse &    9000 &   12 &    4 & $4.77\times 10^{17}$ &  16.6 hours &     19116 \\
           &    reverse &    8000 &   12 &    4 & $3.39\times 10^{17}$ &  12.1 hours &     13918 \\
           &    reverse &    7000 &   12 &    4 & $2.31\times 10^{17}$ &  8.17 hours &      9413 \\
           &    reverse &    6000 &   12 &    4 & $1.49\times 10^{17}$ &  5.30 hours &      6102 \\
           &    reverse &    5000 &   12 &    4 & $8.82\times 10^{16}$ &  3.19 hours &      3671 \\
           & & & & & \multicolumn{2}{r}{\textbf{total CPU hours:}} & \textbf{2768994} \\
     \hline
     \textbf{estimated} & & 393000 & & & $4.54\times 10^{22}$ & & $2.04\times 10^{9}$ \\
    \hline\hline
    \end{tabular}
    \label{tab:sm-dmrg-cost}
\end{table}

\begin{figure}[!htbp]
  \includegraphics[width=\linewidth]{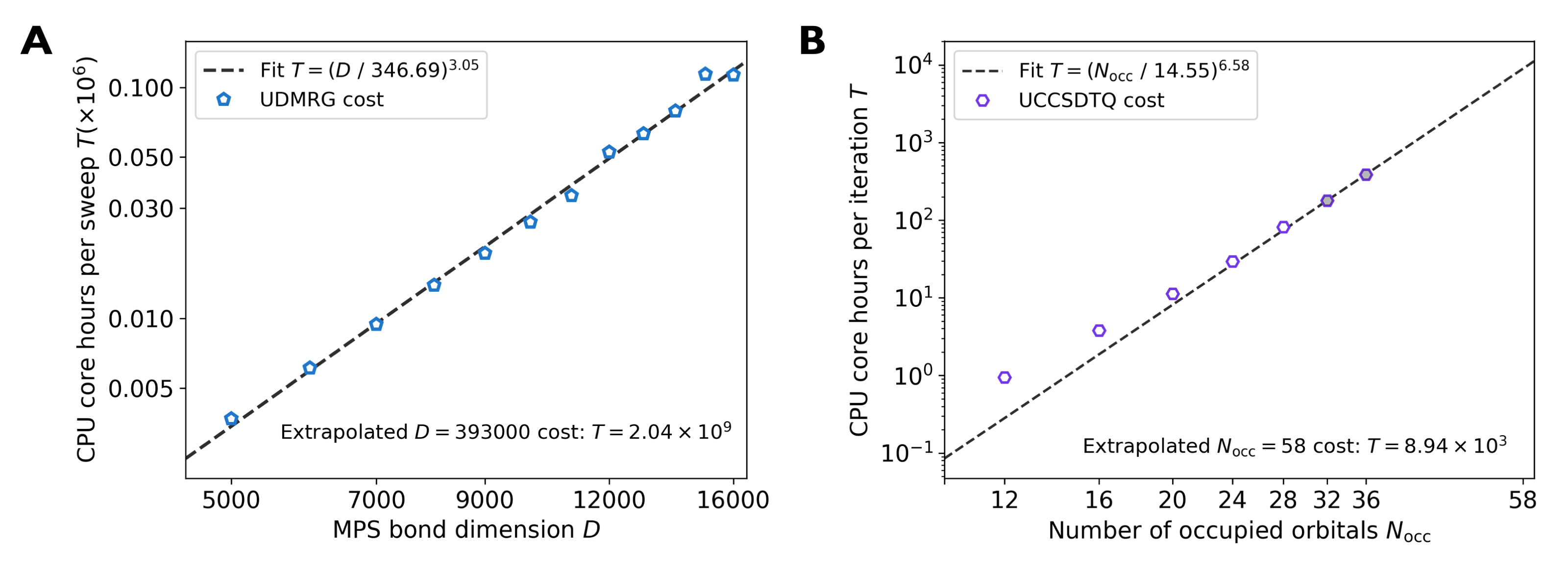}
  \caption{
  \textbf{A} Computational cost and scaling for UDMRG extrapolation, in CPU core hours per sweep, for BS8-E.
  \textbf{B} Computational cost and scaling for UCCSDTQ in the active subspace, in CPU core hours per iteration, for BS8-E. Extrapolation is done using only the data from the 32 and 36 occupied orbital subspaces, as the cost for smaller subspaces has significant overheads.}
  \label{fig:sm-dmrg-cc-cost}
\end{figure}

\textbf{UCC.} We list the FNO-UCCSDTQ computational cost and timings in Supplementary Table~\ref{tab:sm-cc-cost}. Note that the memory cost is the main bottleneck in these calculations, because currently our UCC code does not support distributed storage of amplitudes. We use index-permutation symmetry to reduce the memory cost of amplitudes, which introduces a non-negligible amount of computational overhead for packing and unpacking in the contraction step. Therefore, for estimating FLOPs/iter we consider the situation where there is no usage of amplitudes with index-permutation symmetry. The scaling of the FNO-UCCSDTQ computational cost for BS8-E is shown in Supplementary Figure~\ref{fig:sm-dmrg-cc-cost}B.

\begin{table}[!htbp]
    \small
    \centering
    \caption{Computational cost of FNO-UCCSDTQ per iteration for the LLDUC model. $N_{\mathrm{iter}}$ is the number of iterations to achieve energy convergence of $10^{-8}$ Hartrees. FLOPs/iter is the estimated number without the consideration of index-permutation symmetry (the actual computation was done with index-permutation symmetry to reduce memory costs).}
    \begin{tabular}{
        >{\centering\arraybackslash}p{1.8cm}|
        >{\centering\arraybackslash}p{2.5cm}
        >{\centering\arraybackslash}p{0.9cm}
        >{\centering\arraybackslash}p{0.9cm}
        >{\centering\arraybackslash}p{2.2cm}
        >{\centering\arraybackslash}p{2.1cm}
        >{\centering\arraybackslash}p{2.4cm}
    }
    \hline\hline
    spin isomer & FNO subspace & $N_{\mathrm{nodes}}$ & $N_{\mathrm{iter}}$ & FLOPs/iter & wall time/iter & CPU hours/iter \\
    \hline
     BS7-C & (30o, 21e) &    1 &    46 & $5.83\times 10^{13}$ &   0.02 hours &       0.9 \\
           & (34o, 29e) &    1 &    55 & $3.07\times 10^{14}$ &   0.08 hours &       3.8 \\
           & (38o, 37e) &    1 &    71 & $1.10\times 10^{15}$ &   0.24 hours &      11.3 \\
           & (42o, 45e) &    1 &    73 & $2.84\times 10^{15}$ &   0.62 hours &      29.7 \\
           & (46o, 53e) &    1 &    77 & $6.10\times 10^{15}$ &   0.83 hours &      79.9 \\
           & (50o, 61e) &    1 &    75 & $1.19\times 10^{16}$ &   2.50 hours &     239.6 \\
           & (54o, 69e) &    1 &    80 & $2.14\times 10^{16}$ &   3.96 hours &     380.5 \\
     \hline
     BS8-E & (30o, 21e) &    1 &    72 & $5.83\times 10^{13}$ &   0.02 hours &       0.9 \\
           & (34o, 29e) &    1 &    90 & $3.07\times 10^{14}$ &   0.08 hours &       3.8 \\
           & (38o, 37e) &    1 &    90 & $1.10\times 10^{15}$ &   0.23 hours &      11.2 \\
           & (42o, 45e) &    1 &   109 & $2.84\times 10^{15}$ &   0.61 hours &      29.4 \\
           & (46o, 53e) &    1 &   103 & $6.10\times 10^{15}$ &   0.85 hours &      82.0 \\
           & (50o, 61e) &    1 &    93 & $1.19\times 10^{16}$ &   1.86 hours &     178.8 \\
           & (54o, 69e) &    1 &    98 & $2.14\times 10^{16}$ &   4.04 hours &     388.0 \\
     \hline
     BS8-F & (30o, 21e) &    1 &    61 & $5.83\times 10^{13}$ &   0.02 hours &       0.9 \\
           & (34o, 29e) &    1 &    77 & $3.07\times 10^{14}$ &   0.08 hours &       3.8 \\
           & (38o, 37e) &    1 &    70 & $1.10\times 10^{15}$ &   0.23 hours &      11.3 \\
           & (42o, 45e) &    1 &    81 & $2.84\times 10^{15}$ &   0.62 hours &      29.8 \\
           & (46o, 53e) &    1 &    97 & $6.10\times 10^{15}$ &   0.83 hours &      79.9 \\
           & (50o, 61e) &    1 &   113 & $1.19\times 10^{16}$ &   1.64 hours &     157.0 \\
           & (54o, 69e) &    1 &    99 & $2.14\times 10^{16}$ &   4.38 hours &     420.9 \\
     \hline
     \textbf{estimated} & (76o, 113e) & & & $2.32\times 10^{17}$ & & 8937 \\
    \hline\hline
    \end{tabular}
    \label{tab:sm-cc-cost}
\end{table}

\subsubsection{UDMRG energy extrapolation}
\label{sec:sm-dmrg-extra}

We list the UDMRG energies and discarded weights in Supplementary Table~\ref{tab:sm-fm-dmrg-energy}. The $\exp[-\kappa(\log D)^2]$ fitting is shown in Supplementary Figure~\ref{fig:sm-fm-dmrg-energy}. We also performed the more conventional DMRG extrapolation using discarded weights (using three data points with the largest bond dimensions), shown in Supplementary Figure~\ref{fig:sm-fm-dmrg-dw}. We did not see an overall simple trend in the discarded weight extrapolation, which is possibly caused by the usage of spin-unrestricted orbitals in UDMRG.

We estimate the required bond dimension to reach 1 kcal/mol error using the $\exp[-\kappa(\log D)^2]$ fitting of the BS8-E data listed in Supplementary Table~\ref{tab:sm-fm-dmrg-energy}. The energy at $D=393000$ is predicted to be $-22140.408505$ Hartrees, which is $0.001592$ Hartrees or $1.00$ kcal/mol above the extrapolated BS8-E energy using the $\exp[-\kappa(\log D)^2]$ fit ($-22140.410097$ Hartrees).

\begin{table}[!htbp]
    \centering
    \caption{UDMRG energies and discarded weights for BS7-C and BS8-E of LLDUC model.}
    \begin{tabular}{
        >{\centering\arraybackslash}p{2.0cm}|
        >{\centering\arraybackslash}p{3.4cm}
        >{\centering\arraybackslash}p{1.2cm}
        >{\centering\arraybackslash}p{3.5cm}
        >{\centering\arraybackslash}p{4.1cm}
    }
    \hline\hline
    spin isomer & schedule type & $D$ & energy (Hartree) & max discarded weight \\
    \hline
     BS7-C &          forward &    18000 &   -22140.394853 & $7.10\times 10^{-5}$ \\
           &          reverse &    16000 &   -22140.394216 & $8.52\times 10^{-5}$ \\
           &          reverse &    15000 &   -22140.393634 & $8.88\times 10^{-5}$ \\
           &          reverse &    14000 &   -22140.392959 & $9.15\times 10^{-5}$ \\
           &          reverse &    13000 &   -22140.392205 & $9.38\times 10^{-5}$ \\
           &          reverse &    12000 &   -22140.391360 & $9.59\times 10^{-5}$ \\
           &          reverse &    11000 &   -22140.390407 & $9.84\times 10^{-5}$ \\
           &          reverse &    10000 &   -22140.389322 & $1.01\times 10^{-4}$ \\
           &          reverse &     9000 &   -22140.388070 & $1.04\times 10^{-4}$ \\
           &          reverse &     8000 &   -22140.386616 & $1.07\times 10^{-4}$ \\
           &          reverse &     7000 &   -22140.384879 & $1.11\times 10^{-4}$ \\
           &          reverse &     6000 &   -22140.382754 & $1.20\times 10^{-4}$ \\
           &          reverse &     5000 &   -22140.380060 & $1.40\times 10^{-4}$ \\
           & $\exp[-\kappa(\log D)^2]$ fit & $\infty$ &   -22140.410973 &                      \\
     \hline
     BS8-E &          forward &    18000 &   -22140.396133 & $4.10\times 10^{-5}$ \\
           &          reverse &    16000 &   -22140.395472 & $5.22\times 10^{-5}$ \\
           &          reverse &    15000 &   -22140.394930 & $5.50\times 10^{-5}$ \\
           &          reverse &    14000 &   -22140.394302 & $5.70\times 10^{-5}$ \\
           &          reverse &    13000 &   -22140.393599 & $5.88\times 10^{-5}$ \\
           &          reverse &    12000 &   -22140.392814 & $6.05\times 10^{-5}$ \\
           &          reverse &    11000 &   -22140.391926 & $6.20\times 10^{-5}$ \\
           &          reverse &    10000 &   -22140.390912 & $6.40\times 10^{-5}$ \\
           &          reverse &     9000 &   -22140.389741 & $6.70\times 10^{-5}$ \\
           &          reverse &     8000 &   -22140.388369 & $7.42\times 10^{-5}$ \\
           &          reverse &     7000 &   -22140.386720 & $8.28\times 10^{-5}$ \\
           &          reverse &     6000 &   -22140.384697 & $9.36\times 10^{-5}$ \\
           &          reverse &     5000 &   -22140.382131 & $1.08\times 10^{-4}$ \\
           & $\exp[-\kappa(\log D)^2]$ fit & $\infty$ &   -22140.410097 &                      \\
    \hline\hline
    \end{tabular}
    \label{tab:sm-fm-dmrg-energy}
\end{table}

\begin{figure}[!htbp]
  \includegraphics[width=0.7\linewidth]{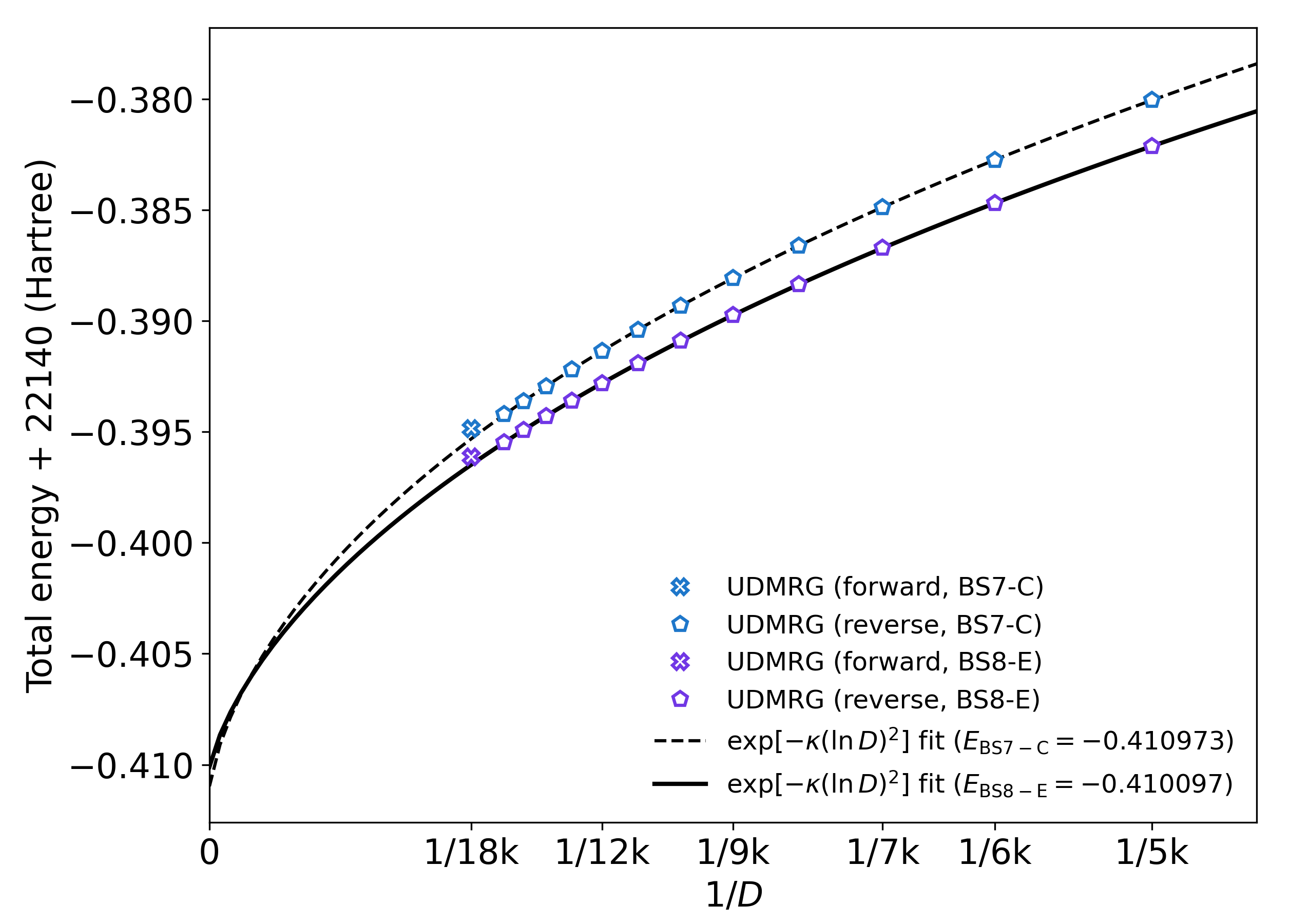}
  \centering
  \caption{The $\exp[-\kappa(\log D)^2]$ fit for BS7-C and BS8-E states for the LLDUC model, using UDMRG energy data obtained from the reverse schedule.}
  \label{fig:sm-fm-dmrg-energy}
\end{figure}

\begin{figure}[!htbp]
  \includegraphics[width=\linewidth]{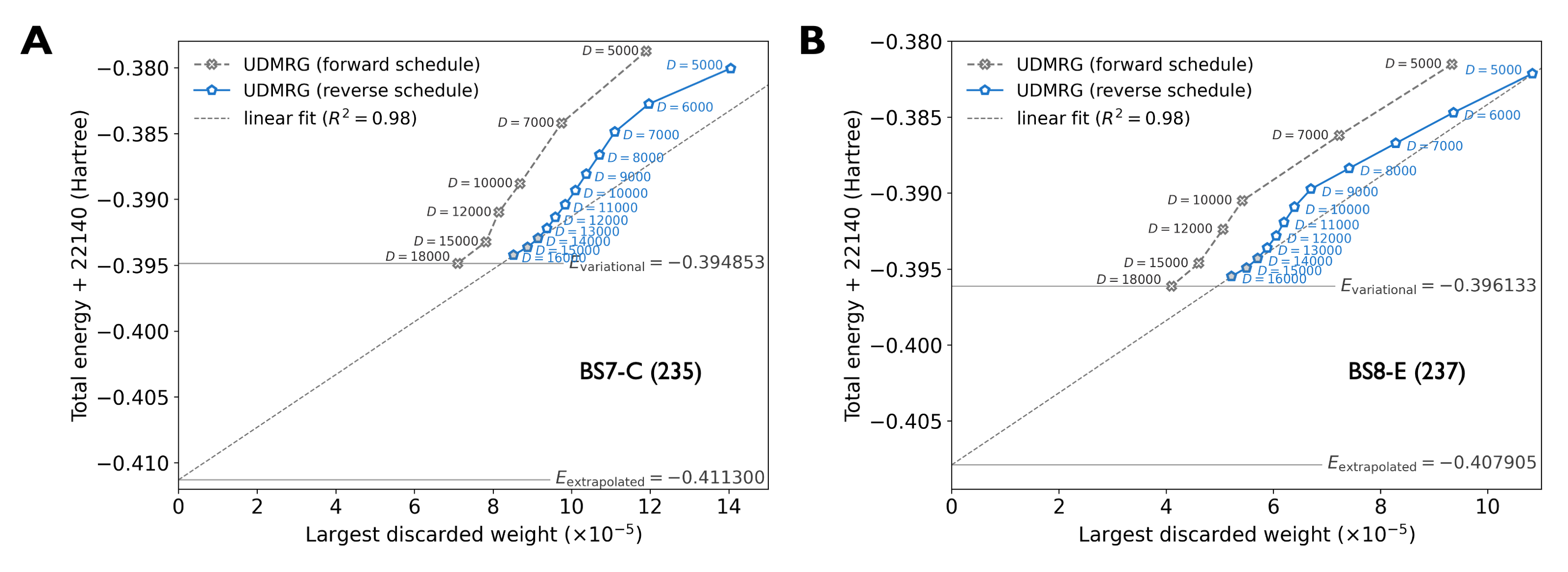}
  \centering
  \caption{The low confidence discarded weight extrapolation for the (A) BS7-C and (B) BS8-E states for the LLDUC model, using UDMRG energy and discarded weights obtained from the reverse schedule. We perform linear extrapolation using the three data points with bond dimension 14000, 15000, and 16000. The energy and discarded weights obtained from the forward schedule are shown in dashed lines.}
  \label{fig:sm-fm-dmrg-dw}
\end{figure}

\subsubsection{Composite energy estimation}
\label{sec:sm-comp}

We estimate the final UCC energy for BS7-C, BS8-E and BS8-F by adding a composite post-Q correction. To estimate the contribution of post quadruples connected correlations, we use an Fe dimer reference system (FeIII-FeII, see Sec.~\ref{sec:sm-spin}) for which we have computed the D, T, Q and post-Q correlation increments, against which we can compare the corresponding D, T, and Q correlation increments for FeMo-co. If FeMo-co can be viewed as an extension of the Fe dimer (i.e. multiple connected copies with the same qualitative correlations), then we can expect a constant ratio of the increments between the systems. (Of course, whether this type of relationship can be found or holds will depend on the system, thus this scaling relation is an empirical one).

In Supplementary Table \ref{tab:sm-fm-ucc-post-q} we show the correlation increments and ratios. For LLDUC model, we use the UHF, UCCSD, and UCCSDT energies with their respective best reference to minimize the fluctuation and reference dependence of the increment ratio at different correlation levels. We see that the ratio of the correlation increments is indeed roughly constant, suggesting that the chosen Fe dimer is a good model system for the correlations in FeMo-co. We estimate the post-Q increment ratio as the average of the D, T, and Q ratios, which gives the post-Q correction as $-0.003963, -0.003980$, and $-0.004134$ Hartrees, for BS7-C, BS8-E and BS8-F, respectively.  Adding these corrections to the composite UCCSDTQ energy, we estimate the exact UCC energy as $-22140.410812, -22140.411097$, and $-22140.409875$ Hartrees, for BS7-C, BS8-E and BS8-F, respectively.

\begin{table}[!htbp]
    \small
    \centering
    \caption{Post quadruples correction for the UCC energy for BS7-C, BS8-E and BS8-F states of the LLDUC model. $\Delta E$ represents the D, T, Q and post-Q correlation increments. Estimated quantities are labeled in bold. Absolute energies are in Hartrees shifted by $-116.0$ Hartrees (dimer) or $-22140.0$ Hartrees (LLDUC).}
    \begin{tabular}{
        >{\centering\arraybackslash}p{1.8cm}|
        >{\centering\arraybackslash}p{1.8cm}
        >{\centering\arraybackslash}p{1.6cm}
        >{\centering\arraybackslash}p{1.6cm}
        >{\centering\arraybackslash}p{1.6cm}
        >{\centering\arraybackslash}p{1.6cm}
        >{\centering\arraybackslash}p{3.5cm}
    }
    \hline\hline
    spin isomer & theory & $E_{\mathrm{dimer}}$ & $\Delta E_{\mathrm{dimer}}$ & $E_{\mathrm{LLDUC}}$ & $\Delta E_{\mathrm{LLDUC}}$ & $\Delta E_{\mathrm{LLDUC}}/\Delta E_{\mathrm{dimer}}$ \\
    \hline
    BS7-C &        UHF &    -0.322099 &            &     0.009608 &            &          \\
           &      UCCSD &    -0.361322 &  -0.039224 &    -0.305389 &  -0.314997 &     8.03 \\
           &     UCCSDT &    -0.372305 &  -0.010983 &    -0.394946 &  -0.089557 &     8.15 \\
           &    UCCSDTQ &    -0.373827 &  -0.001522 &    -0.406849 &  -0.011903 &     7.82 \\
           &      exact &    -0.374322 &  -0.000495 & \textbf{estimated} & \textbf{-0.003963} & \textbf{8.00} \\
     \hline
     BS8-E &        UHF &    -0.322099 &            &     0.016226 &            &          \\
           &      UCCSD &    -0.361322 &  -0.039224 &    -0.306453 &  -0.322678 &     8.23 \\
           &     UCCSDT &    -0.372305 &  -0.010983 &    -0.395255 &  -0.088802 &     8.09 \\
           &    UCCSDTQ &    -0.373827 &  -0.001522 &    -0.407117 &  -0.011862 &     7.79 \\
           &      exact &    -0.374322 &  -0.000495 & \textbf{estimated} & \textbf{-0.003980} & \textbf{8.04} \\
     \hline
     BS8-F &        UHF &    -0.322099 &            &     0.016720 &            &          \\
           &      UCCSD &    -0.361322 &  -0.039224 &    -0.298387 &  -0.315107 &     8.03 \\
           &     UCCSDT &    -0.372305 &  -0.010983 &    -0.392966 &  -0.094580 &     8.61 \\
           &    UCCSDTQ &    -0.373827 &  -0.001522 &    -0.405741 &  -0.012775 &     8.39 \\
           &      exact &    -0.374322 &  -0.000495 & \textbf{estimated} & \textbf{-0.004134} & \textbf{8.35} \\
    \hline\hline
    \end{tabular}
    \label{tab:sm-fm-ucc-post-q}
\end{table}

\subsubsection{Natural orbital occupancy}
\label{sec:sm-llduc-nat-orb}

As an additional metric for the similarity between UCC and UDMRG solutions of the LLDUC model, in Supplementary Figure~\ref{fig:sm-dmrg-cc-nat-occ} we show the comparison between natural orbital occupancy computed at UCCSD, UCCSDT, and UDMRG level, for some representative low-energy spin-isomers. We found good agreement between the UCC and UDMRG theories.

\begin{figure}[!htbp]
  \includegraphics[width=0.97\linewidth]{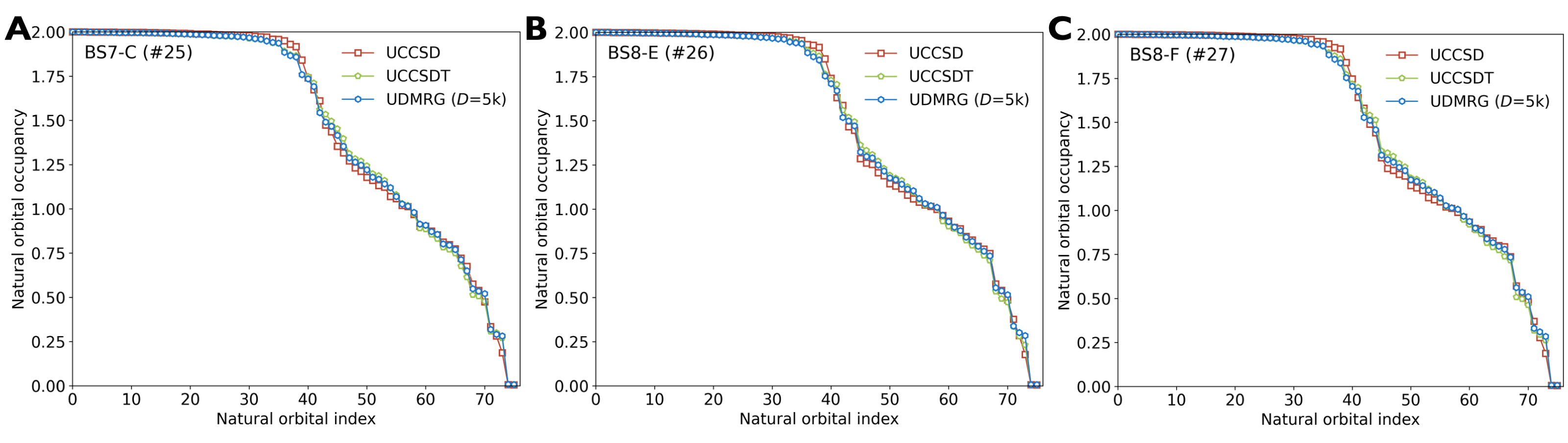}
  \centering
  \caption{UCCSD, UCCSDT and UDMRG natural orbital occupancy computed for spin isomers BS7-C, BS8-E, and BS8-F of the LLDUC model.}
  \label{fig:sm-dmrg-cc-nat-occ}
\end{figure}

\subsection{Large active space models}
\label{sec:sm-large-act}

To investigate the effect of dynamical correlation not captured in the minimal LLDUC model, we construct a series of larger active space models and perform spin unrestricted Hartree-Fock and coupled cluster calculations (up to singles and doubles and perturbative triples) with the same cluster geometry, charge and ligands as the ones used for the LLDUC model, built from PDB 3U7Q.

\subsubsection{Active space setups}

We start with the high-spin UKS solution computed with the B3LYP functional~\cite{becke1988density,lee1988development,becke1993new} (as was used in the construction of the LLDUC model), the TZP-DKH basis~\cite{jorge2009contracted} for Fe, Mo, and S, the def2-SVP basis~\cite{weigend2005balanced} for H, C, O, and N atoms), and the remaining protein outside of the crystal cutout treated using the domain-decomposition COSMO solvation model (dd-COSMO) ($\epsilon = 4.0$)~\cite{cances2013domain,lipparini2013fast,lipparini2014quantum} as implemented in \textsc{PySCF}. We split-localize the spin unrestricted natural orbitals (UNO) computed from the high-spin UKS solution to doubly occupied (PM~\cite{pipek1989fast}), singly occupied (PM), and virtual (SCDM~\cite{damle2015compressed}) subspaces. We then constructed a (994o, 423e) active space including all FeMo-co orbitals (7 Fe atoms + 1 Mo atoms + 1 C atom + 10 S atoms + 2 O atoms + 1 N atom), by selecting 994 localized orbitals that are localized on the FeMo-co core.

To further truncate this active space, we performed high-spin UHF and UMP2~\cite{moller1934note,pople1976theoretical} in the (994o, 423e) active space (without frozen core or density fitting). In Supplementary Figure \ref{fig:sm-act}, we show UMP2 natural orbital occupancies computed in the (994o, 423e) active space and the orbital index ranges used to construct smaller active spaces.

\begin{figure}[!htbp]
  \includegraphics[width=\linewidth]{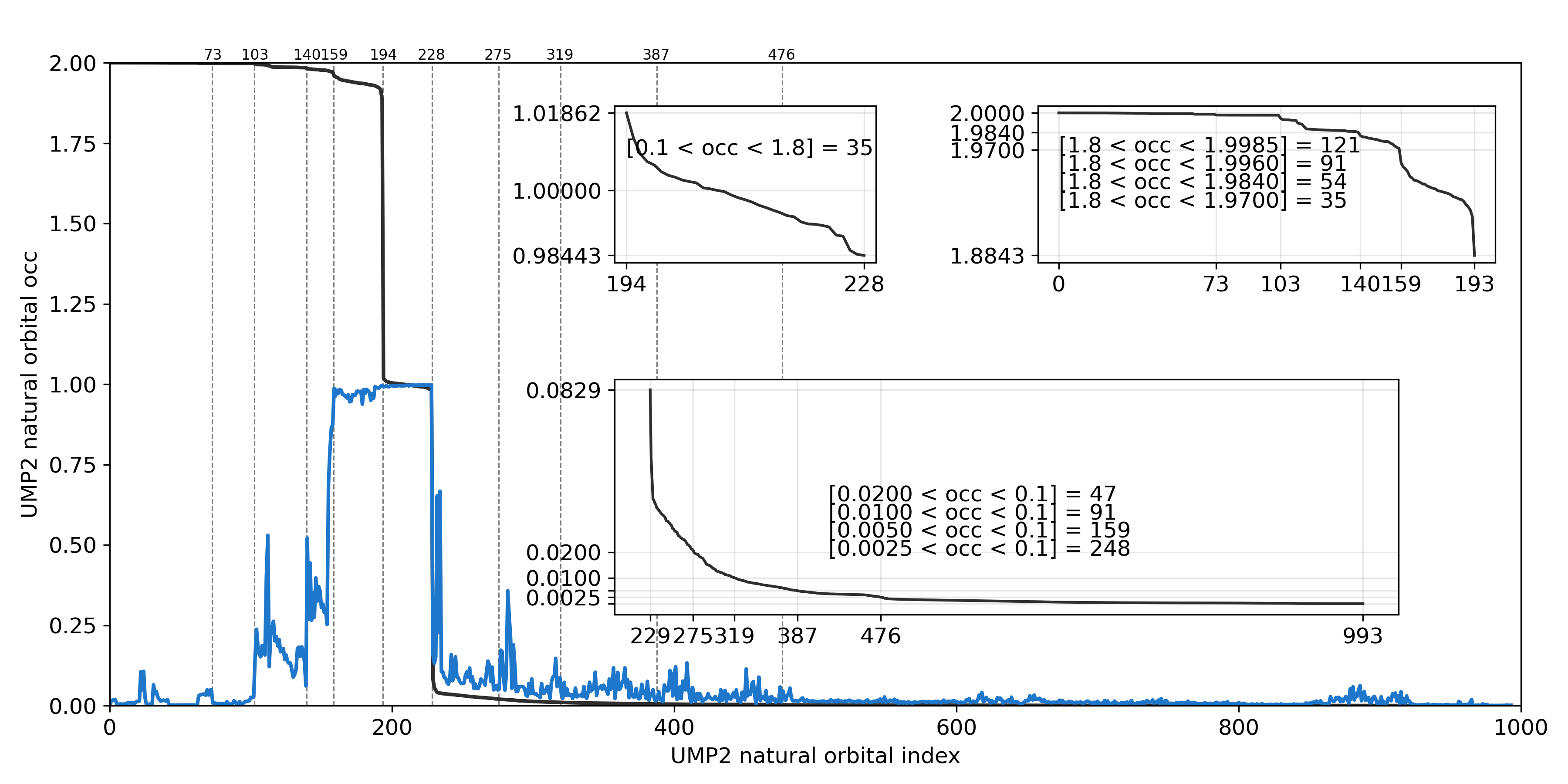}
  \caption{Black lines: UMP2 natural orbital occupation in the (994o, 423e) active space. Blue line: the squared overlap between each UMP2 natural orbital and the LLDUC (76o, 113e) active space.}
  \label{fig:sm-act}
\end{figure}

Based on the UMP2 natural orbital occupancies, we define four truncated active spaces listed in Supplementary Table \ref{tab:sm-act}. the high-spin UHF and UMP2 energies in the 994o active space and the four truncated active spaces are also given in Supplementary Table \ref{tab:sm-act}.

\begin{table}[!htbp]
    \small
    \centering
    \caption{The four truncated active spaces and the full (994o, 423e) active space. Integers are number of orbitals included in each active space. Number of frozen orbitals is counted with respect to the (994o, 423e) active space. Energies are in Hartrees.}
    \begin{tabular}{
        >{\centering\arraybackslash}p{2.0cm}|
        >{\centering\arraybackslash}p{1.3cm}
        >{\centering\arraybackslash}p{1.3cm}
        >{\centering\arraybackslash}p{1.3cm}
        >{\centering\arraybackslash}p{1.2cm}
        >{\centering\arraybackslash}p{2.9cm}
        >{\centering\arraybackslash}p{2.95cm}
    }
    \hline\hline
    active space & $\mathrm{occ} > 1$ & $\mathrm{occ} \approx1$ & $\mathrm{occ} < 1$ & frozen & $E_{\mathrm{UHF}}$ (high spin)
    & $E_{\mathrm{UMP2}}$ (high spin) \\
    \hline
    (117o, 105e) &   35 &   35  &   47  &  159 &    -22127.108167 &    -22127.960360 \\
    (180o, 143e) &   54 &   35  &   91  &  140 &    -22127.119741 &    -22128.792668 \\
    (285o, 217e) &   91 &   35  &  159  &  103 &    -22127.140496 &    -22130.559716 \\
    (404o, 277e) &  121 &   35  &  248  &   73 &    -22127.146440 &    -22132.433149 \\
    \hline
    (994o, 423e) &  194 &   35  &  765  &    0 &    -22127.157903 &    -22136.129826 \\
    \hline\hline
    \end{tabular}
    \label{tab:sm-act}
\end{table}

Note that for all broken-symmetry (low-spin) UHF, UCCSD, and UCCSD(T) computations for these four larger active spaces (see below), the dd-COSMO effect is always computed using the high-spin UHF density matrix.

\subsubsection{Energy of broken-symmetry solutions}

To properly sample UHF solutions in the larger active spaces, we performed UHF calculations using a variety of different initial guess density matrices, followed by a second-order Newton procedure for orbital optimization. For each of the 35 broken-symmetry spin couplings, we consider: (a) 2250 possible initial diagonal density matrices;
(b) around 351 unique UHF solutions in the 76o active space projected to the larger active spaces; and (c) around 24 low-energy UHF solutions in the smaller active space projected to larger active spaces.

For each of the 35 broken-symmetry spin couplings, we identified 24 low-energy UHF solutions from all unique UHF solutions. UCCSD and UCCSD(T) are performed using each of these 24 low-energy UHF as references (some high-energy spin isomers and high-UHF-energy references are skipped due to computational limitations). The lowest UHF energy, the lowest UCCSD energy, and the UCCSD(T) energy corresponding to the lowest UCCSD energy, for BS7-BS10 spin isomers in the 285o and 404o active spaces, are listed in Supplementary Table~\ref{tab:sm-act-energy} and plotted in Supplementary Figure~\ref{fig:sm-act-energy}. We see that the UHF, UCCSD, and UCCSD(T) energy differences between the 285o and 404o active spaces are within a few kcal/mol and BS7, BS8, and BS10 are found to be low-energy spin isomers.

\begin{table}[!htbp]
    \scriptsize
    \centering
    \caption{The lowest UHF energy, the lowest UCCSD energy, and the UCCSD(T) energy corresponding to the lowest UCCSD energy, for BS7-BS10 spin isomers in the (285o, 217e) and (404o, 277e) active spaces.}
    \begin{tabular}{
        >{\centering\arraybackslash}p{1.5cm}|
        >{\centering\arraybackslash}p{1.8cm}
        >{\centering\arraybackslash}p{1.8cm}
        >{\centering\arraybackslash}p{1.8cm}|
        >{\centering\arraybackslash}p{1.8cm}
        >{\centering\arraybackslash}p{1.8cm}
        >{\centering\arraybackslash}p{1.8cm}
    }
    \hline\hline
     & \multicolumn{3}{c|}{(285o, 217e) active space} & \multicolumn{3}{c}{(404o, 277e) active space} \\
    \hline
    spin isomer & $E_{\mathrm{UHF}}$ & $E_{\mathrm{UCCSD}}$ & $E_{\mathrm{UCCSD(T)}}$ & $E_{\mathrm{UHF}}$ & $E_{\mathrm{UCCSD}}$ & $E_{\mathrm{UCCSD(T)}}$ \\
    \hline
     BS9-A &    -22127.219879 &    -22130.861812 &    -22131.063735 &    -22127.232790 &    -22132.645370 &    -22132.903818 \\
    BS10-A &    -22127.225183 &    -22130.870124 &    -22131.082960 &    -22127.234246 &    -22132.654036 &    -22132.914820 \\
    BS10-B &    -22127.224936 &    -22130.871301 &    -22131.083389 &    -22127.234825 &    -22132.654566 &    -22132.915952 \\
    BS10-C &    -22127.216069 &    -22130.859955 &    -22131.072868 &    -22127.230860 &    -22132.649132 &    -22132.908483 \\
     BS9-B &    -22127.215523 &    -22130.851258 &    -22131.061052 &    -22127.229049 &    -22132.639196 &    -22132.897251 \\
    BS10-D &    -22127.216123 &    -22130.865234 &    -22131.072463 &    -22127.227189 &    -22132.652592 &    -22132.908899 \\
    BS10-E &    -22127.222899 &    -22130.864230 &    -22131.074418 &    -22127.236699 &    -22132.651886 &    -22132.910451 \\
    BS10-F &    -22127.222076 &    -22130.867418 &    -22131.076992 &    -22127.237848 &    -22132.654382 &    -22132.912626 \\
     BS9-C &    -22127.214306 &    -22130.856354 &    -22131.063087 &    -22127.224737 &    -22132.642246 &    -22132.898505 \\
     BS8-A &    -22127.208606 &    -22130.867979 &    -22131.077202 &    -22127.227487 &    -22132.654885 &    -22132.913317 \\
     BS8-B &    -22127.215925 &    -22130.870860 &    -22131.079594 &    -22127.231970 &    -22132.654821 &    -22132.913686 \\
     BS7-A &    -22127.212998 &    -22130.873173 &    -22131.082204 &    -22127.230591 &    -22132.660383 &    -22132.918028 \\
     BS8-C &    -22127.209400 &    -22130.864752 &    -22131.072786 &    -22127.225124 &    -22132.652915 &    -22132.909707 \\
     BS7-B &    -22127.221262 &    -22130.866538 &    -22131.078580 &    -22127.237505 &    -22132.655413 &    -22132.914718 \\
     BS8-D &    -22127.212274 &    -22130.868166 &    -22131.077023 &    -22127.221349 &    -22132.652052 &    -22132.911143 \\
     BS7-C &    -22127.218461 &    -22130.873955 &    -22131.086843 &    -22127.236277 &    -22132.660328 &    -22132.921591 \\
     BS8-E &    -22127.216441 &    -22130.875050 &    -22131.086915 &    -22127.228840 &    -22132.658308 &    -22132.916494 \\
     BS8-F &    -22127.214312 &    -22130.874957 &    -22131.085784 &    -22127.222811 &    -22132.656077 &    -22132.917963 \\
    \hline\hline
    \end{tabular}
    \label{tab:sm-act-energy}
\end{table}

\begin{figure}[!htbp]
  \includegraphics[width=\linewidth]{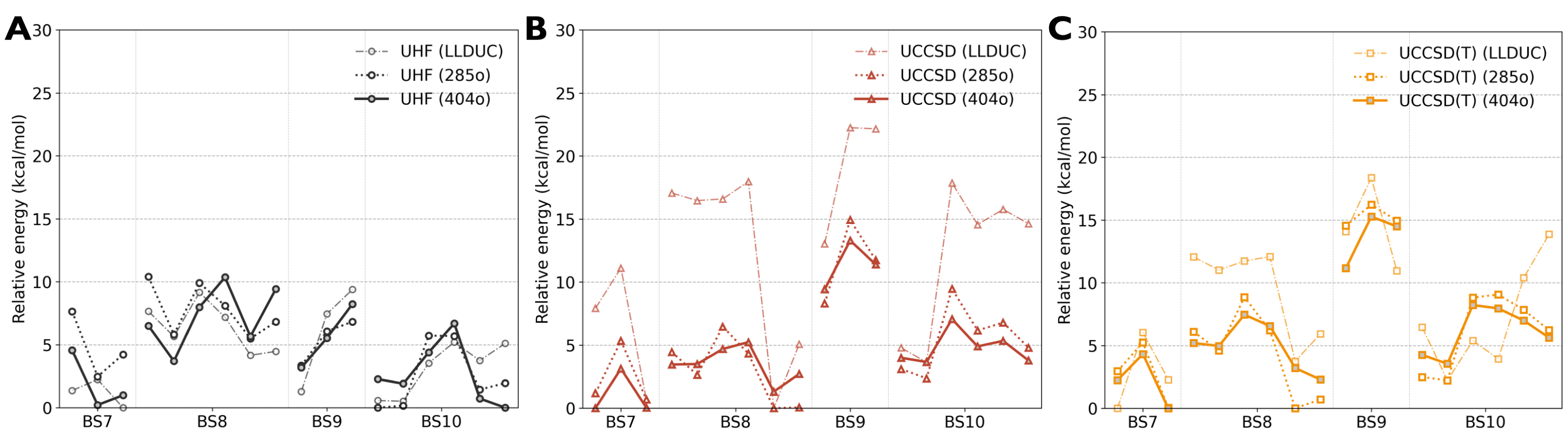}
  \caption{\textbf{A} The lowest UHF energy, \textbf{B} the lowest UCCSD energy, \textbf{C} and the UCCSD(T) energy corresponding to the lowest UCCSD energy, for BS7-BS10 spin isomers in the (285o, 217e) and (404o, 277e) active spaces.}
  \label{fig:sm-act-energy}
\end{figure}

To include the effect of dynamical correction beyond the 404-orbital active space, we compute the composite energy as
\begin{align}
    E_{\mathrm{UCCSD}}^{\text{composite}} =&\ E_{\mathrm{UCCSD}}^{\text{(404o,277e)}} +
    E_{\mathrm{UMP2}}^{\text{(994o,423e)}} - E_{\mathrm{UMP2}}^{\text{(404o,277e)}} \\
    E_{\mathrm{UCCSD(T)}}^{\text{composite}} =&\ E_{\mathrm{UCCSD(T)}}^{\text{(404o,277e)}} +
    E_{\mathrm{UMP2}}^{\text{(994o,423e)}} - E_{\mathrm{UMP2}}^{\text{(404o,277e)}}
\end{align}

To determine the UHF reference for UMP2 in the (994o, 423e) active space we use the projected reference corresponding to the lowest CCSD energy in the (404o, 277e) active space as the initial guess. The UMP2 composite correction and corrected UCCSD and UCCSD(T) energies are listed in Supplementary Table~\ref{tab:sm-act-energy-comp}.

\begin{table}[!htbp]
    \scriptsize
    \centering
    \caption{The UHF and UMP2 energies in the (994o, 423e) active space, the UMP2 energy in the (404o, 277e) active space, and the composite UCCSD and UCCSD(T) energies. $\Delta E_{\mathrm{UMP2}} = E_{\mathrm{UMP2}}^{\text{(994o,423e)}} - E_{\mathrm{UMP2}}^{\text{(404o,277e)}}$.}
    \begin{tabular}{
        >{\centering\arraybackslash}p{1.5cm}|
        >{\centering\arraybackslash}p{1.8cm}
        >{\centering\arraybackslash}p{1.8cm}|
        >{\centering\arraybackslash}p{1.8cm}|
        >{\centering\arraybackslash}p{1.8cm}
        >{\centering\arraybackslash}p{1.8cm}
        >{\centering\arraybackslash}p{1.8cm}
    }
    \hline\hline
     & \multicolumn{2}{c|}{(994o, 423e) active space} & (404o, 277e) & \multicolumn{3}{c}{composite} \\
    \hline
    spin isomer & $E_{\mathrm{UHF}}$ & $E_{\mathrm{UMP2}}$ & $E_{\mathrm{UMP2}}$ & $\Delta E_{\mathrm{UMP2}}$ & $E_{\mathrm{UCCSD}}^{\text{composite}}$ & $E_{\mathrm{UCCSD(T)}}^{\text{composite}}$ \\
    \hline
     BS9-A &    -22127.241126 &    -22136.335626 &    -22132.613577 &    -3.722049 &    -22136.367419 &    -22136.625867 \\
    BS10-A &    -22127.242223 &    -22136.337368 &    -22132.617133 &    -3.720235 &    -22136.374271 &    -22136.635055 \\
    BS10-B &    -22127.239027 &    -22136.337714 &    -22132.616968 &    -3.720746 &    -22136.375311 &    -22136.636698 \\
    BS10-C &    -22127.249909 &    -22136.335728 &    -22132.609092 &    -3.726636 &    -22136.375768 &    -22136.635119 \\
     BS9-B &    -22127.241007 &    -22136.329132 &    -22132.606886 &    -3.722246 &    -22136.361442 &    -22136.619497 \\
    BS10-D &    -22127.252549 &    -22136.338013 &    -22132.615450 &    -3.722563 &    -22136.375155 &    -22136.631462 \\
    BS10-E &    -22127.245289 &    -22136.344946 &    -22132.617834 &    -3.727112 &    -22136.378998 &    -22136.637563 \\
    BS10-F &    -22127.248065 &    -22136.343599 &    -22132.620606 &    -3.722993 &    -22136.377375 &    -22136.635619 \\
     BS9-C &    -22127.248769 &    -22136.328206 &    -22132.607131 &    -3.721075 &    -22136.363321 &    -22136.619580 \\
     BS8-A &    -22127.249320 &    -22136.343744 &    -22132.618995 &    -3.724749 &    -22136.379635 &    -22136.638067 \\
     BS8-B &    -22127.244322 &    -22136.341049 &    -22132.619851 &    -3.721197 &    -22136.376018 &    -22136.634883 \\
     BS7-A &    -22127.255926 &    -22136.344732 &    -22132.620802 &    -3.723930 &    -22136.384314 &    -22136.641958 \\
     BS8-C &    -22127.253498 &    -22136.340116 &    -22132.614982 &    -3.725134 &    -22136.378050 &    -22136.634841 \\
     BS7-B &    -22127.246846 &    -22136.344087 &    -22132.618938 &    -3.725149 &    -22136.380563 &    -22136.639867 \\
     BS8-D &    -22127.240197 &    -22136.335782 &    -22132.614788 &    -3.720994 &    -22136.373047 &    -22136.632138 \\
     BS7-C &    -22127.242612 &    -22136.349251 &    -22132.622766 &    -3.726485 &    -22136.386813 &    -22136.648076 \\
     BS8-E &    -22127.255794 &    -22136.343758 &    -22132.622896 &    -3.720862 &    -22136.379170 &    -22136.637356 \\
     BS8-F &    -22127.240501 &    -22136.336337 &    -22132.618368 &    -3.717969 &    -22136.374047 &    -22136.635932 \\
    \hline\hline
    \end{tabular}
    \label{tab:sm-act-energy-comp}
\end{table}

\subsubsection{Fe oxidation states}
\label{sec:sm-large-act-ox}

To investigate dynamical correlation effects on the assignment of oxidation states, we compute the electron population and nuclear density on Fe atoms in the larger active spaces, and compare the results with those obtained from the LLDUC model. The results are shown in Supplementary Figures \ref{fig:sm-large-pop}. The ordering of Fe centers in large active spaces is shown in Supplementary Figures \ref{fig:sm-large-pop-order}.

\begin{figure}[!htbp]
  \includegraphics[width=\linewidth]{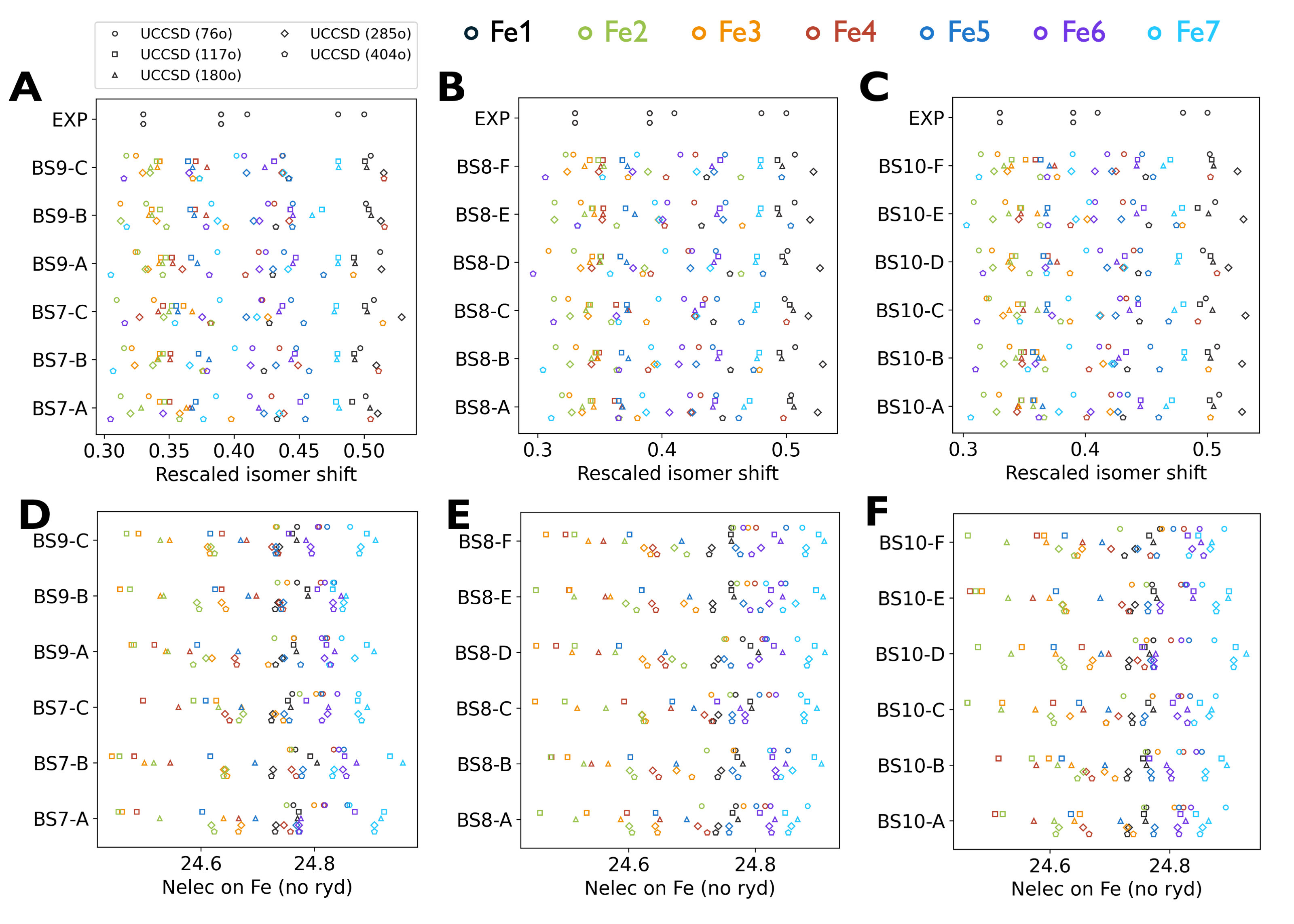}
  \caption{\textbf{A,B,C} Rescaled isomer shift (computed using electron density at Fe nucleus) for selected low-energy spin isomers (BS7-BS10) computed at UCCSD level with different active space models. \textbf{D,E,F} Electron population on Fe (computed using meta-L\"owdin without Rydberg contributions) for selected low-energy spin isomers (BS7-BS10) computed at UCCSD level with different active space models.}
  \label{fig:sm-large-pop}
\end{figure}

\begin{figure}[!htbp]
  \includegraphics[width=\linewidth]{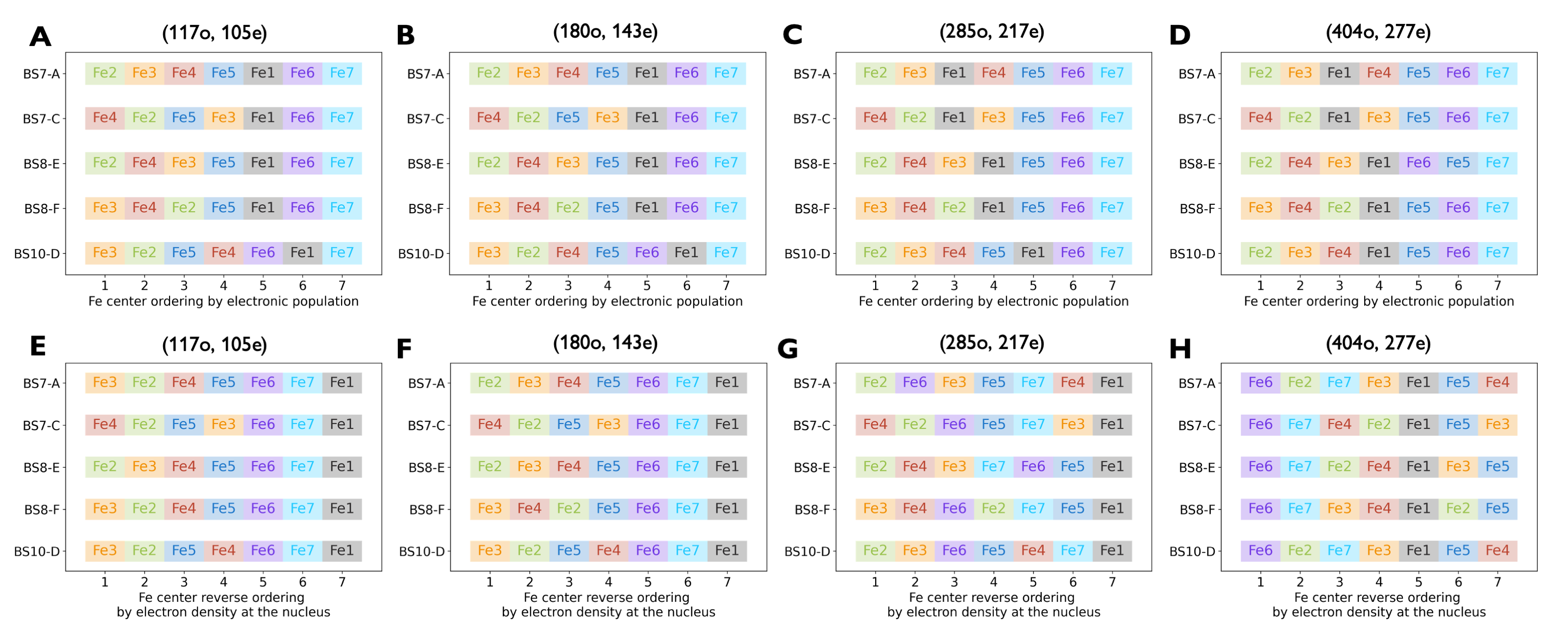}
  \caption{\textbf{A,B,C,D} Ordering of rescaled isomer shift (computed using electron density at Fe nucleus) for selected low-energy spin isomers computed at UCCSD level with different active space models. \textbf{E,F,G,H}  Ordering of electron population on Fe (computed using meta-L\"owdin without Rydberg contributions) for selected low-energy spin isomers computed at UCCSD level with different active space models.}
  \label{fig:sm-large-pop-order}
\end{figure}

\subsubsection{Natural orbital occupancy}
\label{sec:sm-large-nat-orb}

In Supplementary Figure~\ref{fig:m-large-cc-nat-occ} we show the comparison between the natural orbital occupancy computed at the UCCSD level of theory with different active spaces, for some representative low-energy spin-isomers. We found a good agreement for the occupancy in the strongly-correlated region among different active spaces. For less strongly correlated orbitals, the large active spaces show a larger deviation from exact double occupancy due to the inclusion of correlating virtual orbitals into the active space.

\begin{figure}[!htbp]
  \includegraphics[width=0.97\linewidth]{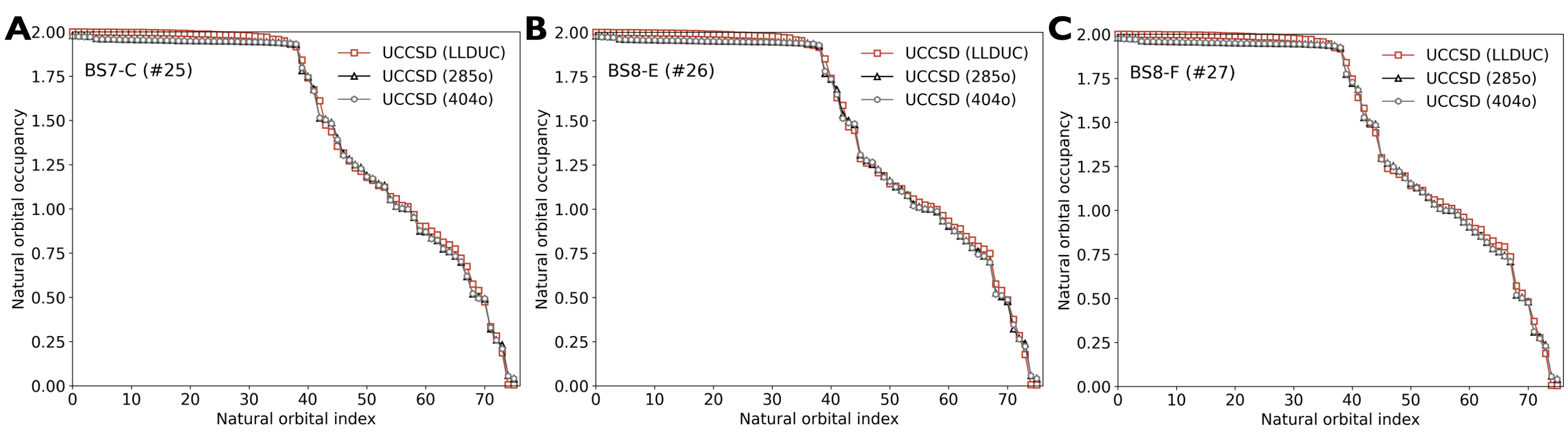}
  \centering
  \caption{UCCSD natural orbital occupancy computed for spin isomers BS7-C, BS8-E, and BS8-F in the LLDUC, (285o, 217e), and (404o, 277e) active spaces. For 285 and 404 orbital active spaces, only occupancy of the 76-orbital subspace corresponding to the LLDUC region is plotted.}
  \label{fig:m-large-cc-nat-occ}
\end{figure}

\subsection{Active space model from QM/MM}
\label{sec:sm-qmmm}
\subsubsection{Molecular dynamics}
The MD model was prepared from the crystal structure (PDB ID: 3U7Q~\cite{spatzal2011evidence}) with OpenMM~\cite{eastman2023openmm} and GFN2-xTB~\cite{bannwarth2019gfn2} through a Python interface provided by ASH~\cite{bjornsson2022ash}. The full heterotetramer of the Mo-Fe protein was included. The imidazole molecules from the buffer were removed, while the co-crystallized Mg$^{2+}$, Ca$^{2+}$, water, P-clusters, and FeMo-co were all kept. The Cys residues that coordinate with the FeMo-co and P-clusters were modeled as deprotonated, and the His residues that are close to the FeMo-co (H195:A and H195:C) were modeled as single-protonated on the $\epsilon$-nitrogen. The homocitrate was modelled in a singly protonated state where a proton was shared between the Mo-coordinating hydroxyl group and one carboxylate oxygen atom. All other residues were assumed to be in normal protonation states under pH 7. The system was solvated in a cubic water box with a side length of $\sim$140~\AA, and 0.1 M of sodium chloride was added.

The force field parameters for the P-cluster and FeMo-co were obtained from the universal force field~\cite{rappe1992uff} with atomic charges derived from GFN2-xTB~\cite{bannwarth2019gfn2}. In xTB calculations, the P-cluster was assumed to be in the resting state, i.e., \ce{Fe8S7^-} with total spin $S=0$, and the FeMo-co was in \ce{Fe7MoS9C^-} with $S=3/2$. The rest of the system was described by the Amber14 force field~\cite{maier2015ff14sb} with TIP3P-FB water~\cite{wang2014building}. 

The system was energy minimized, equilibrated under 1 bar and 300 K for 20 ns, and with a fixed volume under 300 K for 300 ns. During minimization and MD, the geometry of the P-cluster, the FeMo-co, and a few nearby residues, as well as crystal water, were kept frozen and not moved. The frozen atoms around FeMo-co (225 atoms in total) were the same ones used to define a truncated cluster model used in the previous high-spin DFT~\cite{li2019electronic-femoco} to derive the LLDUC model (denoted as LLDUC DFT hereafter). As such, the effects from protein and solvation water can be separated from the influence of a potential geometry change of the cluster, subject to the force field quality. All MD simulations were performed by OpenMM 8.1, using the middle Langevin integrator with a 4 fs timestep, hydrogen-involving bonds constrained, and repartitioned hydrogen masses (1.5 amu). 

\subsubsection{QM/MM potential}

The full periodic QM/MM electrostatic potential on the FeMo-co bound in chain A was computed for the MD-sampled geometries. The MM charges were taken from their force field values, and modelled as Gaussian-distributed charges with radii taken from their covalent~\cite{pyykko2009molecular} and ionic~\cite{shannon1976revised} radius values. The long-range electrostatics was computed with the QM/MM-Multipole approach~\cite{li2025accurate}, using a 25~\AA~cutoff for the short-range electrostatics beyond quadrupoles. The electrostatics generated by the periodic QM images were also included, and their charge distribution was modeled using the LLDUC DFT density~\cite{li2019electronic-femoco}. Due to NPT equilibration, the structure of FeMo-co was slightly rescaled (RMSD$=0.057$~\AA~ for the set of 225 atoms, including FeMo-co and a few nearby residues and water from the crystal structure (and thus from the structure used in the LLDUC model). To maintain consistency, the crystal structure to generate the LLDUC model was substituted in the MD-sampled geometries for the QM/MM potential calculations. The QM/MM potential was computed every 2 ps in the same atomic basis as LLDUC DFT. The potential was first block-averaged in 2-ns-long chunks, and we computed the electrostatic energy of the LLDUC DFT density under the averaged potential to form a time series $\{e_t\}$. The time $t_0$ that minimized the standard error of $\{e_t|t\geq t_0\}$ was 40 ns and determined as the initial equilibration time. The standard error of $e$ on the last 260 ns was 0.66 kcal/mol per atom. The largest absolute-valued matrix element in the standard error of the potential was 1.5 mHa, and the mean absolute value of all elements was $8\times10^{-7}$ Hartrees.

\subsubsection{Active space model}
The same procedure as LLDUC was followed to derive an active space model, with the COSMO potential replaced by the sampled QM/MM potential. A high-spin DFT calculation ($S=35/2$) with spin-free X2C and the B3LYP functional was performed, and the spin-averaged DFT natural orbitals were split-localized using Pipek-Mezey localization~\cite{pipek1989fast}, then all the C 2s, 2p, Fe 3d, Mo 4d, and S 3p orbitals, along with two ligand orbitals, were selected to be in the active space, yielding a 76o active space.

\subsubsection{Energy of broken-symmetry solutions}

To sample UHF solutions for the 76o active space model with the QM/MM potential, for each of the 35 broken-symmetry spin couplings, we consider: (a) 2250 possible initial diagonal density matrices; (b) around 351 unique UHF solutions in the 76o LLDUC active space projected to the 76o active space with QM/MM potential. We identified 24 low-energy UHF solutions from unique UHF solutions obtained from (a) and (b). UCCSD and UCCSD(T) are performed using each the 24 low-energy UHF as references. The lowest UHF energy, the lowest UCCSD energy and the UCCSD(T) energy corresponding to the lowest UCCSD energy, for all spin isomers in the 76o active space model with QM/MM potential, are listed in Supplementary Table~\ref{tab:sm-fm-qmmm-ener} and plotted in Supplementary Figure~\ref{fig:sm-fm-qmmm-ener}. We see that the UHF and UCCSD energies are mostly correlated across the LLDUC and QM/MM models. The UCCSD(T) energies show slightly larger fluctuations than UCCSD, partially caused by the sensitivity in the (T) correction when the $T_1$ diagnostic is large (see Supplementary Sec.~\ref{sec:sm-ref-dep}).

\begin{table}[!htbp]
    \small
    \centering
    \caption{UHF, UCCSD, and UCCSD(T) energies with the best reference for each spin isomer of the 76o active space model with QM/MM potential. Energies are in Hartrees.}
    \begin{tabular}{
        >{\centering\arraybackslash}p{1.8cm}|
        >{\centering\arraybackslash}p{2.4cm}
        >{\centering\arraybackslash}p{2.4cm}
        >{\centering\arraybackslash}p{2.4cm}
        >{\centering\arraybackslash}p{3.2cm}
    }
    \hline\hline
    spin isomer & $E_{\mathrm{UHF}}$ & $E_{\mathrm{UCCSD}}$ & $E_{\mathrm{UCCSD(T)}}$ & CCSD $T_1$ diagnostic \\
    \hline
     BS3-A &     -1128.749828 &     -1129.051423 &     -1129.109900 &      0.08939 \\
     BS3-B &     -1128.747568 &     -1129.050914 &     -1129.110334 &      0.09382 \\
     BS9-A &     -1128.763149 &     -1129.058648 &     -1129.110599 &      0.09186 \\
    BS10-A &     -1128.764822 &     -1129.075438 &     -1129.128793 &      0.09264 \\
    BS10-B &     -1128.759424 &     -1129.075676 &     -1129.126760 &      0.09846 \\
     BS3-C &     -1128.735494 &     -1129.041570 &     -1129.091090 &      0.08658 \\
    BS10-C &     -1128.760007 &     -1129.056808 &     -1129.129654 &      0.13623 \\
     BS9-B &     -1128.752469 &     -1129.047591 &     -1129.108324 &      0.12632 \\
    BS10-D &     -1128.753904 &     -1129.061313 &     -1129.130782 &      0.14911 \\
    BS10-E &     -1128.760105 &     -1129.055396 &     -1129.114975 &      0.10091 \\
    BS10-F &     -1128.758531 &     -1129.068612 &     -1129.128012 &      0.08651 \\
     BS9-C &     -1128.750223 &     -1129.049912 &     -1129.107509 &      0.10196 \\
     BS6-A &     -1128.739521 &     -1129.037462 &     -1129.089403 &      0.09699 \\
     BS6-B &     -1128.734076 &     -1129.035167 &     -1129.089990 &      0.09890 \\
     BS6-C &     -1128.730396 &     -1129.049856 &     -1129.098651 &      0.09151 \\
       BS2 &     -1128.741752 &     -1129.057890 &     -1129.103049 &      0.09519 \\
     BS8-A &     -1128.755322 &     -1129.060500 &     -1129.115902 &      0.09005 \\
     BS8-B &     -1128.756467 &     -1129.058977 &     -1129.117878 &      0.09125 \\
     BS7-A &     -1128.759225 &     -1129.068524 &     -1129.126342 &      0.09288 \\
     BS8-C &     -1128.751086 &     -1129.057622 &     -1129.115664 &      0.10431 \\
     BS7-B &     -1128.761941 &     -1129.068528 &     -1129.127390 &      0.11206 \\
     BS8-D &     -1128.754527 &     -1129.054647 &     -1129.114224 &      0.11251 \\
     BS5-A &     -1128.739661 &     -1129.022878 &     -1129.089226 &      0.19220 \\
     BS5-B &     -1128.739819 &     -1129.024689 &     -1129.080697 &      0.10866 \\
     BS4-A &     -1128.743246 &     -1129.048552 &     -1129.103613 &      0.09069 \\
     BS7-C &     -1128.760625 &     -1129.080084 &     -1129.128564 &      0.08787 \\
     BS8-E &     -1128.756306 &     -1129.075725 &     -1129.125222 &      0.08719 \\
     BS8-F &     -1128.756161 &     -1129.072127 &     -1129.123413 &      0.09122 \\
     BS5-C &     -1128.746947 &     -1129.036154 &     -1129.096812 &      0.15114 \\
     BS4-B &     -1128.754599 &     -1129.054236 &     -1129.142001 &      0.10089 \\
     BS5-D &     -1128.746839 &     -1129.043997 &     -1129.099306 &      0.08231 \\
     BS4-C &     -1128.754485 &     -1129.059842 &     -1129.115663 &      0.14681 \\
     BS5-E &     -1128.748739 &     -1129.048206 &     -1129.123704 &      0.09940 \\
     BS5-F &     -1128.746401 &     -1129.051020 &     -1129.144301 &      0.11775 \\
       BS1 &     -1128.710228 &     -1128.980490 &     -1129.026409 &      0.10471 \\
    \hline\hline
    \end{tabular}
    \label{tab:sm-fm-qmmm-ener}
\end{table}

\begin{figure}[!htbp]
  \includegraphics[width=\linewidth]{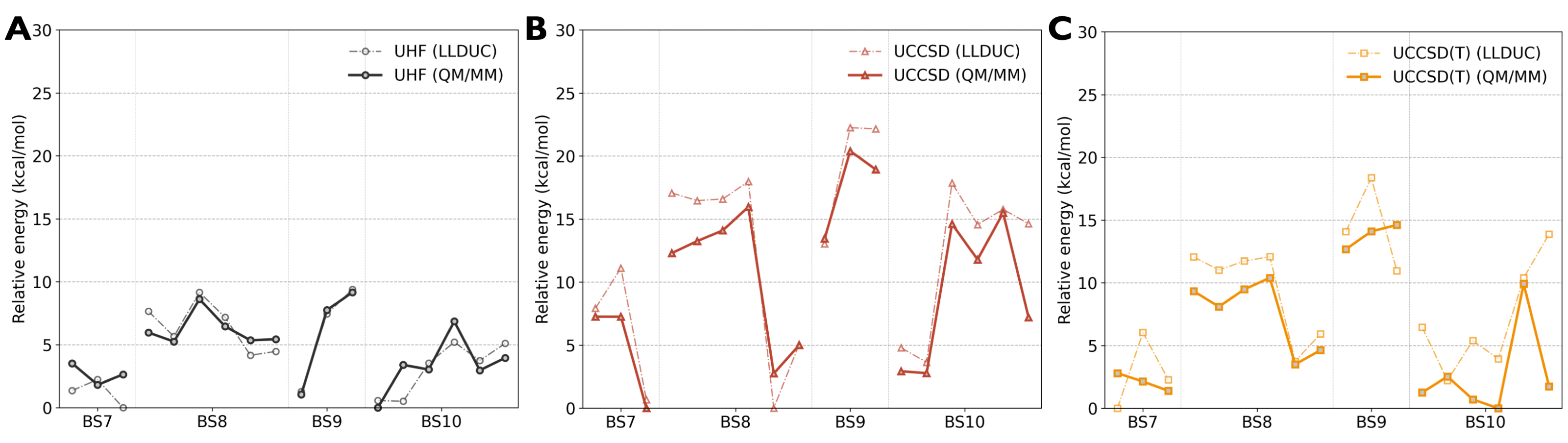}
  \caption{\textbf{A} The lowest UHF energy, \textbf{B} the lowest UCCSD energy, \textbf{C} and the UCCSD(T) energy corresponding to the lowest UCCSD energy, for BS7-BS10 spin isomers in the 76o active space model with QM/MM potential. }
  \label{fig:sm-fm-qmmm-ener}
\end{figure}

\section{4Fe-4S clusters}
\label{sec:sm-fe4}

We examine the accuracy of the spin unrestricted DMRG and high-order CC approaches for FeMo-co models by investigating their performance on the smaller 2Fe(II)-2Fe(III) iron-sulfur cubane cluster. This has Fe atoms in different oxidation states and near-degenerate low-energy spin configurations, representing typical computational challenges in these iron-sulfur systems. The (36o, 54e) active space model was defined in Ref.~\cite{li2017spin} and the integral file can be obtained from \url{https://github.com/zhendongli2008/Active-space-model-for-Iron-Sulfur-Clusters/blob/main/Fe2S2_and_Fe4S4/Fe4S4/fe4s4}. Note that the core energy \\
$-8105.56038966$ Hartrees for this active space model is not included in the reported total energy.

\subsection{Spin-adapted DMRG and extrapolations}

First we compute the ground-state energy for one of the spin isomers of the (36o, 54e) active space model of $\ce{[Fe4S4(SCH3)4]^{2-}}$ using spin-adapted DMRG up to a large bond dimension $D=20000$ (number of spin multiplets) starting from MPS initial guesses generated using spin-projected DMRG,\cite{li2017spin} which gives variational upper bounds of $E_{\mathrm{variational}}^{\text{SA-DMRG}} = -327.244142, -327.246083$, and $-327.246141$ Hartrees, for the BS1, BS2, and BS3 states initial guesses, respectively.
From the one- and two-particle density matrices, we identify the SA-DMRG state starting from the ``BS1'' spin isomer guess as corresponding to the configuration [Fe1$\uparrow$, Fe2$\uparrow$, Fe3$\downarrow$, Fe4$\downarrow$] (see Supplementary Figure \ref{fig:sm-fe4-spin-corr}), which in the subsequent section, we name ``BS1'' even though no symmetry is broken. (This is because the broken symmetry solution is obtained by mixing spin-adapted states in a specific tower of states associated with that spin isomer, and the spin-adapted solution is the ground-state within that tower of states.) The SA-DMRG solution starting from the ``BS2'' and ``BS3'' initial guesses converges to the same state which is a mixture of BS2 and BS3 spin patterns. Hence, in the remainder of this subsection we do not treat ``BS2'' and ``BS3'' as distinct SA-DMRG solutions.

\begin{figure}[!htbp]
  \centering
  \includegraphics[width=0.75\linewidth]{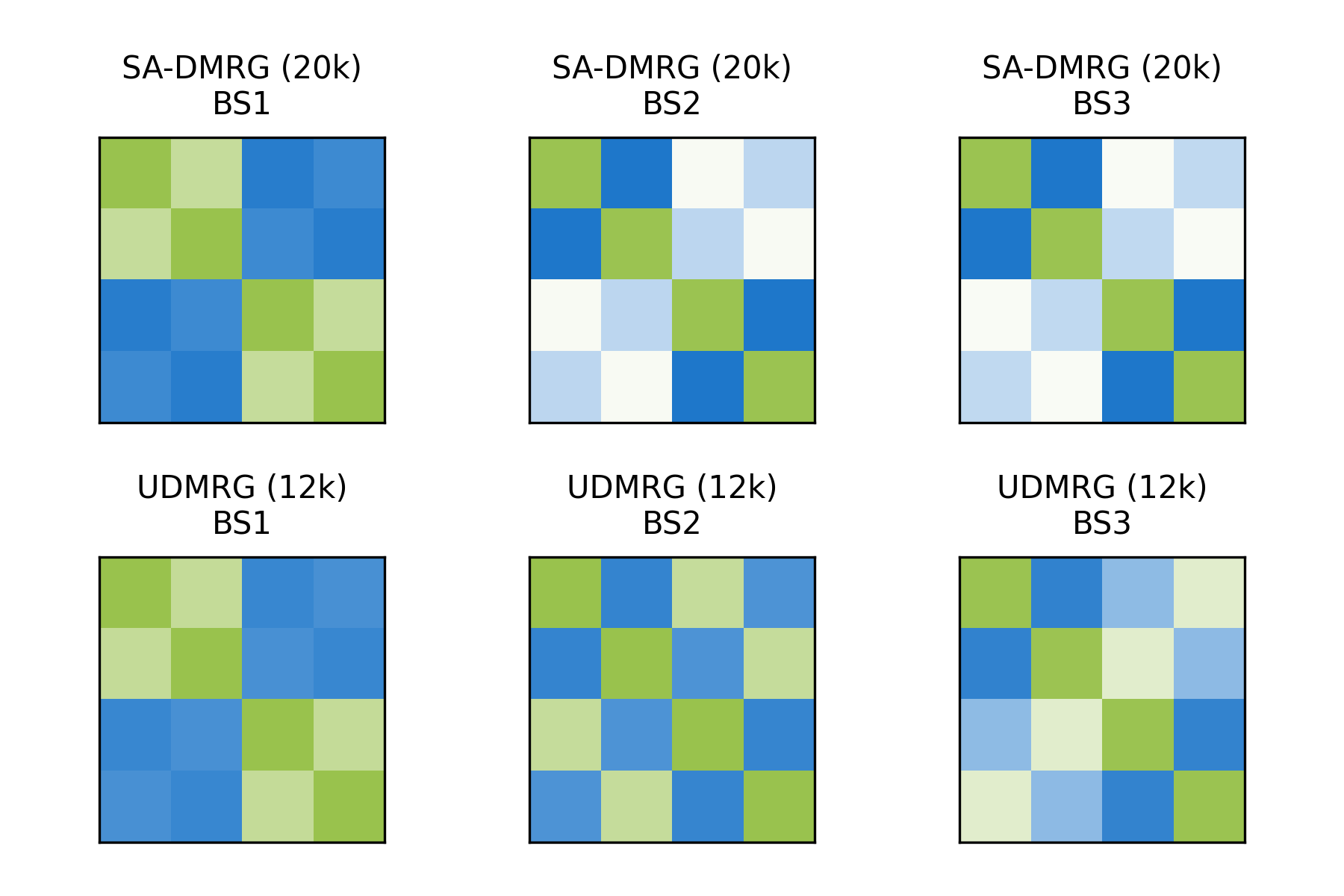}
  \caption{The spin correlation pattern computed for the SA-DMRG state at $D=20000$ starting from different initial guesses, and the three UDMRG states at $D=12000$ for the (36o, 54e) active space model of the 2Fe(II)-2Fe(III) iron-sulfur cubane cluster.}
  \label{fig:sm-fe4-spin-corr}
\end{figure}

Starting from the $D=20000$ solution, we perform discarded weight based extrapolation for energies from a reverse schedule with bond dimensions $D=18000$ to $D=14000$, which gives an extrapolated ground-state energy estimate of $E_{\mathrm{extrap}}^{\text{SA-DMRG}} = -327.245342$ and $-327.248524$ Hartrees for BS1 and BS3, respectively. We list spin-adapted DMRG energies in Supplementary Table \ref{tab:sm-fe4-sa-dmrg}. The energy extrapolation for spin-adapted DMRG is shown in Supplementary Figure \ref{fig:sm-fe4-extra}. We also perform bond dimension extrapolation using $\exp[-\kappa(\log D)^2]$ fitting, to give $E_{\mathrm{extrap}}^{\text{SA-DMRG}} = -327.245287$ and $-327.248841$ Hartrees for BS1 and BS3, which agrees very well with the extrapolated energy using maximal discarded weights. We will compare the energy obtained using spin-adapted DMRG to those from various broken symmetry estimates used in this work.

\begin{table}[!htbp]
    \centering
    \caption{Spin-adapted DMRG energies and discarded weights for the 2FeII-2FeIII iron-sulfur cubane cluster active space model.}
    \begin{tabular}{
        >{\centering\arraybackslash}p{2.1cm}|
        >{\centering\arraybackslash}p{2.7cm}|
        >{\centering\arraybackslash}p{1.2cm}
        >{\centering\arraybackslash}p{3.5cm}
        >{\centering\arraybackslash}p{4.3cm}
    }
    \hline\hline
    initial guess & schedule type & $D$ & energy (Hartree) & max discarded weight \\
    \hline
    \multirow{7}{*}{BS1} &          forward &    20000 &     -327.244142 & $1.52\times 10^{-5}$ \\
    \cline{2-5}
    & \multirow{5}{*}{reverse}      &    18000 &     -327.244025 & $1.99\times 10^{-5}$ \\
    &  &    17000 &     -327.243945 & $2.19\times 10^{-5}$ \\
    &  &    16000 &     -327.243850 & $2.35\times 10^{-5}$ \\
    &  &    15000 &     -327.243744 & $2.47\times 10^{-5}$ \\
    &  &    14000 &     -327.243626 & $2.61\times 10^{-5}$ \\
    \cline{2-5}
    &     extrapolated & $\infty$ &     -327.245342 &                      \\
    \hline
\multirow{1}{*}{BS2} &          forward &    20000 &     -327.246083 & $3.36\times 10^{-5}$ \\
    \hline
\multirow{7}{*}{BS3} &          forward &    20000 &     -327.246141 & $4.07\times 10^{-5}$ \\
    \cline{2-5}
    & \multirow{5}{*}{reverse}      &    18000 &     -327.245860 & $4.24\times 10^{-5}$ \\
    &  &    17000 &     -327.245704 & $4.57\times 10^{-5}$ \\
    &  &    16000 &     -327.245528 & $4.87\times 10^{-5}$ \\
    &  &    15000 &     -327.245332 & $5.16\times 10^{-5}$ \\
    &  &    14000 &     -327.245113 & $5.45\times 10^{-5}$ \\
    \cline{2-5}
    &     extrapolated & $\infty$ &     -327.248524 &                      \\
    \hline\hline
    \end{tabular}
    \label{tab:sm-fe4-sa-dmrg}
\end{table}

\begin{figure}[!htbp]
  \centering
  \includegraphics[width=0.95\linewidth]{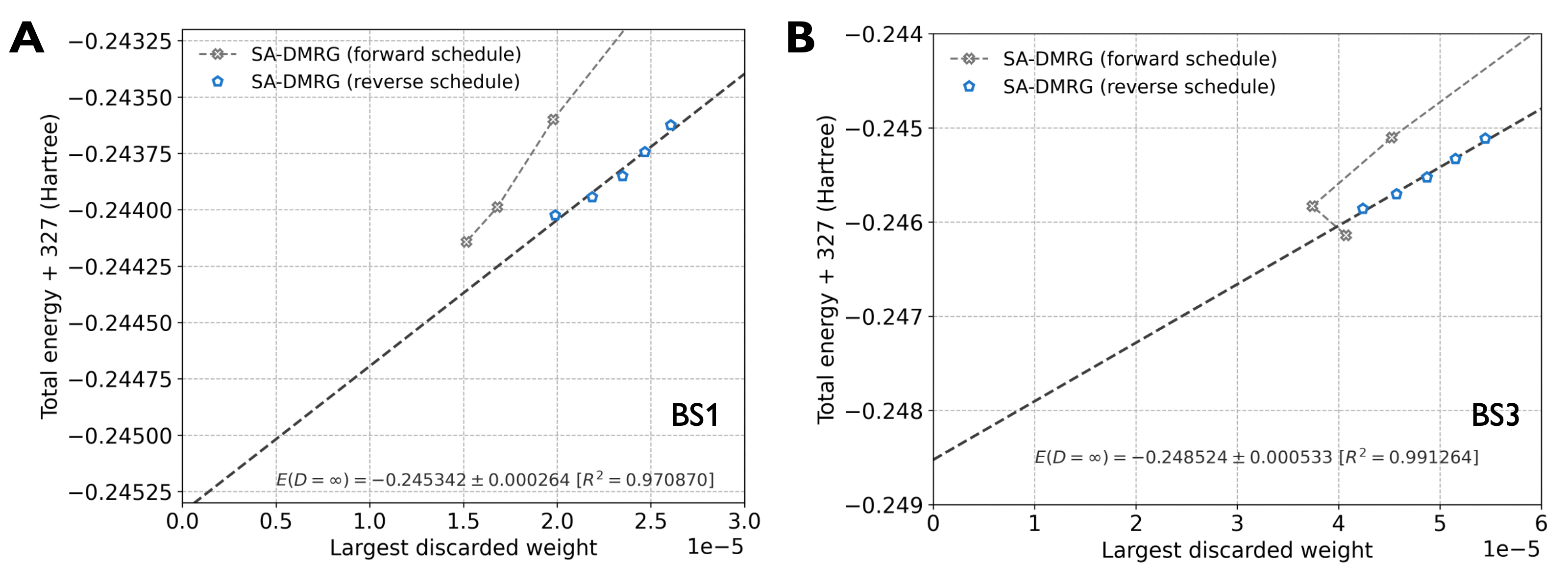}
  \caption{The spin-adapted DMRG energy extrapolation for the ``BS1'' and ``BS3'' spin isomers of 2FeII-2FeIII iron-sulfur cubane cluster active space model. Note that the largest discarded weight from forward-schedule DMRG can typically be uninformative, as the state is not fully converged.}
  \label{fig:sm-fe4-extra}
\end{figure}

\textbf{Entanglement-minimized orbitals.} We also performed spin-adapted DMRG with entanglement-minimized orbitals (EMO)~\cite{li2025entanglement} and a large bond dimension ($D=20000$) to further reduce the error in the ground-state energy estimate. The EMOs were optimized with $D = 100$ for each spin isomer. The EMO-DMRG calculations were performed using the spin-adapted DMRG implementation in the \textsc{FOCUS} package with the support of GPU acceleration~\cite{xiang2024distributed}.
We obtain tighter variational upper bound energies at $D=20000$ as $E_{\mathrm{variational}}^{\text{EMO-DMRG}} = -327.244733, -327.246852$ and $-327.247036$ Hartrees, for BS1, BS2, and BS3, respectively. With the $\exp[-\kappa(\log D)^2]$ extrapolation using the forward schedule sweep energy data (up to bond dimension $D=20000$), we have $E_{\mathrm{extrap}}^{\text{EMO-DMRG}} = -327.246726, -327.248835$ and $-327.248858$ Hartrees, for BS1, BS2, and BS3, respectively.

\subsection{UHF, UCC, and UDMRG energies}

We next consider the broken-symmetry treatment for the 2Fe(II)-2Fe(III) iron-sulfur cubane cluster, following the same protocol discussed in Sec.~\ref{sec:sm-ref-dep} to enumerate UHF solutions and filter references for UCC and UDMRG calculations. The 4Fe--4S cluster has three different spin isomers, denoted BS1, BS2, and BS3. We identified 11, 9, and 6 unique low-energy UHF solutions respectively for the three BS states (denoted BS$n$-$m$), with their UHF, UCCSD, and UCCSD(T) energies listed in Supplementary Table \ref{tab:sm-fe4-bs}. We found that the ranking based on UHF energies is very different from that based on UCCSD energies. Noticeably, the BS2-8 state has the lowest UCCSD and UCCSD(T) energies, but its UHF energy is the highest among all UHF local minima. In contrast, the UCCSD(T) energies mostly track the UCCSD energies for this system.

\begin{table}[!htbp]
    \centering
    \caption{Broken-symmetry UHF energies, and UCCSD and UCCSD(T) energies computed using different references for the 2FeII-2FeIII iron-sulfur cubane cluster active space model. The reference with the lowest UCCSD energies for each spin isomer is shown in bold.}
    \begin{tabular}{
        >{\centering\arraybackslash}p{4.4cm}|
        >{\centering\arraybackslash}p{1.8cm}|
        >{\centering\arraybackslash}p{2.2cm}
        >{\centering\arraybackslash}p{2.2cm}
        >{\centering\arraybackslash}p{2.2cm}
    }
    \hline\hline
    spin configuration & reference & $E_{\mathrm{UHF}}$& $E_{\mathrm{UCCSD}}$ & $E_{\mathrm{UCCSD(T)}}$ \\
    \hline
\multirow{11}{*}{Fe1$\uparrow$, Fe2$\uparrow$, Fe3$\downarrow$, Fe4$\downarrow$}
&      BS1-0 &  -327.084355 &  -327.197168 &  -327.215914 \\
&      BS1-1 &  -327.084217 &  -327.197124 &  -327.215909 \\
&      BS1-2 &  -327.083505 &  -327.199392 &  -327.215847 \\
&      BS1-3 &  -327.083499 &  -327.197465 &  -327.216088 \\
& \textbf{BS1-4} & \textbf{-327.082953} & \textbf{-327.199448} & \textbf{-327.215860} \\
&      BS1-5 &  -327.082836 &  -327.198511 &  -327.217398 \\
&      BS1-6 &  -327.082766 &  -327.197089 &  -327.215605 \\
&      BS1-7 &  -327.082759 &  -327.196287 &  -327.214866 \\
&      BS1-8 &  -327.082038 &  -327.195571 &  -327.214466 \\
&      BS1-9 &  -327.080859 &  -327.193691 &  -327.210521 \\
&     BS1-10 &  -327.079407 &  -327.173790 &  -327.184312 \\
\hline
\multirow{9}{*}{Fe1$\uparrow$, Fe2$\downarrow$, Fe3$\uparrow$, Fe4$\downarrow$}
&      BS2-0 &  -327.084303 &  -327.200126 &  -327.216413 \\
&      BS2-1 &  -327.083894 &  -327.196582 &  -327.215204 \\
&      BS2-2 &  -327.083452 &  -327.199640 &  -327.215888 \\
&      BS2-3 &  -327.083187 &  -327.196116 &  -327.214926 \\
&      BS2-4 &  -327.082796 &  -327.196029 &  -327.215001 \\
&      BS2-5 &  -327.082733 &  -327.195510 &  -327.214555 \\
&      BS2-6 &  -327.081548 &  -327.194452 &  -327.211403 \\
&      BS2-7 &  -327.079922 &  -327.173458 &  -327.184170 \\
& \textbf{BS2-8} & \textbf{-327.079094} & \textbf{-327.209283} & \textbf{-327.221857} \\
\hline
\multirow{6}{*}{Fe1$\uparrow$, Fe2$\downarrow$, Fe3$\downarrow$, Fe4$\uparrow$}
& \textbf{BS3-0} & \textbf{-327.089868} & \textbf{-327.199165} & \textbf{-327.213268} \\
&      BS3-1 &  -327.088966 &  -327.198048 &  -327.211723 \\
&      BS3-2 &  -327.088874 &  -327.198034 &  -327.212002 \\
&      BS3-3 &  -327.088412 &  -327.197841 &  -327.211844 \\
&      BS3-4 &  -327.080834 &  -327.191994 &  -327.209547 \\
&      BS3-5 &  -327.080211 &  -327.192248 &  -327.209510 \\
    \hline\hline
    \end{tabular}
    \label{tab:sm-fe4-bs}
\end{table}

\subsection{Estimation of exact energies}

We now consider higher-order UCC and UDMRG for the three spin isomers with references corresponding to the lowest UCCSD energies, which are BS1-4, BS2-8, and BS3-0. 
Their UHF and UCC (up to UCCSDTQ) energies are listed in Supplementary Table \ref{tab:sm-fe4-ucc-post-q}. We plot the UCC and UDMRG energies (relative to the ``BS3'' SA-DMRG energy $E_{\mathrm{extrap}}^{\text{SA-DMRG}}$) in Supplementary Figure \ref{fig:sm-fe4-ener}. We find that BS2 remains the lowest energy state at the UCCSDT level, but the ordering changes at the UCCSDTQ and UDMRG level. The energy spread of spin isomers decreases with increasing correlation level.

We list the UDMRG energies for the three BS states in Supplementary Table~\ref{tab:sm-fe4-udmrg-ener}, and plot the bond dimension extrapolation in Supplementary Figure \ref{fig:sm-fe4-udmrg-fit}. For BS1, the UDMRG extrapolated energy of $-327.244309$ Hartrees is slightly higher than the SA-DMRG extrapolated energy ($\sim$1 mHa) even with a modest extrapolation distance ($\sim$3 mHa), which gives a sense of the extrapolation accuracy in these challenging systems. In BS2 and BS3, the UDMRG extrapolated energies are 1.0 mHa and 3.6 mHa below the UCCSDTQ energies. This suggests that the pentuples and higher correlations are negative.
As an additional evidence, the extrapolated EMO-DMRG energies are 0.6 mHa, 4.3 mHa, and 2.5 mHa below the UCCSDTQ energies, for BS1, BS2, and BS3, respectively.

We next carry out an estimation of the exact energy from UCC using a Fe dimer reference system as we did for FeMo-co (see Sec.~\ref{sec:sm-comp}).
For this we take the (20o, 31e) Fe(III)–Fe(II) dimer active space model (see Sec.~\ref{sec:sm-spin}),
and we show the correlation energy increments, ratios, and predicted exact energies for BS1, BS2, and BS3, in Supplementary Table~\ref{tab:sm-fe4-ucc-post-q}. We note that the correlation increment ratios are much less uniform for the cubane than they are for FeMo-co. This suggests the dimer is less good of a model system here.

Taking the consensus prediction as the average of the UCC and UDMRG predictions, we finally obtain $-327.246177 \pm 0.001868$ Hartrees, $-327.245781 \pm 0.000240$ Hartrees, $-327.249366 \pm 0.000653$ Hartrees, for BS1, BS2, and BS3, respectively. More conservatively, we can estimate the energy to lie between the UCCSDTQ upper bounds and the lowest energy extrapolations i.e. the ranges
$[-327.246078,-327.248046]$, $[-327.244556,-327.246022]$, and $[-327.246379,-327.250019]$ Hartrees for BS1, BS2, BS3 respectively, or $-327.247062 \pm 0.000984$ Hartrees for BS1, $-327.245289 \pm
0.000733$ Hartrees for BS2, and $-327.248199 \pm 0.001820$ Hartrees for BS3.
The energy difference between the composite UCC/UDMRG estimate and the EMO-DMRG extrapolation is 0.3 mHa, 3.5 mHa, 0.6 mHa for BS1, BS2, and BS3, respectively.
Overall, the close correspondence between the UCC and UDMRG extrapolations, and the tightness of the UCCSDTQ upper bounds, suggests that our protocol predicts the exact energy with estimated errors of $\sim$1 kcal/mol.

\begin{table}[!htbp]
    \small
    \centering
    \caption{UHF, UCC, and post quadruples correction for the UCC energy for BS1, BS2, and BS3 states computed using references with the lowest UCCSD energies (BS1-4, BS2-8, and BS3-0) for the 2FeII-2FeIII iron-sulfur active space model. $\Delta E$ represents the D, T, Q and post-Q correlation increments. Estimated quantities are labeled in bold. Absolute energies are in Hartrees shifted by $-116.0$ Hartrees (dimer) or $-327.0$ Hartrees (2FeII-2FeIII).}
    \begin{tabular}{
        >{\centering\arraybackslash}p{1.8cm}|
        >{\centering\arraybackslash}p{1.8cm}
        >{\centering\arraybackslash}p{1.6cm}
        >{\centering\arraybackslash}p{1.6cm}
        >{\centering\arraybackslash}p{1.6cm}
        >{\centering\arraybackslash}p{1.6cm}
        >{\centering\arraybackslash}p{3.5cm}
    }
    \hline\hline
    spin isomer & theory & $E_{\mathrm{dimer}}$ & $\Delta E_{\mathrm{dimer}}$ & $E_{\mathrm{4Fe}}$ & $\Delta E_{\mathrm{4Fe}}$ & $\Delta E_{\mathrm{4Fe}}/\Delta E_{\mathrm{dimer}}$ \\
    \hline
    BS1 &        UHF &    -0.322099 &            &    -0.082953 &            &          \\
           &      UCCSD &    -0.361322 &  -0.039224 &    -0.199448 &  -0.116494 &     2.97 \\
           &   UCCSD(T) &    -0.367095 &            &    -0.215860 &            &          \\
           &     UCCSDT &    -0.372305 &  -0.010983 &    -0.237763 &  -0.038315 &     3.49 \\
           &    UCCSDTQ &    -0.373827 &  -0.001522 &    -0.246078 &  -0.008314 &     5.46 \\
           &      exact &    -0.374322 &  -0.000495 & \textbf{-0.248046} & \textbf{-0.001968} & \textbf{3.97} \\
     \hline
       BS2 &        UHF &    -0.322099 &            &    -0.079094 &            &          \\
           &      UCCSD &    -0.361322 &  -0.039224 &    -0.209283 &  -0.130189 &     3.32 \\
           &   UCCSD(T) &    -0.367095 &            &    -0.221857 &            &          \\
           &     UCCSDT &    -0.372305 &  -0.010983 &    -0.240411 &  -0.031128 &     2.83 \\
           &    UCCSDTQ &    -0.373827 &  -0.001522 &    -0.244556 &  -0.004145 &     2.72 \\
           &      exact &    -0.374322 &  -0.000495 & \textbf{-0.246022} & \textbf{-0.001466} & \textbf{2.96} \\
     \hline
       BS3 &        UHF &    -0.322099 &            &    -0.089868 &            &          \\
           &      UCCSD &    -0.361322 &  -0.039224 &    -0.199165 &  -0.109298 &     2.79 \\
           &   UCCSD(T) &    -0.367095 &            &    -0.213268 &            &          \\
           &     UCCSDT &    -0.372305 &  -0.010983 &    -0.233939 &  -0.034774 &     3.17 \\
           &    UCCSDTQ &    -0.373827 &  -0.001522 &    -0.246379 &  -0.012440 &     8.17 \\
           &      exact &    -0.374322 &  -0.000495 & \textbf{-0.248712} & \textbf{-0.002332} & \textbf{4.71} \\
    \hline\hline
    \end{tabular}
    \label{tab:sm-fe4-ucc-post-q}
\end{table}

\begin{figure}[!htbp]
  \centering
  \includegraphics[width=0.75\linewidth]{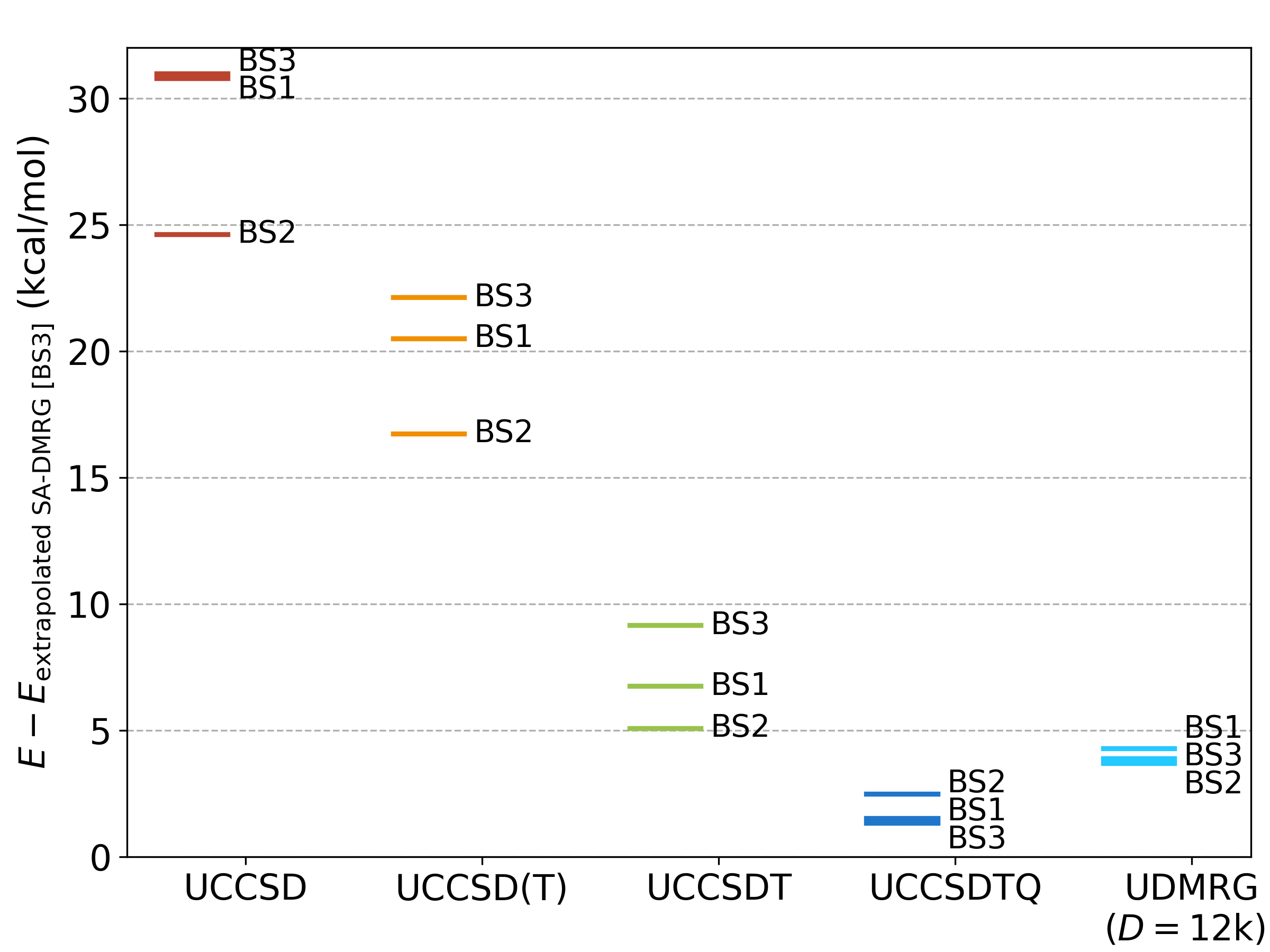}
  \caption{UCC and UDMRG energy error (estimated using the difference from BS3 SA-DMRG energy $E_{\mathrm{extrap}}^{\text{SA-DMRG}}$) for the 2FeII-2FeIII iron-sulfur cubane cluster active space model.}
  \label{fig:sm-fe4-ener}
\end{figure}

\begin{table}[!htbp]
    \small
    \centering
    \caption{UDMRG energies and discarded weights for the 2FeII-2FeIII active space model. $\Delta E_{\text{UDMRG}} = E_{\text{UDMRG}} - E_{\mathrm{extrap}}^{\text{SA-DMRG}}$. $E_{\mathrm{extrap}}^{\text{SA-DMRG}}$ for ``BS2'' is  taken to be the same as that for ``BS3''. Energies are in Hartrees.}
    \begin{tabular}{
        >{\centering\arraybackslash}p{1.8cm}|
        >{\centering\arraybackslash}p{3.2cm}
        >{\centering\arraybackslash}p{1.0cm}
        >{\centering\arraybackslash}p{2.2cm}
        >{\centering\arraybackslash}p{3.4cm}
        >{\centering\arraybackslash}p{2.4cm}
    }
    \hline\hline
    spin isomer & schedule type & $D$ & $E_{\text{UDMRG}}$ & max discarded weight & $\Delta E_{\text{UDMRG}}$ \\
    \hline
BS1
&          forward &    12000 &     -327.241697 & $2.75\times 10^{-5}$ &    +0.003646 \\
&          reverse &    10000 &     -327.241407 & $3.69\times 10^{-5}$ &              \\
&          reverse &     9000 &     -327.241208 & $4.12\times 10^{-5}$ &              \\
&          reverse &     8000 &     -327.240971 & $4.49\times 10^{-5}$ &              \\
&          reverse &     7000 &     -327.240688 & $4.82\times 10^{-5}$ &              \\
&          reverse &     6000 &     -327.240340 & $5.19\times 10^{-5}$ &              \\
&          reverse &     5000 &     -327.239894 & $5.63\times 10^{-5}$ &              \\
&          reverse &     4500 &     -327.239627 & $5.81\times 10^{-5}$ &              \\
&          reverse &     4000 &     -327.239304 & $6.12\times 10^{-5}$ &              \\
&          reverse &     3500 &     -327.238913 & $6.47\times 10^{-5}$ &              \\
&          reverse &     3000 &     -327.238426 & $6.89\times 10^{-5}$ &              \\
& $\exp[-\kappa(\log D)^2]$ fit & $\infty$ &     -327.244309 &                      &    +0.001033 \\
    \hline
BS2
&          forward &    12000 &     -327.242634 & $3.35\times 10^{-5}$ &    +0.005889 \\
&          reverse &    10000 &     -327.242327 & $4.29\times 10^{-5}$ &              \\
&          reverse &     9000 &     -327.242123 & $4.61\times 10^{-5}$ &              \\
&          reverse &     8000 &     -327.241885 & $4.89\times 10^{-5}$ &              \\
&          reverse &     7000 &     -327.241604 & $5.15\times 10^{-5}$ &              \\
&          reverse &     6000 &     -327.241264 & $5.43\times 10^{-5}$ &              \\
&          reverse &     5000 &     -327.240831 & $5.79\times 10^{-5}$ &              \\
&          reverse &     4500 &     -327.240576 & $5.90\times 10^{-5}$ &              \\
&          reverse &     4000 &     -327.240267 & $6.12\times 10^{-5}$ &              \\
&          reverse &     3500 &     -327.239892 & $6.42\times 10^{-5}$ &              \\
&          reverse &     3000 &     -327.239426 & $6.81\times 10^{-5}$ &              \\
& $\exp[-\kappa(\log D)^2]$ fit & $\infty$ &     -327.245541 &                      &    +0.002983 \\
    \hline
BS3
&          forward &    12000 &     -327.242319 & $8.29\times 10^{-5}$ &    +0.006204 \\
&          reverse &    10000 &     -327.241750 & $9.50\times 10^{-5}$ &              \\
&          reverse &     9000 &     -327.241403 & $9.74\times 10^{-5}$ &              \\
&          reverse &     8000 &     -327.241009 & $1.00\times 10^{-4}$ &              \\
&          reverse &     7000 &     -327.240555 & $1.03\times 10^{-4}$ &              \\
&          reverse &     6000 &     -327.240017 & $1.07\times 10^{-4}$ &              \\
&          reverse &     5000 &     -327.239355 & $1.12\times 10^{-4}$ &              \\
&          reverse &     4500 &     -327.238975 & $1.12\times 10^{-4}$ &              \\
&          reverse &     4000 &     -327.238523 & $1.15\times 10^{-4}$ &              \\
&          reverse &     3500 &     -327.237985 & $1.20\times 10^{-4}$ &              \\
&          reverse &     3000 &     -327.237326 & $1.25\times 10^{-4}$ &              \\
& $\exp[-\kappa(\log D)^2]$ fit & $\infty$ &     -327.250019 &                      &    -0.001496 \\
    \hline\hline
    \end{tabular}
    \label{tab:sm-fe4-udmrg-ener}
\end{table}

\begin{figure}[!htbp]
  \centering
  \includegraphics[width=0.9\linewidth]{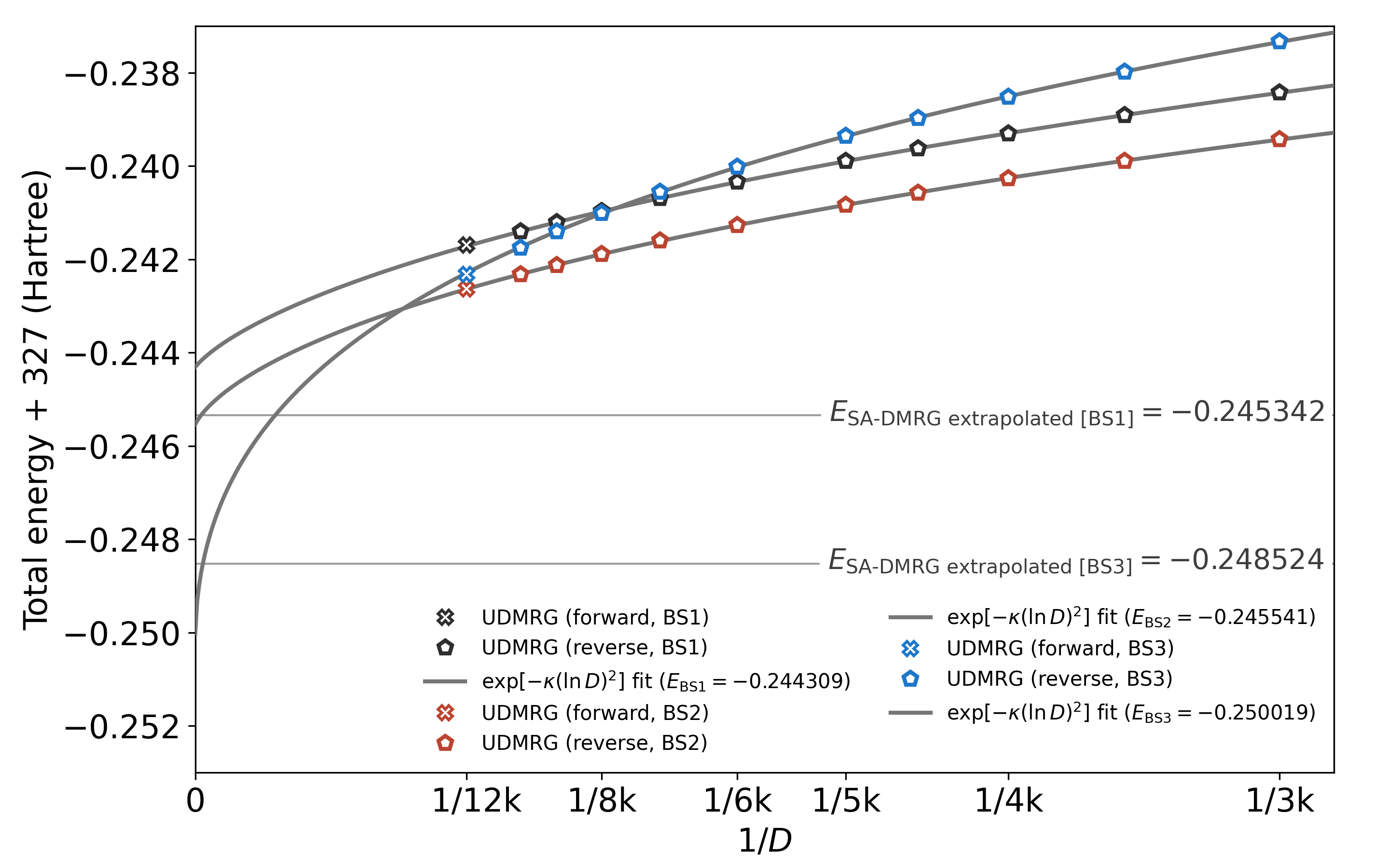}
  \caption{The $\exp[-\kappa(\log D)^2]$ fitting for UDMRG energies for the (36o, 54e) active space model of the 2Fe(II)-2Fe(III) iron-sulfur cubane cluster.}
  \label{fig:sm-fe4-udmrg-fit}
\end{figure}

\section{Oxidation state calibration}
\label{sec:sm-ox-calib}

We performed broken-symmetry DFT and correlated wavefunction calculations on 2Fe-2S clusters (\ce{Fe2S2(SCH3)4^{3-}}) and a model of the P-cluster to establish a scale to correlate iron oxidation states with the electronic structure, and to test the sensitivity to computational settings. For Mo(III) we use $\ce{[MoFe3S4(SH)6]^{3-}}$ to set the scale for the oxidation states.

\subsection{Description of Meta-L\"owdin procedure}

Meta-L\"owdin population analysis is a form of L\"owdin population analysis where the atomic basis is first projected (using a reference set of atomic natural orbitals) so that individual atomic orbitals have core, valence, and Rydberg character. The L\"owdin analysis is then carried out in the core, valence, and Rydberg spaces separately~\cite{sun2014exact}. 

\subsection{Meta-L\"owdin population sensitivity}

We tested the robustness of the meta-L\"owdin population on 2Fe-2S clusters by simulating an extensive combination of total charge and spin states, computational basis sets, and active space sizes (controlled by the number of frozen core orbitals). Meta-L\"owdin analysis shows that the iron populations are sensitive to the computational basis, which can be eliminated by only counting the electrons in the core and valence space (and thus not the Rydberg contributions). For example, in the high-spin Fe(III)-Fe(II) state (total charge = -3 and $S=9/2$), meta-L\"owdin analysis on MP2/aug-cc-pwCVTZ density gives 24.98 and 24.93 electrons on the two irons, while giving 24.29 and 24.77 for the MP2/TZP-DKH density. After removing Rydberg contributions, meta-L\"owdin analysis gives 23.96 and 24.52 on MP2/aug-cc-pwCVTZ and 24.00 and 24.55 on MP2/TZP-DKH, so the basis dependency is largely suppressed and the Fe(III)-Fe(II) charge ordering is consistently retained. 

\subsection{Fe electron population scale}
We first tested if electron populations on the 2Fe-2S cluster are transferable to the P-cluster using DFT. Calculations were performed in ORCA 6.1.0 with the following input:
{\small
\begin{verbatim}
! UKS B3LYP NORI old-TZVP DKH2 DEFGRID2 TightSCF NOCOSX SlowConv CPCMC
\end{verbatim}
}
to best reproduce the settings in Ref.~\cite{bjornsson2017revisiting} using an older version of ORCA. Our meta-L\"owdin analysis, without Rydberg contributions, gives 24.94 for both irons in high-spin, both ferrous 2Fe-2S (total charge = -4 and $S=4$). The same DFT and population analysis gives close to 24.97 on average over the eight irons in the P-cluster normal state (total charge = -4 and $S=4$; geometry obtained from Supplementary Section 4.1 of Ref.~\cite{li2019electronic}).

To establish the scales of \ce{Fe^{2+}} and \ce{Fe^{3+}} electron populations, we performed UCCSD on the 2Fe-2S cluster in both Fe(II)-Fe(II) (total charge = -4) and Fe(III)-Fe(III) (total charge = -2) states. We performed calculations using different basis sets, with and without frozen core, both with high-spin and low-spin, and obtained the average \ce{Fe^{3+}} and \ce{Fe^{2+}} populations and standard deviations: $24.53\pm0.04$ and $24.83\pm0.02$, respectively. All the calculations used the geometry from supplementary Table 1 of Ref.~\cite{sharma2014low}. For \ce{Fe^{2.5+}}, we first performed high-spin ($S=9/2$) calculations with a total charge of -3 on the same geometry. Due to the small geometric asymmetry, the populations on the two irons are not exactly the same. To make the two irons strictly equivalent, we additionally optimized the geometry using B3LYP with ddCOSMO, and then performed UCCSD with a total charge of -3 and $S=9/2$, both with enforced C2h symmetry. The resulting population statistics over all the \ce{Fe^{2.5+}} data is $24.66\pm0.04$. These values are then scaled by the ratio between the DFT P-cluster charge and the 2Fe-2S charge (24.97/24.94) to better connect to FeMo-co oxidation states.

\subsection{Nucleus electron density sensitivity}
In general, the electron density values at the Fe nuclei are sensitive to the computational basis and active space sizes. However, we found that the changes correspond to constant shifts and scaling factors, and thus do not change the ordering between irons. We demonstrate this in Supplementary Figure~\ref{fig:sm-nuc-rho}.

\begin{figure}
    \centering
    \includegraphics[width=0.8\linewidth]{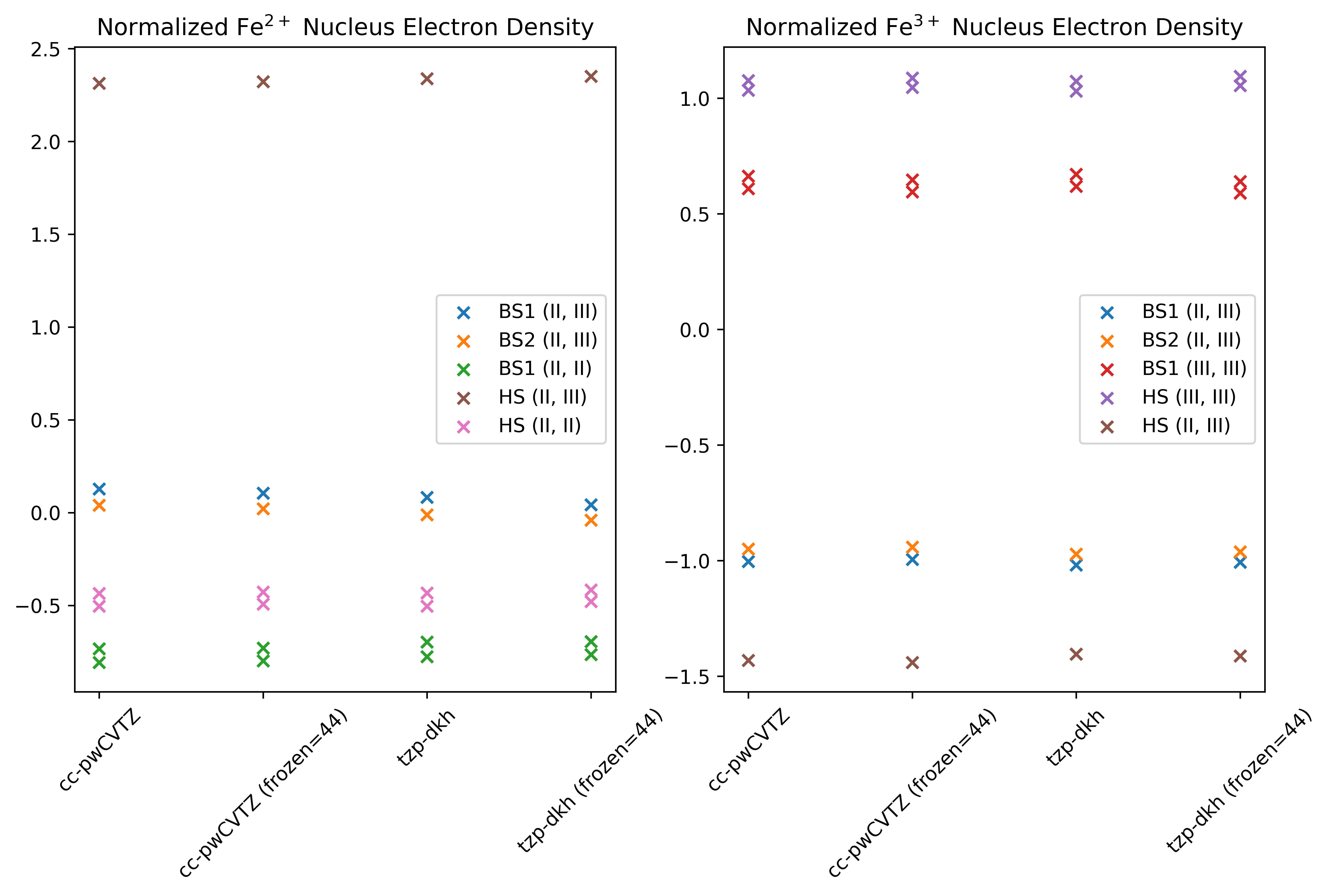}
    \caption{Electron density values for \ce{Fe^{2+}} and \ce{Fe^{3+}} at the iron nuclei in 2Fe-2S clusters from UCCSD calculations in the high spin state (labeled ``HS") or broken spin symmetry states (labeled ``BS$n$" with $n$ indicating the iron that is spin-flipped), with different iron basis sets (def2-SVP basis fixed for other atoms), with and without 44 frozen core orbitals, and different total charges (-2 for ``III, III", -3 for ``II, III" and -4 for ``II, II"). The values are normalized by subtracting the mean and scaled by the standard deviation to show only the iron ordering. }
    \label{fig:sm-nuc-rho}
\end{figure}

We further checked if details of the relativistic treatment, in particular, the relativistic picture change, can affect the iron ordering. We performed the test on FeMo-co using broken-symmetry DFT. We used the same ORCA input as above while setting the dielectric constant for the implicit solvent to 4.0 to represent the protein environment. The DFT was first converged for high spin ($S_z=35/2$) and then spin-flipped to generate initial guesses for symmetry-broken solutions targeting $S_z=3/2$. Calculations were performed on a DFT-optimized geometry previously used for computing M\"ossbauer isomer shifts~\cite{bjornsson2017revisiting}. We focused on a few low-lying spin isomers, namely BS7-A, BS7-B, BS7-C, and BS8-E. Again, the effect of the picture change  only introduces a global shift and scaling, as shown in Supplementary Figure~\ref{fig:sm-nuc-rho-dft}.

\begin{figure}
    \centering
    \includegraphics[width=0.5\linewidth]{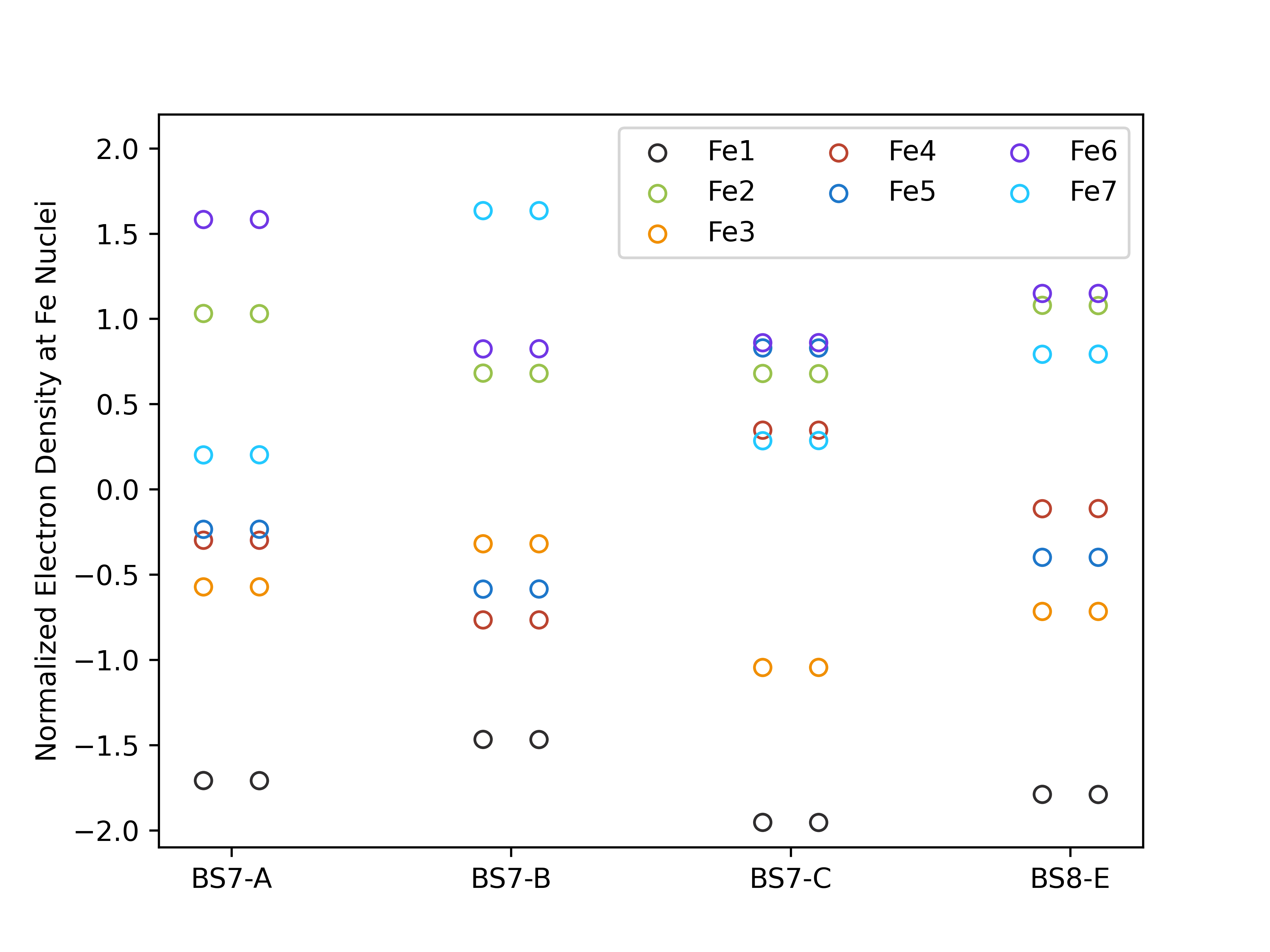}
    \caption{Electron density values at the iron nuclei in FeMo-co from broken-symmetry DFT calculations, with and without the relativistic picture change. The values are normalized by subtracting the mean and scaled by the standard deviation to show only the iron ordering. }
    \label{fig:sm-nuc-rho-dft}
\end{figure}

\subsection{Mo electron population scale}
\label{sec:sm-mo-ox-calib}

To establish the scale of the \ce{Mo^{3+}} electron population, we performed BS-DFT (with the B3LYP functional~\cite{becke1988density,lee1988development,becke1993new}), UHF, and UHF/UCCSD for $\ce{[MoFe3S4(SH)6]^{3-}}$ ($S=3/2$), which is believed to have the Mo(III) oxidation state~\cite{cook1985electronic,bjornsson2014identification}. We use the same basis sets for Mo, Fe, S, and H as those used for the construction of LLDUC model~\cite{li2019electronic-femoco}. UCCSD was performed with core orbitals frozen and using density fitting with an even tempered Gaussian density fitting basis as implemented in PySCF~\cite{sun2018pyscf}. The computed Meta-L\"owdin populations without Rydberg contributions and the comparison with electron populations in FeMo-co are listed in Supplementary Table~\ref{tab:sm-mo-oxi}. We find that at the UCCSD level of theory, the electron populations on Mo in $\ce{[MoFe3S4(SH)6]^{3-}}$ and the FeMo-co model are very similar.

\begin{table}[!htbp]
    \small
    \centering
    \caption{The Meta-L\"owdin populations without Rydberg contributions at the Fe and Mo sites for $\ce{[MoFe3S4(SH)6]^{3-}}$ and BS7-C, BS8-E, and BS8-F states for the FeMo-co (404o, 277e) active space model.}
    \begin{tabular}{
        >{\centering\arraybackslash}p{1.5cm}|
        >{\centering\arraybackslash}p{1.2cm}|
        >{\centering\arraybackslash}p{1.5cm}|
        >{\centering\arraybackslash}p{0.85cm}
        >{\centering\arraybackslash}p{0.85cm}
        >{\centering\arraybackslash}p{0.85cm}
        >{\centering\arraybackslash}p{0.85cm}
        >{\centering\arraybackslash}p{0.85cm}
        >{\centering\arraybackslash}p{0.85cm}
        >{\centering\arraybackslash}p{0.85cm}
        >{\centering\arraybackslash}p{0.85cm}
    }
    \hline\hline
    \multicolumn{2}{c|}{model} & theory & Mo & Fe1 & Fe2 & Fe3 & Fe4 & Fe5 & Fe6 & Fe7 \\
    \hline
\multicolumn{2}{c|}{\multirow{3}{*}{$\ce{[MoFe3S4(SH)6]^{3-}}$}}
&   UKS &    40.75 &    25.04 &    25.04 &    24.97 &          &          &          &          \\
\multicolumn{2}{c|}{} &   UHF &    40.44 &    24.72 &    24.19 &    24.66 &          &          &          &          \\
\multicolumn{2}{c|}{} & UCCSD &    40.60 &    24.84 &    24.75 &    24.73 &          &          &          &          \\
    \hline
    \multirow{3}{*}{FeMo-co}
&  BS7-C & \multirow{3}{*}{UCCSD} &    40.58 &    24.72 &    24.67 &    24.75 &    24.65 &    24.76 &    24.81 &    24.88 \\
&  BS8-E &       &    40.58 &    24.73 &    24.61 &    24.71 &    24.64 &    24.81 &    24.79 &    24.84 \\
&  BS8-F &       &    40.58 &    24.73 &    24.69 &    24.63 &    24.64 &    24.78 &    24.79 &    24.88 \\
    \hline\hline
    \end{tabular}
    \label{tab:sm-mo-oxi}
\end{table}

\section{Cluster geometry dependence}
\label{sec:sm-geom}

The QM/MM MD studies primarily studied the effect of protein fluctuations. We further 
use broken-symmetry DFT to check the dependence of the energies, iron electron populations, and nucleus electron densities on the FeMo-co cluster geometry. We used the same broken-symmetry DFT protocol as above. Two geometries were considered, a crystal structure cutout (which was used to derive the LLDUC model), and a DFT-optimized geometry (using the TPSSh functional), both taken from Section 4 of the supporting information of Ref.~\cite{bjornsson2017revisiting}.

\begin{figure}
    \centering
    \includegraphics[width=0.5\linewidth]{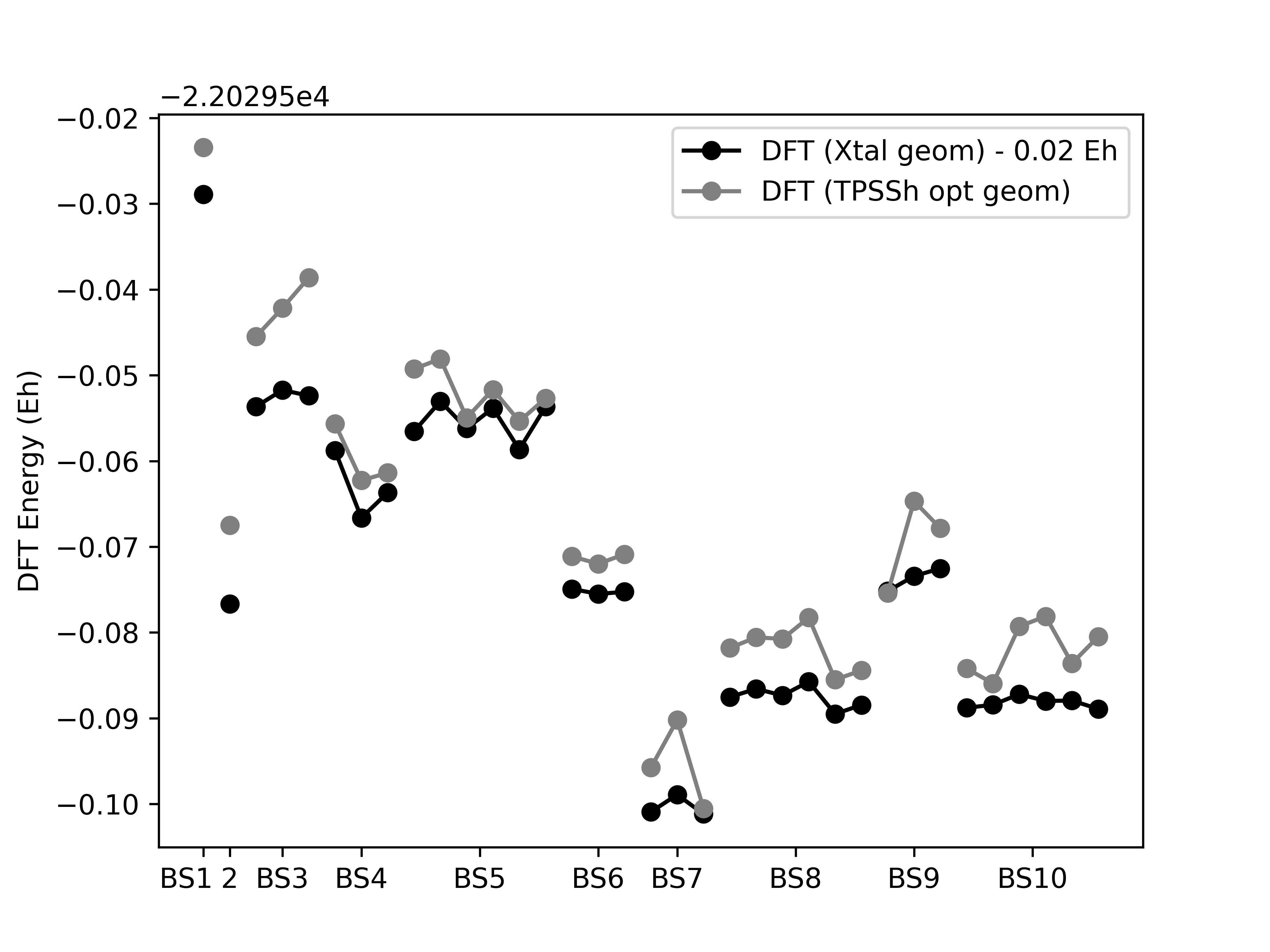}
    \caption{Cluster geometry dependence of BS-DFT energies of spin isomers.}
    \label{fig:sm-dft-energy-geom}
\end{figure}

As shown in Supplementary Figure~\ref{fig:sm-dft-energy-geom}, the qualitative energy landscape of spin isomers is consistent between the two geometries, while geometry optimization on the BS7-C (BS7-235) potential energy surface as performed in Ref.~\cite{bjornsson2017revisiting} increases the energy gap between BS7-C and other isomers.

We further examine the iron ordering dependence on geometry, according to the population and nuclear electron density. As shown in Supplementary Figure~\ref{fig:sm-dft-geom-pop}, the general ordering is consistent between the two geometries, particularly in the BS7-C isomer, used for oxidation assignments previously~\cite{bjornsson2017revisiting}.

\begin{figure}
    \centering
    \includegraphics[width=\linewidth]{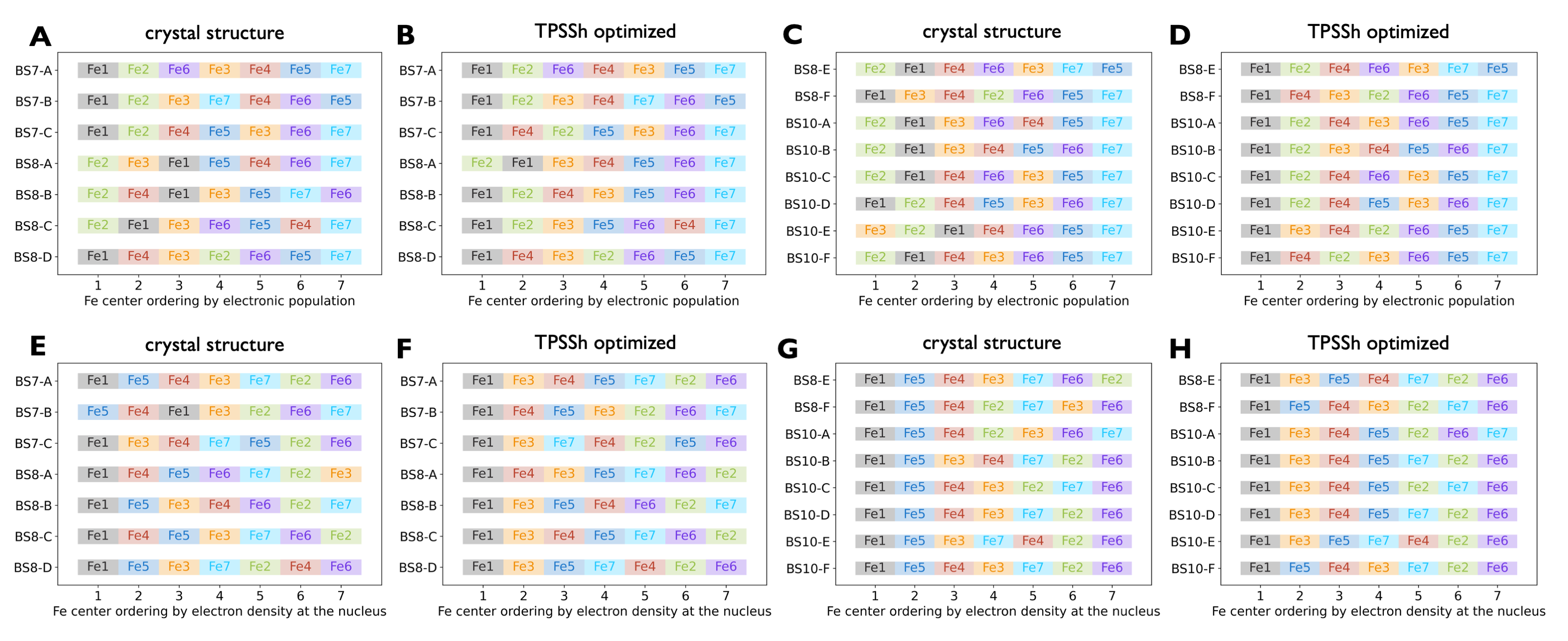}
    \caption{Geometry dependence of the Fe center ordering: \textbf{A,B} Fe center ordering by meta-L\"owdin populations for BS7 and BS8-A/B/C/D, computed using the crystal and TPSSh optimized structures, respectively.
    \textbf{C,D} Fe center ordering by meta-L\"owdin populations for BS8-E/F and BS10, computed using the crystal and TPSSh optimized structures, respectively.
    \textbf{E,F} Fe center ordering by electron density at the nucleus for BS7 and BS8-A/B/C/D, computed using the crystal and TPSSh optimized structures, respectively.
    \textbf{G,H} Fe center ordering by electron density at the nucleus for BS8-E/F and BS10, computed using the crystal and TPSSh optimized structures, respectively.
    }
    \label{fig:sm-dft-geom-pop}
\end{figure}

\section{Model Hamiltonians and qualitative electronic structure}

\subsection{Model for mixed valence 2Fe-2S clusters}
\label{sec:sm-model-dimer}

We use a multi-orbital Kanamori model to examine the stabilization
of spin states in mixed-valence [2Fe-2S] clusters.
The many-body Hamiltonian reads
\begin{align}
    H &= \sum_{i}\sum_{\mu}\sum_{\sigma} \varepsilon_{i\mu} n_{i\mu\sigma} +
    \sum_{i\neq j}\sum_{\mu\nu}\sum_{\sigma} t_{ij}^{\mu\nu} c_{i\mu\sigma}^\dagger c_{i\nu\sigma} \nonumber\\ &+
    \sum_{i} \Big(U\sum_{\mu} n_{i\mu\uparrow} n_{i\mu\downarrow}
    +(U-2J)\sum_{\mu<\nu}\sum_{\sigma\sigma'} n_{i\mu\sigma} n_{i\nu\sigma'} - J\sum_{\mu\neq\nu}S_{i\mu}\cdot S_{i\nu}+
    J\sum_{\mu\neq\nu} c_{i\mu\uparrow}^\dagger 
    c_{i\mu\downarrow}^\dagger c_{i\nu\downarrow} c_{i\nu\uparrow}\Big) \;,
\end{align}
where orbital energies $\varepsilon$ and hopping integrals $t$ are derived from DFT calculations for a model system taken from Ref.~\citenum{sharma2014low}, and the Hund's coupling $J$ is set to 0.3 eV. By varying the on-site Coulomb repulsion $U$ from 3 eV to 10 eV, we observe stabilization of all possible spin states, as shown in Supplementary Figure~\ref{fig:sm-dimer}.
The result for $U=3$~eV is consistent with the DMRG calculation for a 32-orbital active space model~\cite{sharma2014low}, which predicts an $S=1/2$ ground state, however increasing $U$ leads to intermediate- and high-spin ground states. 

Typical $U$ values for Fe range from 3 to 6 eV. While double exchange in mixed valence dimers is sometimes used to suggest a ferromagnetic alignment of neighbouring Fe centers, we see that the ground-state spin configuration is governed by a delicate balance among competing interactions, namely, antiferromagnetic coupling, double exchange, and vibronic coupling, etc. In FeMo-co, the delocalization over multiple Fe centers further complicates a deduction of the relative alignment of neighbouring Fe centers simply from the mixed valence nature.

\begin{figure}
    \centering
    \includegraphics[width=.5\linewidth]{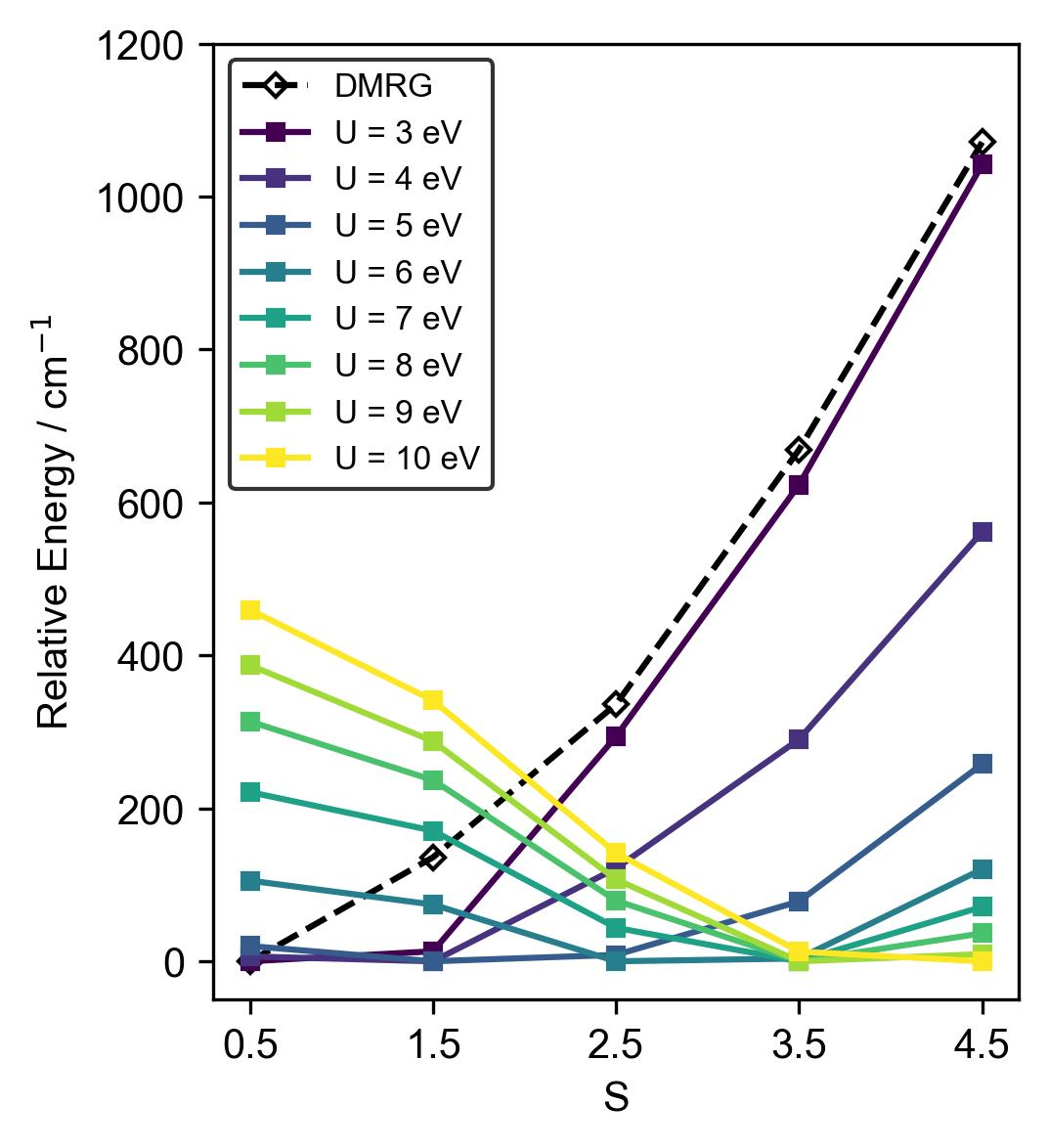}
    \caption{
    Relative energies of different spin states obtained from a 10-orbital model Hamiltonian for on-site Coulomb repulsion $U$ ranging from 3 eV to 10 eV.
    The reference DMRG result is taken from Ref.~\citenum{sharma2014low} for a 32-orbital active space.}
    \label{fig:sm-dimer}
\end{figure}

\subsection{Toy Heisenberg model for FeMo-co}
\label{sec:sm-model-heis}

\subsubsection{Hierarchy of energies in the Heisenberg-like models}
\label{sec:sm-hierarchy}

We first briefly recall the theoretical terminology used to describe different excitations and the theoretical basis of spontaneous symmetry breaking in Heisenberg-like models, as first elucidated by Anderson's treatment of the antiferromagnetic state~\cite{anderson1952approximate,tasaki2019long}.

It is known that the ground-states of spin systems on certain lattices can have long-range spin-order and that such ground-states can be described by a spontaneously symmetry broken state that is not an eigenstate of $S^2$. The basis for this was first described by Anderson. In such systems, there is a set of low-lying excitations whose energy spacing goes like $1/V$, where $V$ is the volume of the system. As $V\to \infty$, this gap vanishes, and it is therefore valid to consider a symmetry broken eigenstate that is a mixture of such excitations. To give rise to a long-range spin order (such as antiferromagnetic long-range order) the set of states that are mixed together must have the same correlations as $V \to \infty$. This set of states is known as the `tower of states' and the symmetry broken ground-state lies in the manifold of this tower of states.

There is another set of excitations in the spin system that is associated with the Goldstone mode, or spin-wave, excitations. These are also gapless in a system with spontaneous symmetry breaking, but they can be distinguished from the above tower of states because the gap scales like $1/L$, the linear dimension of the problem. These excitations are referred to as magnons.

As discussed in Ref.~\cite{tasaki2019long}, the magnon excitations modify the order of the ground-state. They are the excitations that are measured in spectroscopy. The tower of states can be better thought of as a degeneracy of the ground-state. By mixing the tower of states to obtain a spontaneously symmetry broken state, one obtains a physically realistic ground-state.

In this language, in FeMo-co the manifold of spin-pure eigenstates that mix to form a given broken symmetry isomer is the `tower of states', while the gap between spin isomer energies is analogous to (one or more) magnon excitations.

\subsubsection{Numerical simulations}

\textbf{Hamiltonian.} To achieve a better qualitative understanding of the spin coupling in FeMo-co and the consequences of the broken-symmetry treatment, we consider a toy model for FeMo-co with a Heisenberg type interaction
\begin{equation}
    \hat{H} = \sum_{\langle ij \rangle}^n J_{ij}\ \hat{S}_i\cdot \hat{S}_j + E_0
\end{equation}
where $E_0$ is an energy constant (for fitting purposes), $i, j = 1, 2,\cdots, 7$ representing 7 Fe sites (all with a local spin $S=2$) , and $j=8$ representing 1 Mo site (with a local spin $S=\frac{1}{2}$). In this setup, the charge state is different from the $E_0$ state, and charge fluctuations and electron hopping for Fe sites are ignored. However, the use of a spin model allows the exact solutions to be obtained by simple diagonalization, and this is sufficient to illustrate the hierarchy of states in the model. We further assume a topology with $C_3$ symmetry, with the six independent $J_{ij}$ parameters listed in Supplementary Table~\ref{tab:sm-heis-j}.

\begin{table}[!htbp]
    \small
    \centering
    \caption{The independent $J_{ij}$ parameters used in the toy Heisenberg model for FeMo-co.}
    \begin{tabular}{
        >{\centering\arraybackslash}p{3cm}|
        >{\centering\arraybackslash}p{1.5cm}|
        >{\centering\arraybackslash}p{0.8cm}
        >{\centering\arraybackslash}p{0.8cm}
        >{\centering\arraybackslash}p{0.8cm}
        >{\centering\arraybackslash}p{0.8cm}
        >{\centering\arraybackslash}p{0.8cm}
        >{\centering\arraybackslash}p{0.8cm}
    }
    \hline\hline
    type & notation & \multicolumn{6}{c}{$(i, j)$ pairs} \\
    \hline
            left terminal & $J_{\mathrm{LT}}$ & (1,2) & (1,3) & (1,4) &       &       &       \\
             left cubane  & $J_{\mathrm{LC}}$ & (2,3) & (2,4) & (3,4) &       &       &       \\
             right cubane & $J_{\mathrm{RC}}$ & (5,6) & (5,7) & (6,7) &       &       &       \\
      right (Mo) terminal & $J_{\mathrm{RT}}$ & (5,8) & (6,8) & (7,8) &       &       &       \\
          central (axial) & $J_{\mathrm{CA}}$ & (2,6) & (3,7) & (4,5) &       &       &       \\
          central (cross) & $J_{\mathrm{CC}}$ & (2,5) & (2,7) & (3,5) & (3,6) & (4,6) & (4,7) \\
    \hline\hline
    \end{tabular}
    \label{tab:sm-heis-j}
\end{table}

\textbf{Parameter fitting.} We determine the $J_{ij}$ and $E_0$ parameters using least squares fitting of the UCCSD(T) energies (and alternately UHF or UCCSD energies) computed in the LLDUC model for the 35 spin isomers. To compute the Heisenberg model Hamiltonian energy for fitting purposes, we assume that, regardless of the level of theory (i.e. UHF or UCC), every BS state corresponds to a pure Ising state. This means, 
we set $\langle \hat{S}_i\cdot \hat{S}_j\rangle = 4.0$ and $-6.0$ for FM and AFM Fe--Fe coupling respectively, and $\langle \hat{S}_i\cdot \hat{S}_j\rangle = 1.0$ and $-1.5$ for FM and AFM Fe--Mo coupling respectively. The FM or AFM coupling is determined using the 1- and 2-particle density matrices of the UHF states. We list the Heisenberg model parameters determined using different levels of theory in Supplementary Table~\ref{tab:sm-heis-param}. The energy fitting error is shown in Supplementary Figure~\ref{fig:sm-heis-ener}. The small energy fluctuation within each BS family for the Heisenberg model is due to the Fe--Mo coupling sign mismatch between the BS initial guess and the actual UHF states (this only appears in a few spin isomers). Since the Heisenberg model assumes $C_3$ symmetry, the fitting error mostly measures the variation of the LLDUC model energies within each of the BS families.

\begin{table}[!htbp]
    \small
    \centering
    \caption{The Heisenberg model $J_{ij}$ (in $\mathrm{cm}^{-1}$) and $E_0$ (in mHa, shifted by $-22140$ Hartrees) parameters determined using the UHF, UCCSD, or UCCSD(T) energies computed in the LLDUC model. Root mean square error (RMSE) of the fitting is given in mHa.}
    \begin{tabular}{
        >{\centering\arraybackslash}p{1.8cm}|
        >{\centering\arraybackslash}p{0.9cm}
        >{\centering\arraybackslash}p{0.9cm}
        >{\centering\arraybackslash}p{1.1cm}
        >{\centering\arraybackslash}p{0.9cm}
        >{\centering\arraybackslash}p{0.9cm}
        >{\centering\arraybackslash}p{0.9cm}
        >{\centering\arraybackslash}p{1.2cm}
        >{\centering\arraybackslash}p{1.1cm}
    }
    \hline\hline
    theory & $J_{\mathrm{LT}}$ & $J_{\mathrm{LC}}$ & $J_{\mathrm{RC}}$ & $J_{\mathrm{RT}}$ & $J_{\mathrm{CA}}$ & $J_{\mathrm{CC}}$ & $E_0$ & RMSE \\
    \hline
       UHF &   219.03 &   304.86 &    54.29 &   247.49 &    77.91 &    13.43 &    41.83 &    24.52 \\
     UCCSD &   564.78 &   538.70 &   -14.90 &   135.02 &   185.82 &    48.45 &  -242.23 &    44.99 \\
  UCCSD(T) &   581.95 &   586.48 &    32.16 &   117.97 &   168.21 &    23.91 &  -295.32 &    37.01 \\
    \hline\hline
    \end{tabular}
    \label{tab:sm-heis-param}
\end{table}

\begin{figure}[!htbp]
  \centering
  \includegraphics[width=0.85\linewidth]{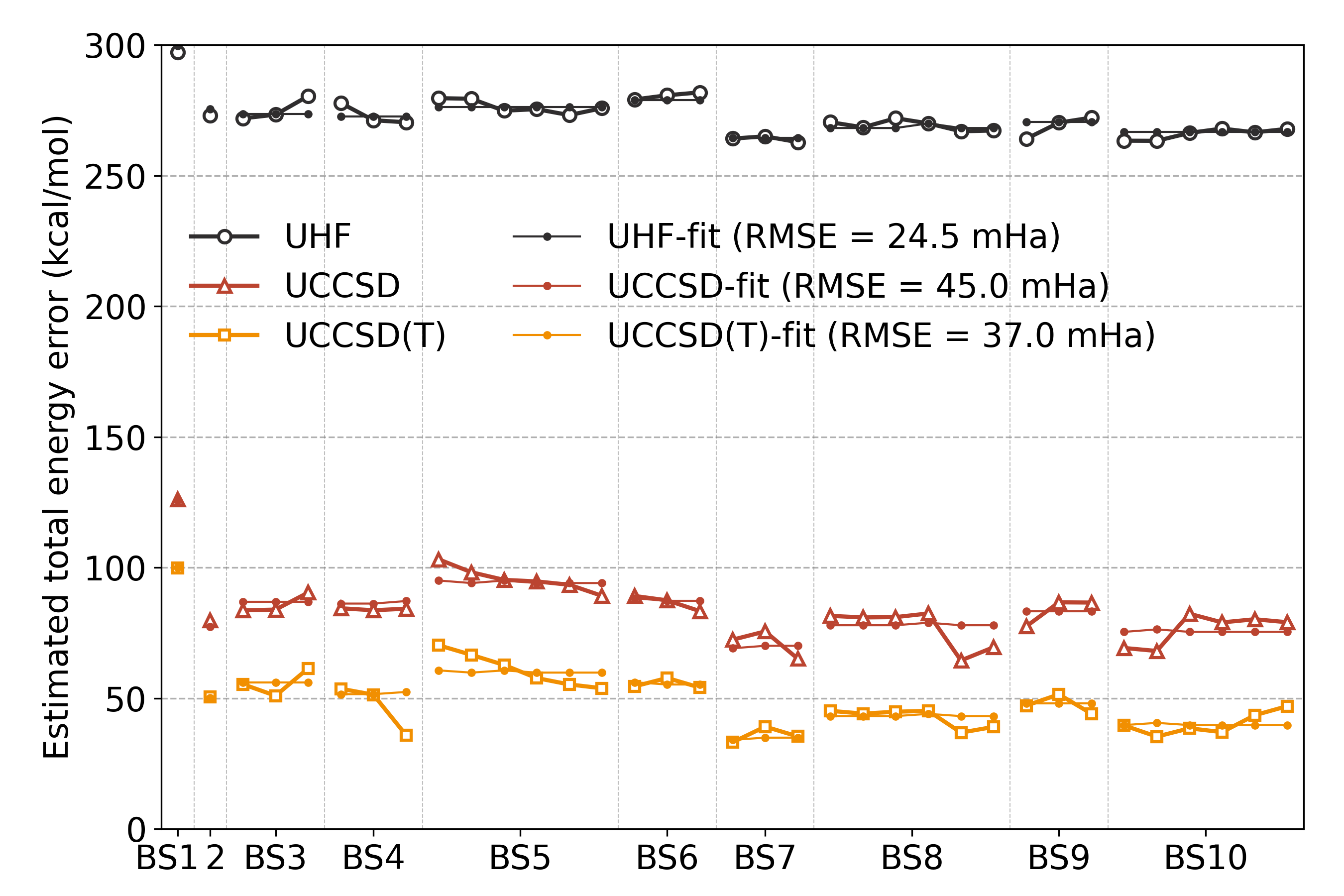}
  \caption{Comparison between the predicted and the actual energies, for the Heisenberg model fitted using UHF, UCCSD, and UCCSD(T) energies.}
  \label{fig:sm-heis-ener}
\end{figure}

\textbf{Ground and excited states.} In the following, we focus on solving for the ground and excited states of the model with parameters fitted to UCCSD(T) energies. In Supplementary Table~\ref{tab:sm-heis-eigen} we list the energies of the lowest eight eigenstates found using state-averaged spin-adapted DMRG (with maximal bond dimension $D = 800$, which produces the exact result for this problem) for $S=1/2,3/2,\cdots, 13/2$, respectively. The ground state energy for the model is -22140.335969 Hartrees, with $S=3/2$. The $S=1/2$ and $S=5/2$ states are found to be 0.63 and 0.39 mHa higher in energy, respectively, indicating a small spin gap in the model. Note that as the Heisenberg model does not correspond to the same charge state as the E0 state, the spin of the ground-state here does not imply anything about the true ground-state spin of FeMo-co in the LLDUC model.

\begin{table}[!htbp]
    \small
    \centering
    \caption{The Heisenberg model eigenstate energies (in mHa, relative to the $S=3/2$ ground state energy -22140.335969 Hartrees) determined using state-averaged spin-adapted DMRG.}
    \begin{tabular}{
        >{\centering\arraybackslash}p{2cm}|
        >{\centering\arraybackslash}p{0.9cm}
        >{\centering\arraybackslash}p{0.9cm}
        >{\centering\arraybackslash}p{0.9cm}
        >{\centering\arraybackslash}p{0.9cm}
        >{\centering\arraybackslash}p{0.9cm}
        >{\centering\arraybackslash}p{0.9cm}
        >{\centering\arraybackslash}p{0.9cm}
        >{\centering\arraybackslash}p{0.9cm}
    }
    \hline\hline
    $S$ & \multicolumn{8}{c}{eigenstate energies} \\
    \hline
   1/2 &    0.626 &    0.626 &    0.943 &    1.244 &    1.244 &    1.362 &    1.630 &    1.630 \\
   3/2 &    0.000 &    0.246 &    0.246 &    0.747 &    1.058 &    1.058 &    1.105 &    1.432 \\
   5/2 &    0.389 &    0.389 &    0.892 &    1.193 &    1.301 &    1.345 &    1.345 &    1.528 \\
   7/2 &    1.220 &    1.220 &    1.292 &    2.238 &    2.238 &    2.290 &    2.290 &    2.726 \\
   9/2 &    2.500 &    2.666 &    2.775 &    2.775 &    3.588 &    3.588 &    3.675 &    4.421 \\
  11/2 &    4.818 &    4.818 &    5.198 &    5.198 &    5.249 &    5.338 &    5.536 &    5.654 \\
  13/2 &    8.046 &    8.046 &    8.481 &    8.481 &    8.542 &    9.025 &    9.025 &    9.115 \\
    \hline\hline
    \end{tabular}
    \label{tab:sm-heis-eigen}
\end{table}

\begin{table}[!htbp]
    \small
    \centering
    \caption{The energy and $\langle S_z \rangle$ expectation at Fe and Mo sites of the low-energy Heisenberg model eigenstates determined using non-spin-adapted DMRG. Energies are in mHa relative to the $S=3/2$ ground state energy.}
    \begin{tabular}{
        >{\centering\arraybackslash}p{1.5cm}|
        >{\centering\arraybackslash}p{0.9cm}|
        >{\centering\arraybackslash}p{0.85cm}
        >{\centering\arraybackslash}p{0.85cm}
        >{\centering\arraybackslash}p{0.85cm}
        >{\centering\arraybackslash}p{0.85cm}
        >{\centering\arraybackslash}p{0.85cm}
        >{\centering\arraybackslash}p{0.85cm}
        >{\centering\arraybackslash}p{0.85cm}
        >{\centering\arraybackslash}p{0.85cm}
    }
    \hline\hline
    $(S,S_z)$ & energy & Fe1 & Fe2 & Fe3 & Fe4 & Fe5 & Fe6 & Fe7 & Mo8 \\
    \hline
(3/2,1/2) &    0.000 &   0.44 &  -0.15 &  -0.15 &  -0.15 &   0.20 &   0.20 &   0.20 &  -0.10 \\
(3/2,1/2) &    0.246 &   0.40 &  -0.07 &  -0.31 &  -0.01 &   0.05 &   0.13 &   0.41 &  -0.10 \\
(3/2,1/2) &    0.246 &   0.40 &  -0.19 &   0.05 &  -0.25 &   0.34 &   0.26 &  -0.02 &  -0.10 \\
\hline
(3/2,3/2) &    0.000 &   1.32 &  -0.44 &  -0.44 &  -0.44 &   0.60 &   0.60 &   0.60 &  -0.30 \\
(3/2,3/2) &    0.246 &   1.21 &   0.16 &  -0.71 &  -0.60 &   0.84 &  -0.07 &   0.98 &  -0.30 \\
(3/2,3/2) &    0.246 &   1.21 &  -0.93 &  -0.05 &  -0.17 &   0.32 &   1.23 &   0.19 &  -0.30 \\
    \hline\hline
    \end{tabular}
    \label{tab:sm-heis-bs-spin}
\end{table}

\textbf{Approximate solutions.} Because the gap between different pure $S$ eigenstates is small, then in an approximate MPS (which is not spin-adapted), different $S$ eigenstates will mix for a given $S_z$ state. This mixing is expected to persist until we reach a bond dimension compatible with resolving the spin gap. Thus we study the deviation of $\langle S^2\rangle$ from $S(S+1)$ in approximate broken-symmetry solutions using non-spin-adapted DMRG with a series of low MPS bond dimensions. The results are shown in Supplementary Figure~\ref{fig:sm-heis-bs}. We find that the amount of spin contamination (measured by the difference between $\langle S^2\rangle$ and $S(S+1)$) decreases rapidly once the energy error is close to or smaller than the spin gap, which is 0.389 mHa in this model. When the MPS bond dimension is large enough, the pure-spin is recovered for the ground state.

\begin{figure}[!htbp]
  \centering
  \includegraphics[width=0.85\linewidth]{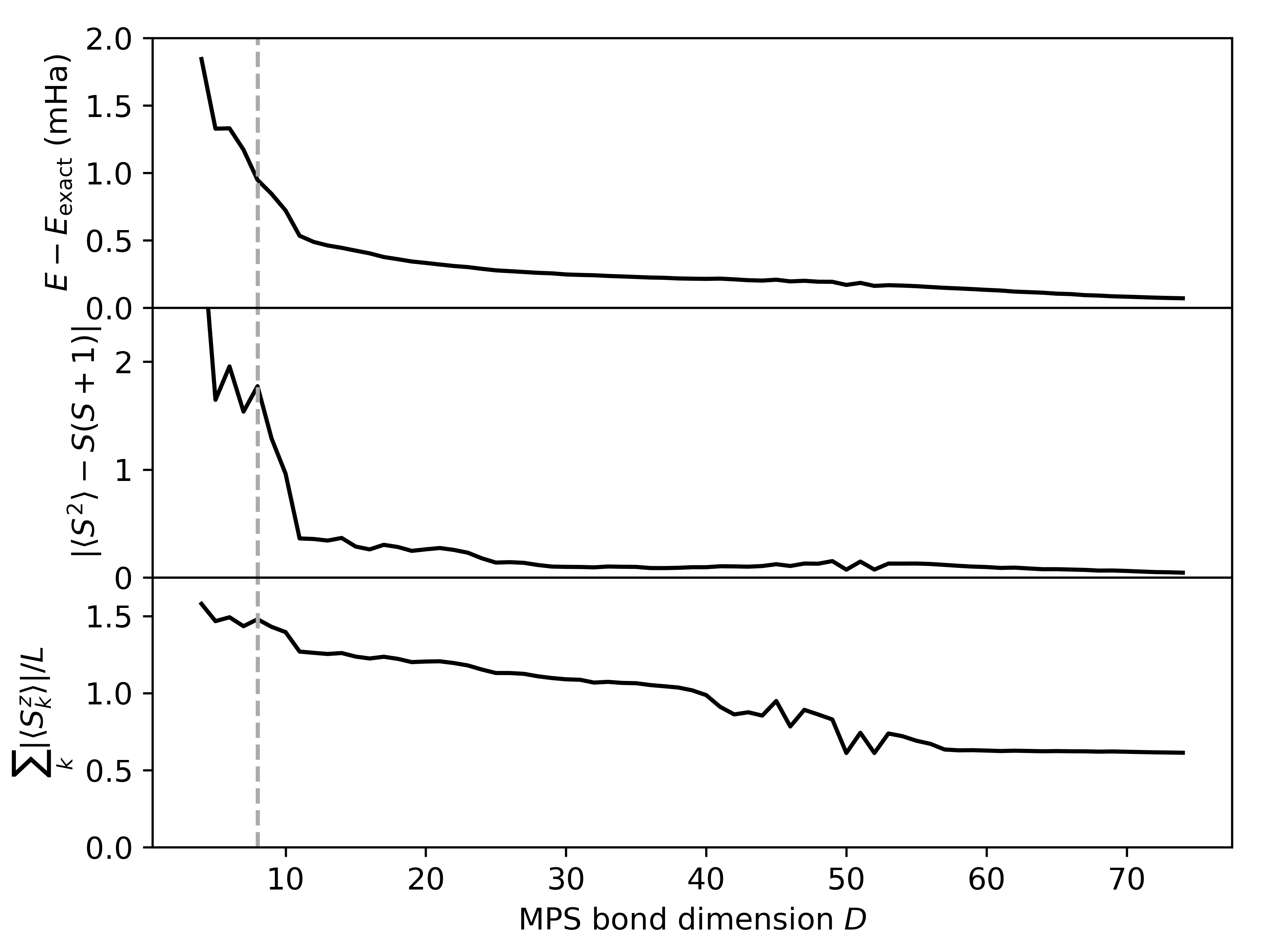}
  \caption{The energy error, difference between $\langle S^2\rangle$ and $S(S+1)$ (with $S=3/2$), and average $\langle S_z\rangle$ on each site for the Heisenberg model ground state computed using non-spin-adapted DMRG at finite MPS bond dimensions. The dashed gray line indicates the MPS bond dimension with 1 mHa energy error.}
  \label{fig:sm-heis-bs}
\end{figure}

\begin{figure}[!htbp]
  \centering
  \includegraphics[width=0.92\linewidth]{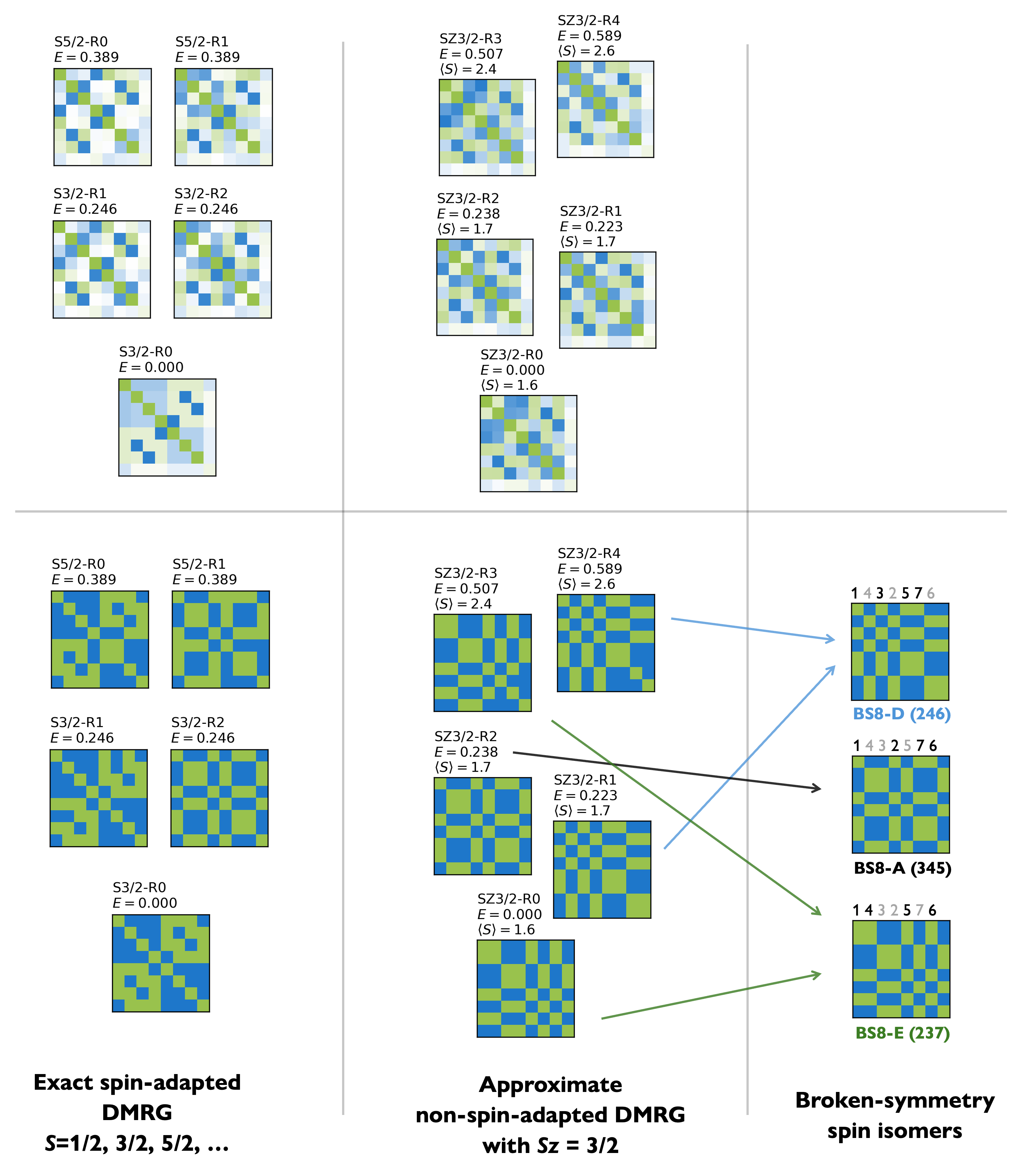}
  \caption{Spin correlation (upper panel) and its sign (lower panel) between Fe and Mo sites in low-energy states computed using exact spin-adapted DMRG and exact and approximate non-spin-adapted DMRG. Site ordering in spin correlation plots is the computational ordering: Fe1-Fe4-Fe3-Fe2-Fe5-Fe7-Fe6-Mo8, to clearly show the one-dimensional entanglement structure. The $i$th eigenstate (counting from zero) of $S=3/2$ and $S_z=3/2$ states are denoted as ``S3/2-R$i$'' and ``SZ3/2-R$i$'', respectively. Energies are in millihartrees, relative to the ground state.}
  \label{fig:sm-heis-spin-corr}
\end{figure}

\textbf{The nature of spin isomers.} Based on the eigenstates for this Heisenberg model, we now briefly discuss the nature of spin isomers in FeMo-co models. Supplementary Figure~\ref{fig:sm-heis-spin-corr} shows the spin correlation computed for the low-energy states, using exact spin-adapted DMRG (SA-DMRG) and
approximate non-spin-adapted DMRG (NSA-DMRG, with small MPS bond dimension).
When we use approximate NSA-DMRG, two types of symmetry breaking appear: (i) total spin symmetry, as can be seen by the non-half-integer $\langle S\rangle$ values; and (ii) $C_3$ spatial symmetry, as the energy of the $C_3$-equivalent BS8-A, BS8-D, and BS8-E states are no longer degenerate. The origin of (i) is the fact that in an approximate non-spin-adapted theory, states with different $S$, but with otherwise similar correlations, mix to overcome the representation limitation of the ansatz. The origin of 
(ii) is the one-dimensional asymmetric ordering of $C_3$-equivalent sites in DMRG (but note that in FeMo-co this $C_3$ symmetry is actually broken anyway). 

The relationship between the correlations in the spin-adapted eigenstates, the approximate non-spin-adapted eigenstates, and the mean-field broken symmetry spin isomers can be seen from the correlation functions and the colors. While the correlations are slightly hard to interpret in the pure spin eigenstates due to mixing from the exact $C_3$ symmetry, one can see some of the BS8 correlations already (e.g. BS8A). These become very clear in the non-spin-adapted approximate eigenstates with fixed $S_z$, which are then clearly connected to the mean-field spin isomers. Thus the approximate MPS calculations of spin isomers can be viewed as representing a ``tower'' of pure spin eigenstates with a similar spin coupling pattern, in a small energy window, analogous to the theoretical structure of states in Heisenberg models originally proposed by Anderson~\cite{tasaki2019long}.

\end{document}